\documentclass{aa}  

\usepackage{graphicx}
\usepackage{txfonts}
\usepackage[hyperindex,breaklinks=true, colorlinks, citecolor=blue]{hyperref}   
\usepackage{natbib}

\newcommand{\ud}{\,\mathrm{d}}
\newcommand{\Planck}{\textit{Planck}}

\newcommand{\B}{\boldsymbol{B}}
\newcommand{\nd}{n_\mathrm{d}}
\newcommand{\nH}{n_\mathrm{H}}
\newcommand{\NH}{N_\mathrm{H}}

\providecommand{\sorthelp}[1]{}

\begin{document}

   \title{On the statistics of the polarized submillimetre emission maps from thermal dust in the turbulent, magnetized, diffuse ISM}
   \titlerunning{Statistics of polarized thermal dust emission}

\author{F.~Levrier\inst{\ref{inst1}} \and J.~Neveu\inst{\ref{inst1},\ref{inst2}} \and E.~Falgarone\inst{\ref{inst1}} \and F.~Boulanger\inst{\ref{inst3}} \and A.~Bracco\inst{\ref{inst3},\ref{inst4},\ref{inst5}} \and T.~Ghosh\inst{\ref{inst6}} \and F.~Vansyngel\inst{\ref{inst3}}}

\institute{{Sorbonne Universit\'e, Observatoire de Paris, Universit\'e PSL, \'Ecole normale sup\'erieure, CNRS, LERMA, F-75005, Paris, France\label{inst1}} \and {Universit\'e Paris-Sud, LAL, UMR 8607, F-91898 Orsay Cedex, France \& CNRS/IN2P3, F-91405 Orsay, France\label{inst2}} \and {Institut d'Astrophysique Spatiale, CNRS, Univ. Paris-Sud, Universit\'e Paris-Saclay, B\^at. 121, 91405 Orsay cedex, France\label{inst3}} \and {Laboratoire AIM, IRFU/Service d'Astrophysique - CEA/DSM - CNRS - Universit\'e Paris Diderot, B\^at. 709, CEA-Saclay, F-91191 Gif-sur-Yvette Cedex, France\label{inst4}} \and {Nordita, KTH Royal Institute of Technology and Stockholm University, Roslagstullsbacken 23, 10691 Stockholm, Sweden\label{inst5}} \and {School of Physical Sciences, National Institute of Science Education and Research, HBNI, Jatni 752050, Odissa, India\label{inst6}}}

   \date{Received 19 October 2017; accepted 23 February 2018}

  \abstract
   {The interstellar medium (ISM) is now widely recognized to display features ascribable to magnetized turbulence. With the public release of {\Planck} data and the current balloon-borne and ground-based experiments, the growing amount of data tracing the polarized thermal emission from Galactic dust in the submillimetre provides choice diagnostics to constrain the properties of this magnetized turbulence.}
   {We aim to constrain these properties in a statistical way, focusing in particular on the power spectral index $\beta_B$ of the turbulent component of the interstellar magnetic field in a diffuse molecular cloud, the Polaris Flare.}
   {We present an analysis framework which is based on simulating polarized thermal dust emission maps using model dust density (proportional to gas density $\nH$) and magnetic field cubes, integrated along the line of sight, and comparing these statistically to actual data. The model fields are derived from fractional Brownian motion (fBm) processes, which allow a precise control of their one- and two-point statistics. The parameters controlling the model are (1)-(2) the spectral indices of the density and magnetic field cubes, (3)-(4) the RMS-to-mean ratios for both fields, (5) the mean gas density, (6) the orientation of the mean magnetic field in the plane of the sky (POS), (7) the dust temperature, (8) the dust polarization fraction, and (9) the depth of the simulated cubes. We explore the nine-dimensional parameter space through a Monte-Carlo Markov Chain analysis, which yields best-fitting parameters and associated uncertainties.}
   {We find that the power spectrum of the turbulent component of the magnetic field in the Polaris Flare molecular cloud scales with wavenumber as $k^{-\beta_B}$ with a spectral index $\beta_B=2.8\pm 0.2$. It complements a uniform field whose norm in the POS is approximately twice the norm of the fluctuations of the turbulent component, and whose position angle with respect to the North-South direction is $\chi_0 \approx -69^\circ$. The density field $\nH$ is well represented by a log-normally distributed field with a mean gas density $\left\langle\nH\right\rangle \approx 40\,\mathrm{cm}^{-3}$, a fluctuation ratio $\sigma_{\nH}/\langle\nH\rangle\approx 1.6$, and a power spectrum with an index $\beta_n=1.7^{+0.4}_{-0.3}$. We also constrain the depth of the cloud to be $d\approx 13\,\mathrm{pc}$, and the polarization fraction $p_0 \approx 0.12$. The agreement between the {\Planck} data and the simulated maps for these best-fitting parameters is quantified by a $\chi^2$ value that is only slightly larger than unity.}
   {We conclude that our fBm-based model is a reasonable description of the diffuse, turbulent, magnetized ISM in the Polaris Flare molecular cloud, and that our analysis framework is able to yield quantitative estimates of the statistical properties of the dust density and magnetic field in this cloud.}


   \maketitle

\section{Introduction}

In recent years, a number of experiments have dramatically increased the amount of data pertaining to the polarized thermal emission from Galactic dust in the submillimetre~\citep[e.g.][]{matthews-et-al-2009,ward-thompson-et-al-2009,dotson-et-al-2010,bierman-et-al-2011,vaillancourt-matthews-2012,hull-et-al-2014,koch-et-al-2014,fissel-et-al-2016}. Chief among these is {\Planck}\,\footnote{{\Planck} (\url{http://www.esa.int/Planck}) is a project of the European Space Agency (ESA) with instruments provided by two scientific consortia funded by ESA member states and led by Principal Investigators from France and Italy, telescope reflectors provided through a collaboration between ESA and a scientific consortium led and funded by Denmark, and additional contributions from NASA (USA).}, which provided the first full-sky map of this emission, leading to several breakthrough results. It was thus found that the polarization fraction $p$ in diffuse regions of the sky can reach values above 20\%~\citep{planck2014-XIX}, confirming results previously obtained over one fifth of the sky by the {\it Archeops} balloon-borne experiment~\citep{benoit-et-al-2004,ponthieu-et-al-2005}. Furthermore, the polarization fraction is anti-correlated with the local dispersion $\mathcal{S}$ of polarization angles $\psi$~\citep{planck2014-XIX,planck2014-XX}, and the decrease of the maximum observed $p$ with increasing gas column density $\NH$ may be fully accounted for, at the scales probed by {\Planck} (5{\arcmin} at 353\,GHz), by the tangling of the magnetic field on the line of sight (LOS)~\citep{planck2014-XX}. Similar anti-correlations were found by the BLASTPol experiment~\citep{fissel-et-al-2016} at higher angular resolution (a few tens of arcseconds) towards a single Galactic molecular cloud (Vela C). At 10{\arcmin} scales, and over a larger sample of clouds, {\Planck} data showed that the relative orientation of the interstellar magnetic field $\B$ and filamentary structures of dust emission is consistent with simulated observations derived from numerical simulations of sub- or trans-Alfv\'enic MHD turbulence~\citep{planck2015-XXXV}, and starlight polarization data in extinction yield similar diagnostics~\citep{soler-et-al-2016}. In diffuse regions, the preferential alignment of filamentary structures with the magnetic field~\citep{planck2014-XXXII,planck2015-XXXVIII} is linked to the measured asymmetry between the power spectral amplitudes of the so-called E- and B-modes of polarized emission. Finally, measurements of the spatial power spectrum of polarized dust emission showed that it must be taken into account in order to obtain reliable estimates of the cosmological polarization signal~\citep{planck2014-XXX}.

With this wealth of data, we may be able to put constraints on models of magnetized turbulence in the interstellar medium (ISM), provided we can extract the relevant information from polarization maps. Of particular interest are the statistical properties of the Galactic magnetic field (GMF) $\B$. Let us write it as a sum $\B=\B_0+\B_t$ of a uniform, large-scale component $\B_0$, and a turbulent component $\B_t$ with a null spatial average, $\langle\B_t\rangle=\boldsymbol{0}$. The statistical properties in question are then essentially modelled by two quantities, i) the ratio of the turbulent component to the mean, $y_B=\sigma_B/B_0$, where $\sigma_B^2=\langle \B_t^2\rangle$ and $B_0=||\B_0||$, and ii) the spectral index $\beta_B$, which characterizes the distribution of power of $\B_t$ across spatial scales, through the relationship $P(k)\propto k^{-\beta_B}$, where $k$ is the wavenumber and $P(k)$ is the power spectrum\,\footnote{In all generality, several spectral indices may be defined, as one may consider the power spectrum of any one of the three cartesian components of $\B_t$, or of the modulus $|\B_t|$. Assuming that $\B_t$ is isotropic, which we will, all of these spectral indices are identical.}. 

As already mentioned, \cite{planck2015-XXXV} studied the relative orientation between the magnetic field, probed by polarized thermal dust emission, and filaments of matter in and around nearby molecular clouds. They found that this relative orientation changes, from mostly parallel to mostly perpendicular, as the total gas column density $\NH$ increases, which is a trend observed in simulations of trans-Alfv\'enic or sub-Alfv\'enic MHD turbulence~\citep{soler-et-al-2013}. Using the Davis-Chandraskehar-Fermi method~\citep{chandrasekhar-fermi-1953} improved by~\cite{falceta-goncalves-et-al-2008} and~\cite{hildebrand-et-al-2009}, and from their results, we can estimate the ratio $y_B$ to be in the range 0.3-0.7. \cite{planck2014-XXXII} studied that same relative orientation in the diffuse ISM at intermediate and high Galactic latitudes, and their estimate of $y_B$ is in the range 0.6-1.0 with a preferred value at 0.8. These estimates are confirmed in \cite{planck2016-XLIV}, which presents a fit of the distributions of polarization angles and fractions observed by {\Planck} towards the southern Galactic cap. They use a model of the GMF involving a uniform large-scale field $\B_0$ and a small number ($N_l\simeq 7$) of independent ``polarization layers'' on the line of sight, each of which accounts for a fraction $1/N_l$ of the total unpolarized emission. Within each layer, the turbulent component of the magnetic field $\B_t$, which is used to compute synthetic Stokes $Q$ and $U$ maps, is an independent realization of a Gaussian 2D random field with a prescribed spectral index $\beta_B$. Through these fits, they confirm the near equipartition of large-scale and turbulent components of $\B$, with $y_B\simeq 0.9$. They also provide a rough estimate of the magnetic field's spectral index $\beta_B$ in the range 2-3. This work was complemented in~\cite{vansyngel-et-al-2017}, using the same framework, but including observational constraints on the power spectra of polarized thermal dust emission. These authors were able to constrain $\beta_B\simeq 2.5$, an exponent which is compatible with the rough estimate of \cite{planck2016-XLIV}, close to that measured for the total intensity of the dust emission. We note that their exploration of the parameter space does not allow for an estimation of the uncertainty on $\beta_B$.

In~\cite{planck2014-XXXII},~\cite{planck2016-XLIV} and~\cite{vansyngel-et-al-2017}, the description of structures, in both dust density and magnetic field, along the LOS is reduced to the bare minimum, while statistical properties in the plane of the sky (POS) are modelled through $y_B$ and $\beta_B$. Orthogonal approaches have also been pursued~\citep[e.g.][]{miville-deschenes-et-al-2008,odea-et-al-2012}, in which the turbulent component of the magnetic field is modelled along each LOS independently from the neighbouring ones, as a realization of a one-dimensional Gaussian random field with a power-law power spectrum. In this type of approach there is no correlation from pixel to pixel on the sky, and such studies seek to exploit the depolarization along the LOS, rather than spatial correlations in the POS, to constrain statistical properties of the interstellar magnetic field.

We seek to explore another avenue, taking into account statistical correlation properties of $\B$ in all three dimensions, as well as properties of the dust density field, building on methods developed in~\cite{planck2014-XX} to compare {\Planck} data with synthetic polarization maps. In that paper, the synthetic maps were computed from data cubes of dust density $\nd$ and magnetic field $\B$ produced by numerical simulations of MHD turbulence. One could think to generalize this approach, taking advantage of the ever-increasing set of such simulations~\citep[see, e.g.,][]{hennebelle-et-al-2008,hennebelle-2013,hennebelle-iffrig-2014,inutsuka-et-al-2015,seifried-walch-2015}. However, this would be impractical for two main reasons : i) these simulations often have a limited inertial range over which the power spectrum has a power-law behaviour, and ii) a systematic study exploring a wide range of physical parameters with sufficient sampling is not possible due to the computational cost of each simulation.

We therefore propose an alternative approach, which is to build simple, approximate, three-dimensional models for the dust density $\nd$ and the magnetic field $\B$, allowing us to perfectly control the statistical properties of these 3D fields, and to fully explore the space of parameters characterizing them. With this approach, we are able to perform a statistically significant number of simulated polarization maps for each set of parameters. Actual observations may then be compared to these simulated maps, using least-square analysis methods, to extract best-fitting parameters, in particular the spectral index of the magnetic field, $\beta_B$, and the ratio of turbulent to regular field, $y_B$.

The paper is organized as follows : Sec.~\ref{sec:buildingmaps} presents the method used to build simulated thermal dust polarized emission maps using prescribed statistical properties for $\nd$ and $\B$. Observables derived from these maps, serving as statistical diagnostics of the input parameters, are presented in Sec.~\ref{sec:observables}. In Sec.~\ref{sec:method}, we describe the analysis method devised to explore the space of input parameters for a given set of polarization maps. The validation of the method and its application to actual observations of polarized dust emission from the Polaris Flare molecular cloud observed by {\Planck} are given in Sec.~\ref{sec:results}. Finally, Sec.~\ref{sec:discussion} discusses our results and offers conclusions. Several appendices complement our work. Appendix~\ref{sec:appendix:nHproperties} presents further statistical properties of the model dust density fields. Appendix~\ref{sec:appendix:L2distance} details the likelihood used in the MCMC analysis. Finally, appendix~\ref{sec:appendix:chi2} details the $\chi^2$ parameter used to estimate the goodness-of-fit.

\section{Building synthetic polarized emission maps}
\label{sec:buildingmaps}

In this section, we first describe the synthetic dust density and magnetic field cubes we use in our analysis, then explain how simulated polarized emission maps are built from these cubes. 

\subsection{Fractional Brownian motions}
The basic ingredients to synthetise polarized thermal dust emission maps are three-dimensional cubes of dust density $\nd$ and magnetic field $\B$, which we build using fractional Brownian motions (fBm)~\citep{falconer-1990}. An $N$-dimensional fBm $X$ is a random field defined on $\mathbb{R}^N$ such that $\langle\left[X\left(\boldsymbol{r}_2\right)-X\left(\boldsymbol{r}_1\right)\right]^2\rangle\propto{||}\boldsymbol{r}_2-\boldsymbol{r}_1{||}^{2H}$, for any pair of points $(\boldsymbol{r}_1,\boldsymbol{r}_2)$. $H$ is called the Hurst exponent. These fBm fields are usually built in Fourier space\footnote{In all of this paper, for any field $F$ the notation $\widetilde{F}$ represents its Fourier transform.}, 
\begin{equation}
\label{eq:Xfbm}
\widetilde{X}(\boldsymbol{k})=A(\boldsymbol{k})\exp{\left[i\phi_X(\boldsymbol{k})\right]},
\end{equation}
by specifying amplitudes that scale as a power-law of the wavenumber $k=||\boldsymbol{k}||$, 
$$
A(\boldsymbol{k})=A_0k^{-\beta_X/2},
$$
with $\beta_X=2H+N$ the spectral index, and phases drawn from a uniform random distribution in $[-\pi,\pi]$, subject to the constraint $\phi_X(-\boldsymbol{k})=-\phi_X(\boldsymbol{k})$ so that $X$ is real-valued. Their power spectra are therefore power laws, 
$$
P_X(k)=\left<\left|\widetilde{X}(\boldsymbol{k})\right|^2\right>_{||\boldsymbol{k}||=k}\propto k^{-\beta_X},
$$
where the average is taken over the locus of constant wavenumber $k$ in Fourier space. Such fields have been used previously as toy models for the fractal structure of molecular clouds, in both density and velocity space~\citep{stutzki-et-al-1998,brunt-heyer-2002,miville-deschenes2003,correia-et-al-2016}.

\subsection{Dust density}
\label{sec:modeldensity}

\begin{figure}[htbp]
\includegraphics[width=9cm]{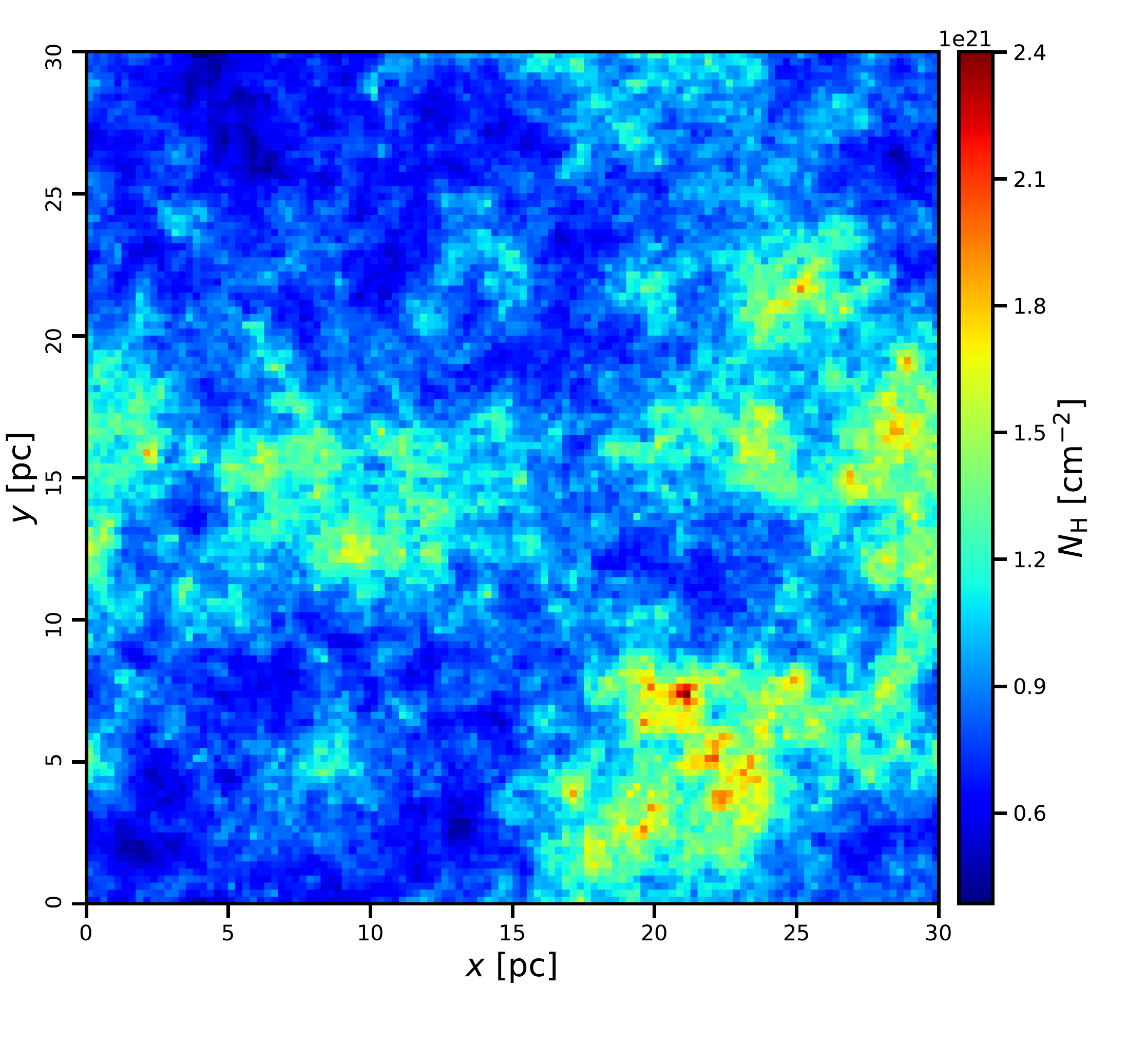}
\caption{Total gas column density $\NH$, derived from a synthetic density field $\nH$ built by exponentiation of a fBm field with spectral index $\beta_X=2.6$ and size $120 \times 120 \times 120$ pixels. The volume density fluctuation level is $y_n=1$, and the column density fluctuation level is $y_{\NH}\simeq0.25$.}  
\label{fig:densitymaps}
\end{figure}

In our approach, the dust density $\nd$ is taken to be proportional to the total gas density $\nH$, so that the dust optical depth within each cell is also proportional to $\nH$ (see the derivation of polarization maps in Sec.~\ref{sec:polarmaps}). Therefore, we mean to model $\nH$ from numerical realizations of three-dimensional fBm fields built in Fourier space. These have means defined by the value chosen for the null-wavevector amplitude $A(\boldsymbol{0})$, so if one wished to use such a synthetic random field $X$ directly as a model for the positive-valued $\nH$, one would be required to choose $\nH=X'=X-a$ with $a\geqslant\mathrm{min}(X)$ a constant. However, since the distributions of these fields in 3D are close to Gaussian, their ratio of standard deviation to mean is typically $\sigma_{X'}/\langle X'\rangle\lesssim 0.3$, which is much too small compared to observational values. For instance, the total gas column density fluctuation ratios $\sigma_{\NH}/\left<\NH\right>$ in the ten nearby molecular clouds selected for study in~\cite{planck2014-XX} are in the range 0.3-1, and one should keep in mind that these are only lower bounds for fluctuation ratios in the three-dimensional density field $\nH$. 

We remedy this shortcoming by taking $X$ to represent the log-density, i.e., $\nH$ is given by 
\begin{equation}
\label{eq:expfbm}
\nH=n_0\exp{\left(\frac{X}{X_r}\right)},
\end{equation}
where $X$ is a three-dimensional fBm field with spectral index $\beta_X$, and $X_r$ and $n_0$ are positive parameters. The $\nH$ fields built in this fashion have simple and well-controlled statistical properties. First, their probability distribution functions (PDF) are log-normal, which allows, through an adequate choice of $X_r$, to set the fluctuation level of the density field $y_n=\sigma_{\nH}/\langle\nH\rangle$ to any desired value. Second, their power spectra, as azimuthal averages in Fourier space, retain the power-law behaviour of the original fBm $X$, 
$$
P_{\nH}(k)=\left<\left|\widetilde{\nH}(\boldsymbol{k})\right|^2\right>_{||\boldsymbol{k}||=k} \propto k^{-\beta_n},
$$
although the spectral indices $\beta_n$ may deviate significantly from $\beta_X$. An example of such a field is shown in Fig. \ref{fig:densitymaps}, which represents the total gas column density $\NH$ derived from a gas volume density $\nH$ built as the exponential of a $120 \times 120 \times 120$ fractional Brownian motion with zero mean, unit variance, and spectral index $\beta_X=2.6$. The parameters of the exponentiation are $X_r=1.2$ and $n_0=20\,\mathrm{cm}^{-3}$, and the grid is chosen so that the extent of the cube is 30\,pc on each side, corresponding to a pixel size of 0.25\,pc. More details on the properties of these fields are given in Appendix~\ref{sec:appendix:nHproperties}. 

The density fields built in this fashion are of course only a rough statistical approximation for actual interstellar density fields. For instance, they are unable to reproduce the filamentary structures observed in dust emission maps~\citep{andre-et-al-2010,miville-deschenes-et-al-2010}. These structures cannot be captured by one- and two-point statistics such as those used here, and require a description involving higher-order moments, or equivalently of the Fourier phases~\citep[see, e.g.,][]{levrier-et-al-2004,burkhart-lazarian-2016}.

\subsection{Magnetic field}

\begin{figure}[h!]
\includegraphics[width=9cm]{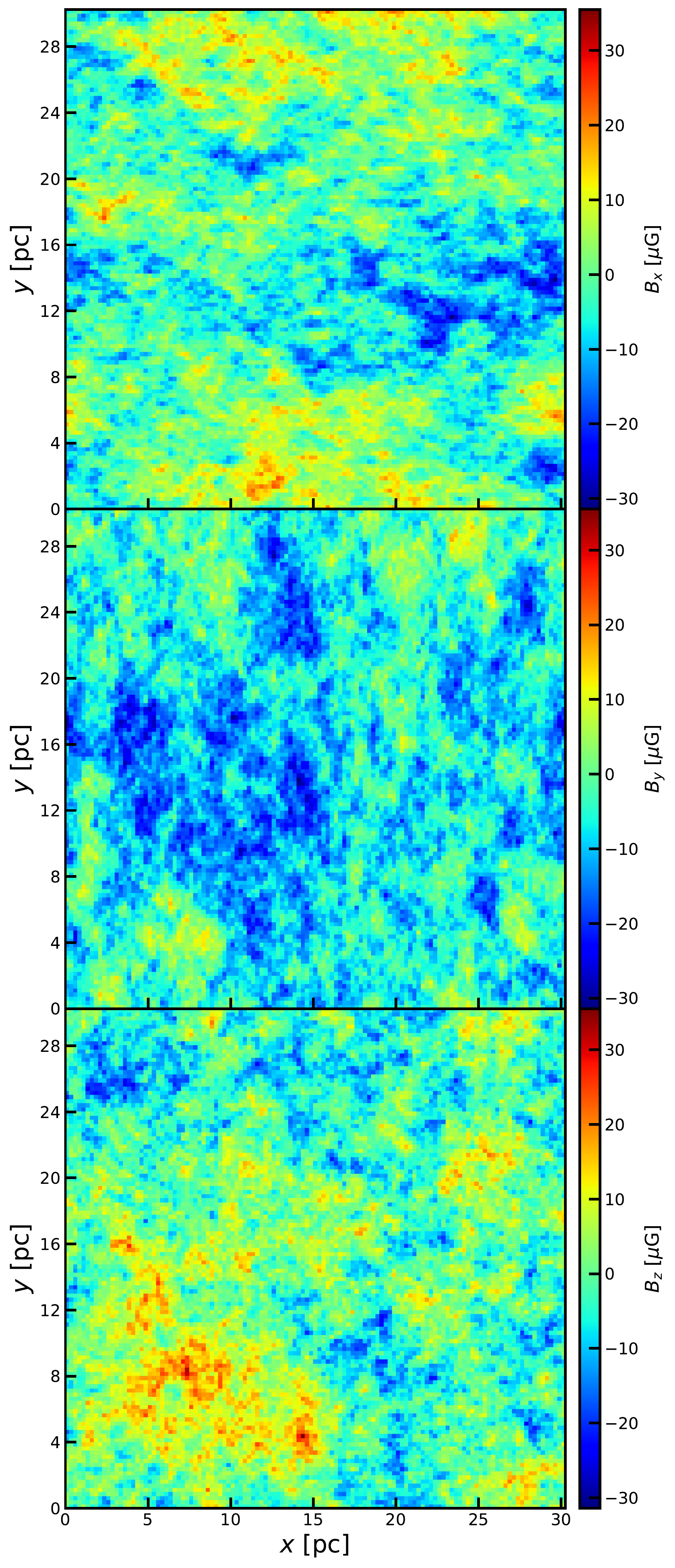}
\caption{Synthetic magnetic field $\B$ built using Eq.~\ref{eq:Btlambda}. The spectral index of the vector potential is $\beta_A=5$ and the size of the cubes is $120 \times 120 \times 120$ pixels, corresponding to 30\,pc on each side. Shown are 2D slices through the cubes of the components $B_{x}$ ({\it top}), $B_{y}$ ({\it middle}), and $B_{z}$ ({\it bottom}). The ratio of the fluctuations of the turbulent component $\B_t$ to the norm of the uniform component $\B_0$ is $y_B=1$ in this particular case, with angles $\chi_0=\gamma_0=0^\circ$. }
\label{fig:Bmaps}
\end{figure}

\begin{figure}[htbp]
\resizebox{\hsize}{!}{
\includegraphics[width=9cm]{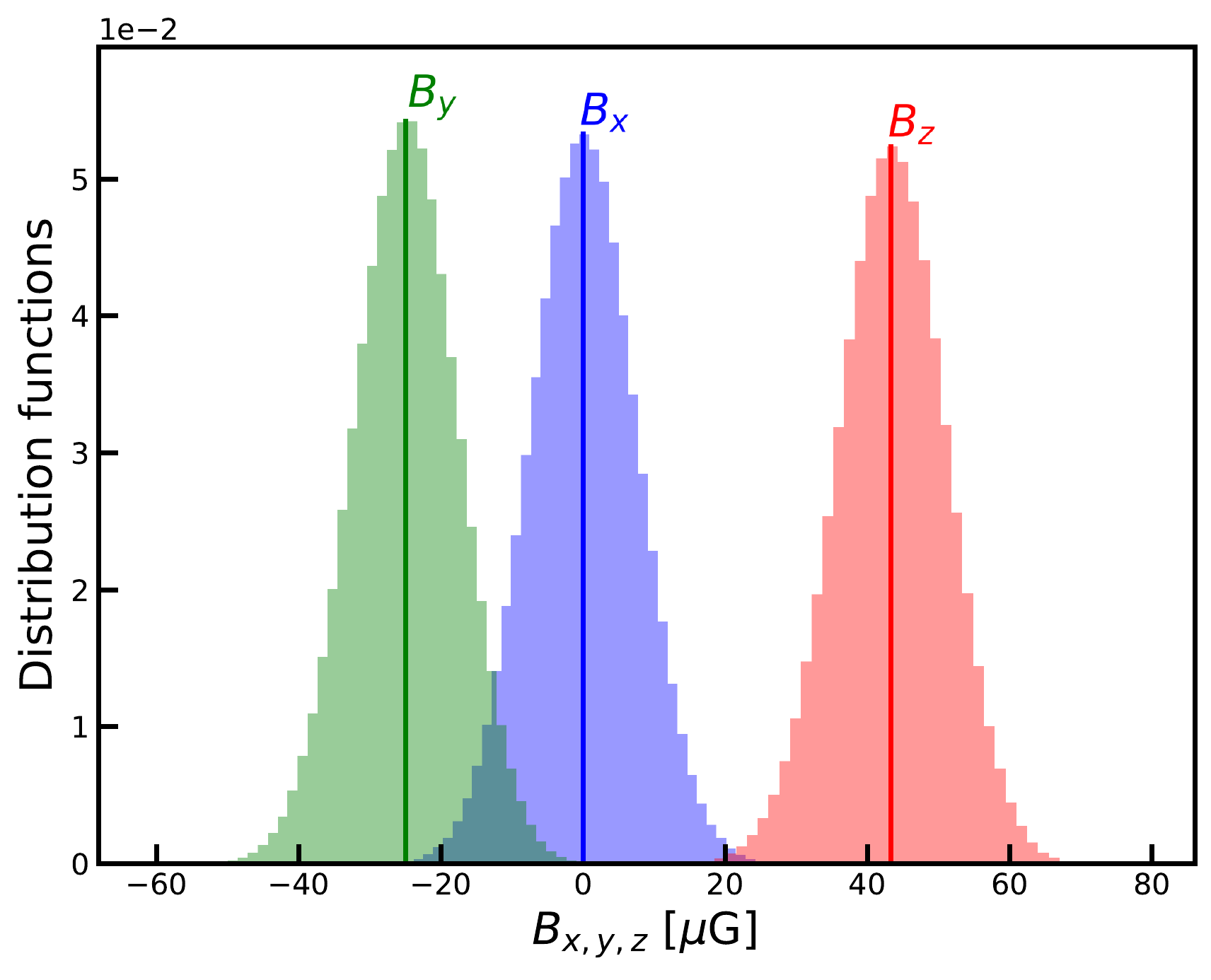}
}
\caption{Distribution functions of the components $B_x$, $B_y$ and $B_z$ of a model magnetic field $\B=\B_0+\B_t$ built on a grid $120 \times 120 \times 120$ pixels using Eq.~\ref{eq:Btlambda} with $\beta_A=5$, and a mean, large-scale magnetic field $\B_0$ defined by the angles $\chi_0=0^\circ$ and $\gamma_0=60^\circ$, and a norm $B_0=50\,\mu{\mathrm{G}}$ such that the fluctuation level is $y_B=0.1$. The vertical lines represent the projected values of the large scale magnetic field $B_{0x}=B_0 \sin \chi_0 \cos\gamma_0$, $B_{0y}=-B_0 \cos \chi_0 \cos\gamma_0$ and $B_{0z} = B_0 \sin\gamma_0$. See figure 14 in~\cite{planck2014-XX} for the definition of angles.}
\label{fig:Bpdfs}
\end{figure}

\begin{figure}[htbp]
\resizebox{\hsize}{!}{
\includegraphics[width=9cm]{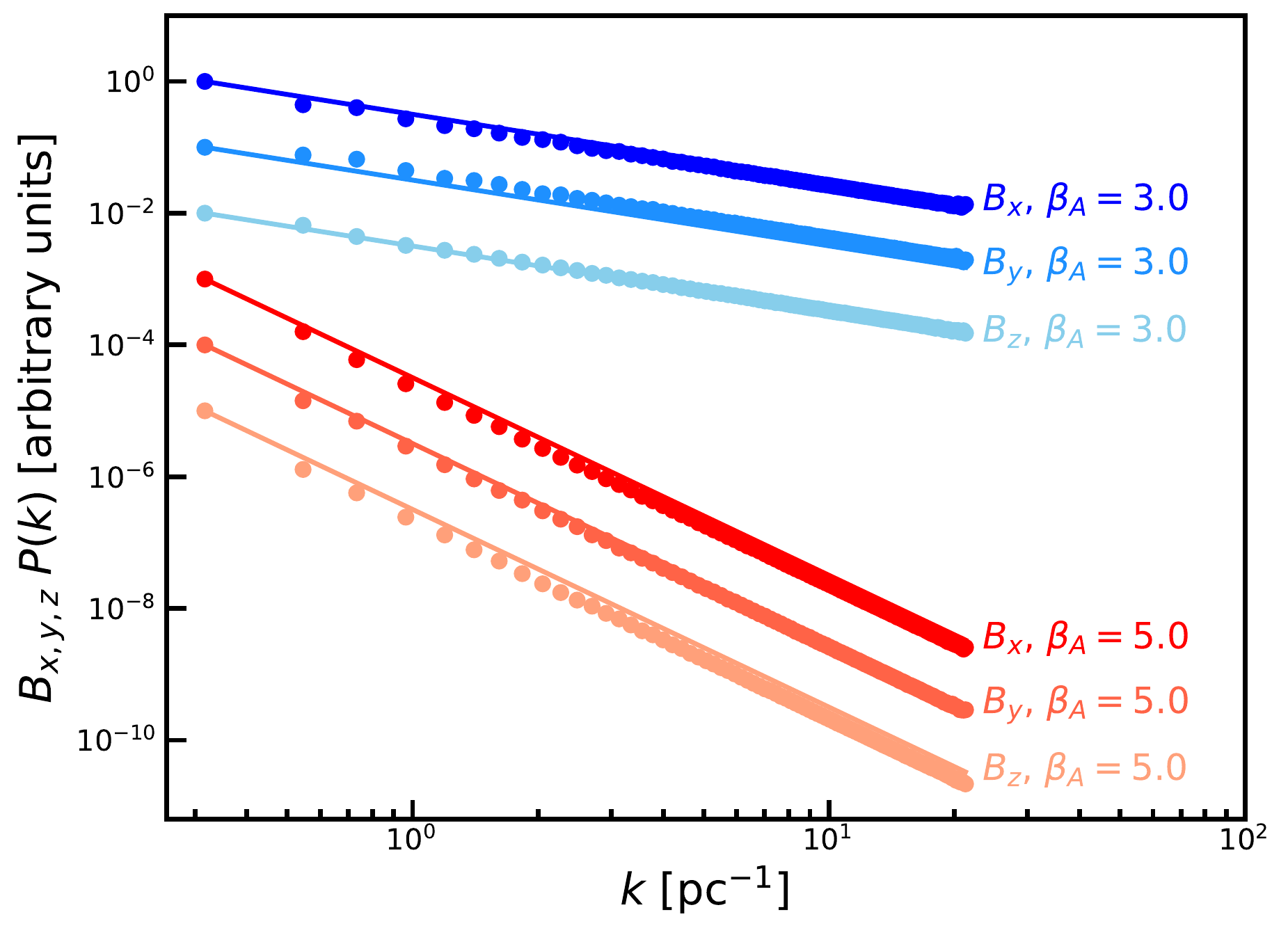}
}
\caption{Power spectra of the components $B_x$, $B_y$ and $B_z$ of a model magnetic field $\B=\B_0+\B_t$ built on a grid $120 \times 120 \times 120$ pixels using Eq.~\ref{eq:Btlambda} with $\beta_A=3$ (different shades of blue for the three components) and $\beta_A=5$ (different shades of red for the three components). The power spectra are normalized differently so as to allow comparison between them. The fitted power-laws shown as solid lines yield spectral indices $\beta_B=1$ and $\beta_B=3$, in agreement with Eq.~\ref{eq:PB}. These are the power spectra of the same particular realizations shown in Fig.~\ref{fig:Bmaps}.}
\label{fig:Bpowerspectra}
\end{figure}

To obtain a synthetic turbulent component of the magnetic field $\B_t$ with null divergence and controlled power spectrum, we start from a vector potential $\boldsymbol{A}$ built as a three-dimensional fractional Brownian motion. To be more precise, each Cartesian component $A_\lambda$ of $\boldsymbol{A}$ is a fBm field,
\begin{equation*}
\widetilde{A_\lambda}(\boldsymbol{k})=\mathcal{A}_0k^{-\beta_A/2}\exp\left[i\phi_{A_\lambda}(\boldsymbol{k})\right],
\end{equation*}
where the spectral index $\beta_A$ and the overall normalization parameter $\mathcal{A}_0$ are independent of the Cartesian component $\lambda=x,y,z$ considered. Using the definition of the magnetic field from the vector potential $B_{t,\lambda}=\epsilon_{\lambda\mu\nu}\partial_\mu A_\nu$, where $\epsilon_{\lambda\mu\nu}$ is the Levi-Civita tensor, and the derivation relation in Fourier space 
$$
\widetilde{\partial_\lambda F}=ik_\lambda \widetilde{F}
$$ 
we have the expression of the components of $\B_t$ in Fourier space
\begin{equation}
\label{eq:Btlambda}
\widetilde{B_{t,\lambda}}(\boldsymbol{k})=\mathcal{A}_0\epsilon_{\lambda\mu\nu}ik_\mu k^{-\beta_A/2}\exp\left[i\phi_{A_\nu}(\boldsymbol{k})\right]
\end{equation}
As it should, this expression corresponds to a divergence-free turbulent magnetic field, 
$$
ik_\lambda\widetilde{B_{t,\lambda}}=0,
$$
using the Einstein notation. Writing $k_\mu=kf_\mu$, with $\boldsymbol{f}=\left(\sin\vartheta\cos\varphi,\sin\vartheta\sin\varphi,\cos\vartheta\right)$, the power spectrum of each component of $\B_t$ is then
\begin{equation*}
P_{B_{t,\lambda}}(k)=\mathcal{A}_0^2k^{2-\beta_A}\left<\left|\epsilon_{\lambda\mu\nu}f_\mu\exp\left[i\phi_{A_\nu}(\boldsymbol{k})\right]\right|^2\right>_{||\boldsymbol{k}||=k}.
\end{equation*}
The last factor is essentially independent of the wavenumber $k$, so the spectral index of each component of $\B_t$ is $\beta_{B_t}=\beta_A-2$. After Fourier-transforming back to real space, $\B_t$ is shifted and scaled so that it has zero mean and a standard deviation $\sigma_B$ of $5\,\mu\mathrm{G}$, a value typical of the interstellar magnetic field~\citep[see, e.g.,][and references therein]{haverkorn-et-al-2008}. 

The model magnetic field $\B$ is obtained by adding a uniform\footnote{Note that we do not consider an ordered random or striated random component of the field~\citep{jaffe-et-al-2010,jansson-farrar-2012}, which we justify by the smallness of the field-of-view considered.} vector field $\B_0$ to that turbulent magnetic field $\B_t$. The effect in Fourier space is limited to a modification for $\boldsymbol{k}=\boldsymbol{0}$ only, so the spectral index of each component $B_\lambda$ of the total magnetic field is the same as that of $B_{t,\lambda}$, i.e.,
\begin{equation}
\label{eq:PB}
P_{B_{\lambda}}(k)\propto k^{2-\beta_A}.
\end{equation}
Note that this means that the resulting magnetic fields thus only display anisotropy in the $\boldsymbol{k}=\boldsymbol{0}$ mode, and not at the other scales. This is a limitation of our model, which is thus not fully consistent with observations of the magnetic field structure~\citep{planck2015-XXXV}, but it is sufficient for our purposes.

The uniform field $\B_0$ which is added to the turbulent field $\B_t$ is defined by its norm $B_0$ and a pair of angles, $\gamma_0$ and $\chi_0$, which are respectively the angle between the magnetic field and the POS, and the position angle of the projection of $\B_0$ in the POS, counted positively clockwise from the north-south direction~\citep[see figure 14 of][]{planck2014-XX}. The total magnetic field's direction in three-dimensional space is characterized by angles $\gamma$ and $\chi$ defined in the same way. The ratio of the turbulent to mean magnetic field strengths is then defined by 
\begin{equation*}
y_B=\frac{\sigma_B}{B_0}=\frac{\sqrt{\left<\B_t^2\right>-\left<\B_t^{\phantom{2}}\right>^2}}{||\B_{0}||}=\frac{\sqrt{\left<\B_t^2\right>}}{||\B_{0}||}.
\end{equation*}

Fig.~\ref{fig:Bmaps} shows an example of a synthetic magnetic field $\B$ generated in this fashion, and defined on the same 120 $\times$ 120 $\times$ 120 pixels grid that was used for the gas density model described in Sec.~\ref{sec:modeldensity}. The parameters used for this specific realization are $\beta_A=5$, $y_B=1$, and $\chi_0=\gamma_0=0^\circ$. The PDFs of the components of such model magnetic fields are Gaussian, as shown in Fig.~\ref{fig:Bpdfs}, and their power spectra are power laws of the wavenumber, as shown in Fig.~\ref{fig:Bpowerspectra}, with a common spectral index $\beta_B$ that is related to the input $\beta_A$ by $\beta_B = \beta_A - 2$.

\subsection{Polarization maps}
\label{sec:polarmaps}
Once cubes of total gas density $\nH$ and magnetic field $\B$ are available, maps of Stokes parameters $I$, $Q$, and $U$ at 353\,GHz (the frequency of the {\Planck} channel with the best signal-to-noise ratio in polarized thermal dust emission) are built by integrating along the line of sight through the simulation cubes, following the method in~\cite{planck2014-XX} :
\begin{eqnarray}
\label{eq:defI}
I_0&=&\int S_\nu\left[1-p_0\left(\cos^2\gamma-\frac{2}{3}\right)\right]\sigma_{353}\,\nH\ud z;\\
\label{eq:defQ}
Q_0&=&\int p_0\,S_\nu\cos\left(2\phi\right)\cos^2\gamma\,\sigma_{353}\,\nH\ud z;\\
\label{eq:defU}
U_0&=&\int p_0\,S_\nu\sin\left(2\phi\right)\cos^2\gamma\,\sigma_{353}\,\nH\ud z.
\end{eqnarray}
In these equations, we take the intrinsic polarization fraction parameter $p_0$ to be uniform, and the source function $S_\nu=B_\nu(T_\mathrm{d})$ to be that of a blackbody with an assumed uniform dust temperature $T_\mathrm{d}$. The dust opacity at this frequency $\sigma_{353}$ is taken to vary with $\NH$, following figure 20 from~\cite{planck2013-p06b} for $X_\mathrm{CO}=2\,10^{20}\,\mathrm{H_2\,cm^{-2}\,K^{-1}\,km^{-1}\,s}$, and propagating the associated errors. The order of magnitude of the dust opacity is around $\sigma_{353}\approx 10^{-26}\,\mathrm{cm}^{2}$. The values of $\NH$ considered in our study are typically at most a few $10^{21}\,\mathrm{cm}^{-2}$, so the optically thin limit applies in the integrals of Eqs.~\ref{eq:defI}-\ref{eq:defU}. The angle $\phi$ is the local polarization angle, which is related to the position angle\footnote{Not to be confused with the corresponding position angle $\chi_0$ of the uniform component of the magnetic field $\B_0$.} $\chi$ of the magnetic field's projection on the POS at each position on the LOS by a rotation of $90^\circ$~\citep[see definitions of angles in][]{planck2014-XX}.

The $\nH$ and $\B$ cubes are built on a grid which is $132\times132$ pixels in the POS and $N_z$ pixels in the $z$ direction (that of the LOS). The cells have a physical size $\delta=0.24\,\mathrm{pc}$ in each direction, so the depth $d=N_z\,\delta$ of the cloud is a free parameter in our analysis, and the Stokes maps built from Eq.~\ref{eq:defI}-\ref{eq:defU} are $32\,\mathrm{pc}$ in both $x$ and $y$ directions.

\subsection{Noise and beam convolution}

In order to proceed with the analysis of observational data, one cannot use these model Stokes maps directly : it is necessary to properly take into account noise and beam convolution. Anticipating somewhat on the description of the \Planck\ data we shall use as an application of the method, the 353\,GHz noise covariance matrix maps are taken directly from the {\it \Planck\ Legacy Archive}\footnote{\tt http://pla.esac.esa.int/pla/} and are part of the 2015 public release of {\Planck} data~\citep{planck2014-a01}, 
\begin{equation}
\boldsymbol{\Sigma}=\left(\begin{array}{ccc}
\sigma_{II} & \sigma_{IQ} & \sigma_{IU} \\
\sigma_{QI} & \sigma_{QQ} & \sigma_{QU} \\
\sigma_{UI} & \sigma_{UQ} & \sigma_{UU} \\
\end{array}\right).
\end{equation}
Noise is added to the model Stokes maps pixel by pixel, as
\begin{equation}
\label{eq:addnoise}
I_n=I_0+n_I \qquad Q_n=Q_0+n_Q \qquad U_n=U_0+n_U,
\end{equation}
where $n_I$, $n_Q$, and $n_U$ are random values drawn from a three-dimensional Gaussian distribution with zero mean and characterized by the noise covariance matrix $\boldsymbol{\Sigma}$. To preserve the properties of {\Planck} noise, the random values are directly drawn from the {\tt Healpix}~\citep{gorski_et_al_2005} covariance matrix maps and added to the simulated maps after a gnomonic projection of the region under study, in our case the Polaris Flare molecular cloud.

The resulting Stokes $I_n$, $Q_n$, and $U_n$ maps are then placed at a distance\footnote{A more recent determination of the distance to the Polaris Flare places it at 350-400\,pc~\citep{schlafly2014}. For the demonstration of the method presented here, this is not a critical issue, as the power-law power spectra underline  self-similar behaviours, so that a change of the distance can be compensated by a change in pixel size.} $D=140\,\mathrm{pc}$, so that the angular size of each pixel is about $6\arcmin$, and then convolved by a circular 15{\arcmin} full-width at half maximum (FWHM) Gaussian beam $\mathcal{B}$. To avoid edge effects, only the central $120\times120$ pixels of the convolved maps $I_\mathrm{m}=\mathcal{B}\otimes I_n$, $Q_\mathrm{m}=\mathcal{B}\otimes Q_n$, and $U_\mathrm{m}=\mathcal{B}\otimes U_n$ are retained, corresponding to a field-of-view (FoV) of approximately 12$^\circ$. With this approach, we ensure that these model maps $(I_\mathrm{m},Q_\mathrm{m},U_\mathrm{m})$ are fit to be compared to actual {\Planck} data, which we discuss in Sec.~\ref{sec:Polaris}.

\begin{table*}[htb]
\caption[]{Parameter space explored in the grid of model polarization maps.}\label{tab:priors}
\centering
\begin{tabular}{cccc} \hline \hline \\ [-1ex]
Parameter   & Prior\,\tablefootmark{a}  &  Definition \\  [1ex] \hline \\ [-1ex]  
$\beta_B$  &  $\left[1,4\right]$  & Spectral index of the 3D turbulent magnetic field  \\ 
$\beta_n$  & $\left[1,5\right]$   & Spectral index of the 3D dust density field \\ 
$\log_{10}{y_n}$  & $\left[-1,1\right]$   & Log of the RMS-to-mean ratio of dust density\\ 
$\log_{10}{y_B^{\rm POS}}$ & $\left[-1,1\right]$    & Log of the ratio of the turbulent magnetic field RMS to the mean magnetic field in the POS\\ 
$\chi_0$ & $\left[-90^\circ,90^\circ\right]$   & Position angle of the mean magnetic field in the POS\\ 
$\log_{10}\left(d/1\,\mathrm{pc}\right)  $ & $\left[-0.3,1.5\right]$\,\tablefootmark{b}  & Depth of the simulated cube \\ 
$\log_{10}\left(\langle\nH\rangle/1\,\mathrm{cm}^{-3}\right)  $ & $\left[1,2.7\right]$\,\tablefootmark{c}  & Mean dust density \\ 
$T_\mathrm{d} $ & $\left[5\,\mathrm{K},200\,\mathrm{K}\right]$  & Dust temperature \\ 
$p_0 $ & $\left[0.01,0.5\right]$  & Intrinsic polarisation fraction parameter \\ 
 [1ex]  \hline \\ [-1ex]
\end{tabular}
\tablefoot{\tablefoottext{a}{Priors are assumed to be flat in the given range for the parametrization given in this table, and zero outside, except $\chi_0$, for which a $180^\circ$ periodicity is applied when the Metropolis algorithm draws values outside the given range.}\tablefoottext{b}{Corresponding to a cube depth in the interval $\left[0.5\,\mathrm{pc},32.5\,\mathrm{pc}\right]$.}\tablefoottext{c}{Corresponding to a density in the interval $\left[10\,\mathrm{cm^{-3}},500\,\mathrm{cm^{-3}}\right]$.}}
\end{table*}

\section{Observables}
\label{sec:observables}

\begin{figure*}[htbp]
\centerline{
\includegraphics[width=0.43\textwidth,trim=0 0 0 20,clip=true]{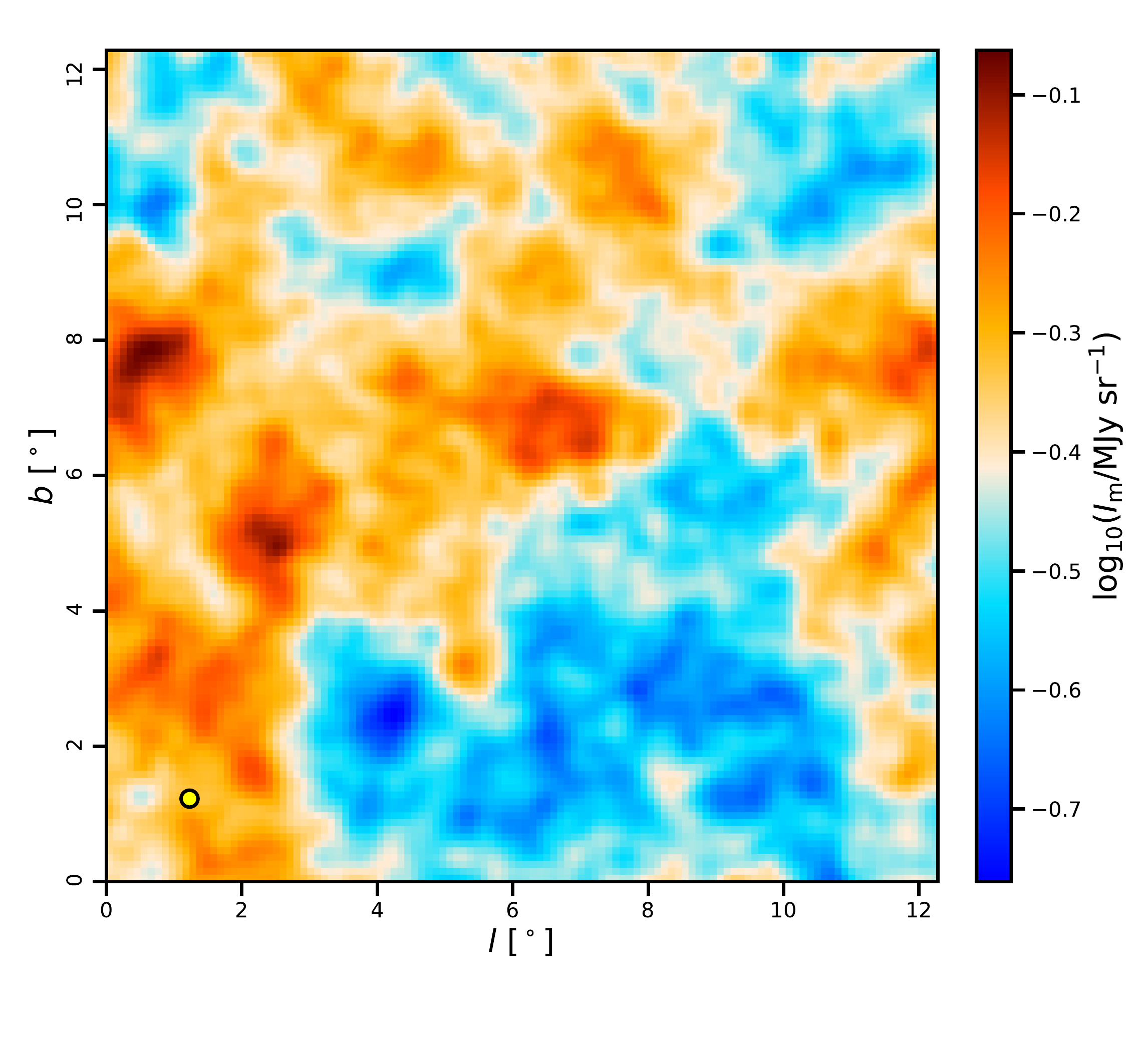}
\includegraphics[width=0.49\textwidth,trim=-20 -10 0 0,clip=true]{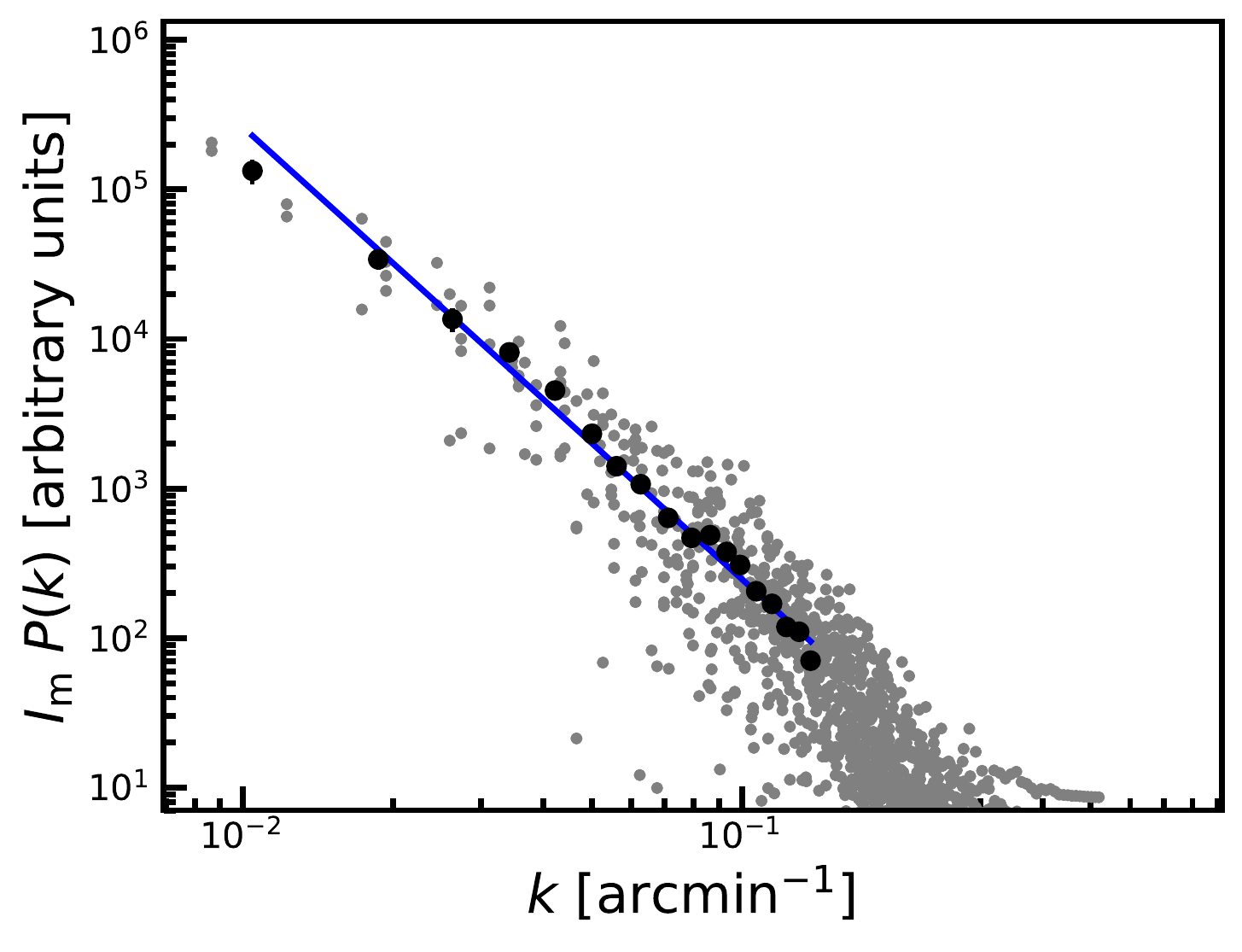}
}
\centerline{
\includegraphics[width=0.43\textwidth,trim=0 0 0 20,clip=true]{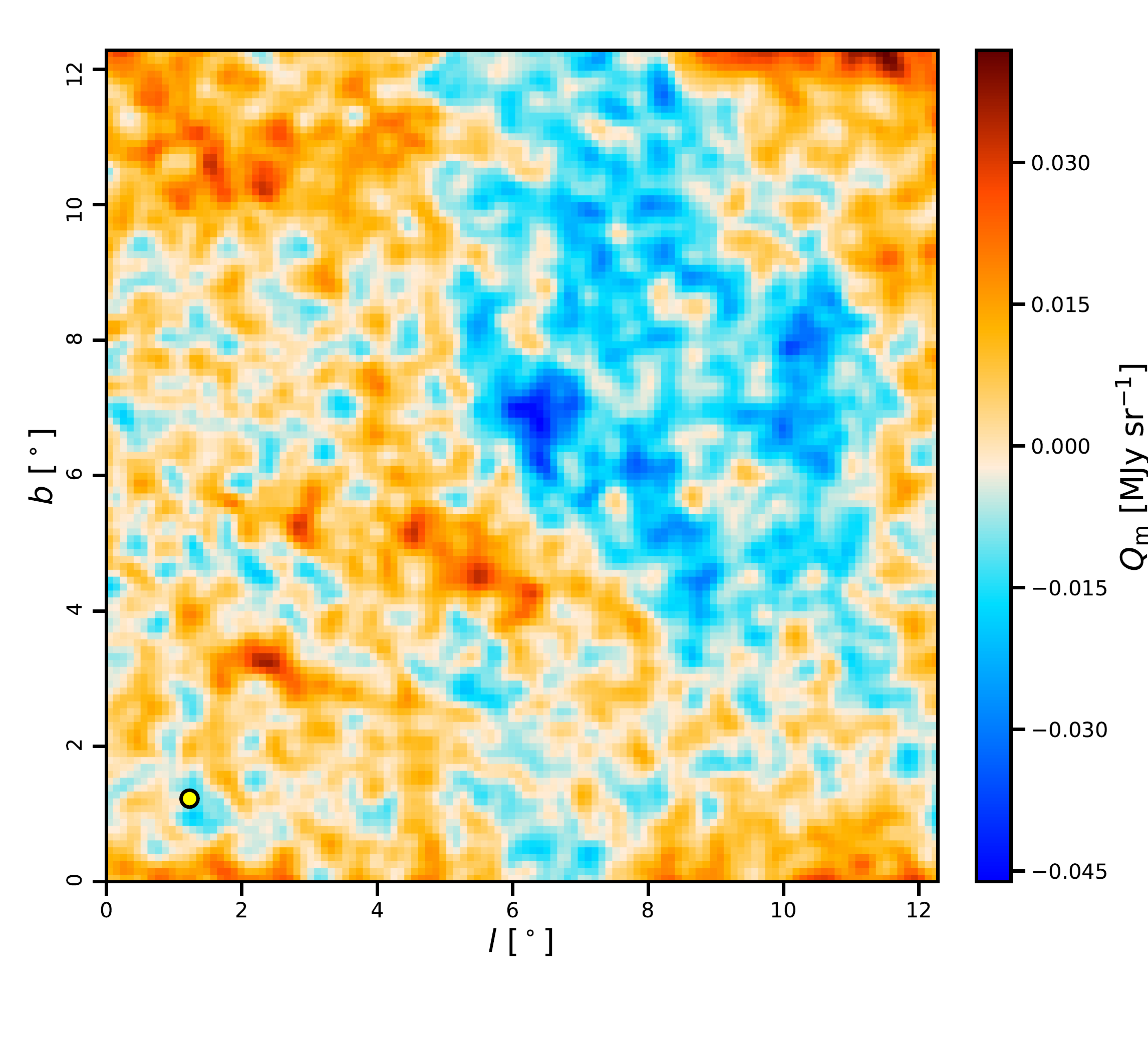}
\includegraphics[width=0.49\textwidth,trim=-20 -10 0 0,clip=true]{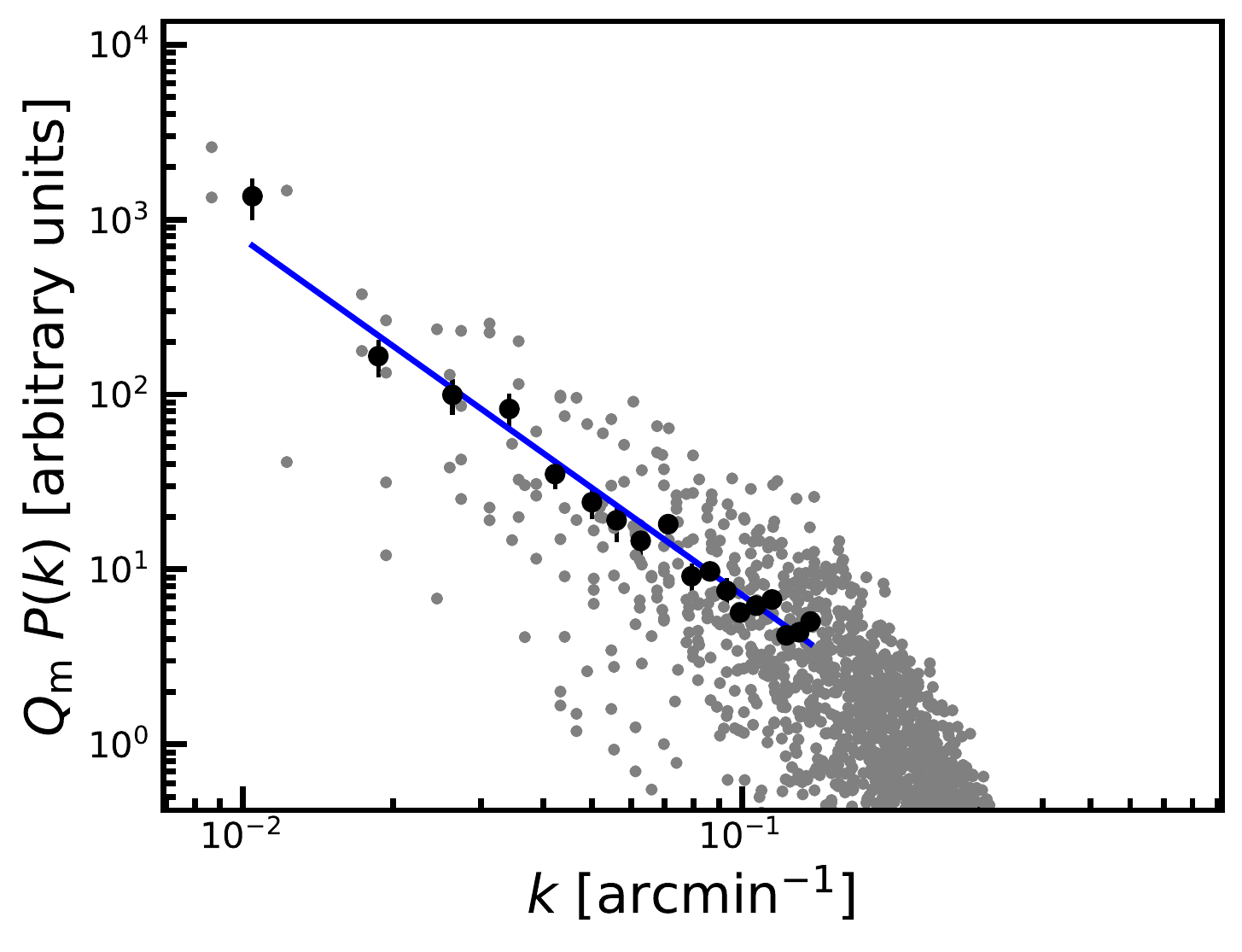}
}
\centerline{
\includegraphics[width=0.43\textwidth,trim=0 0 0 20,clip=true]{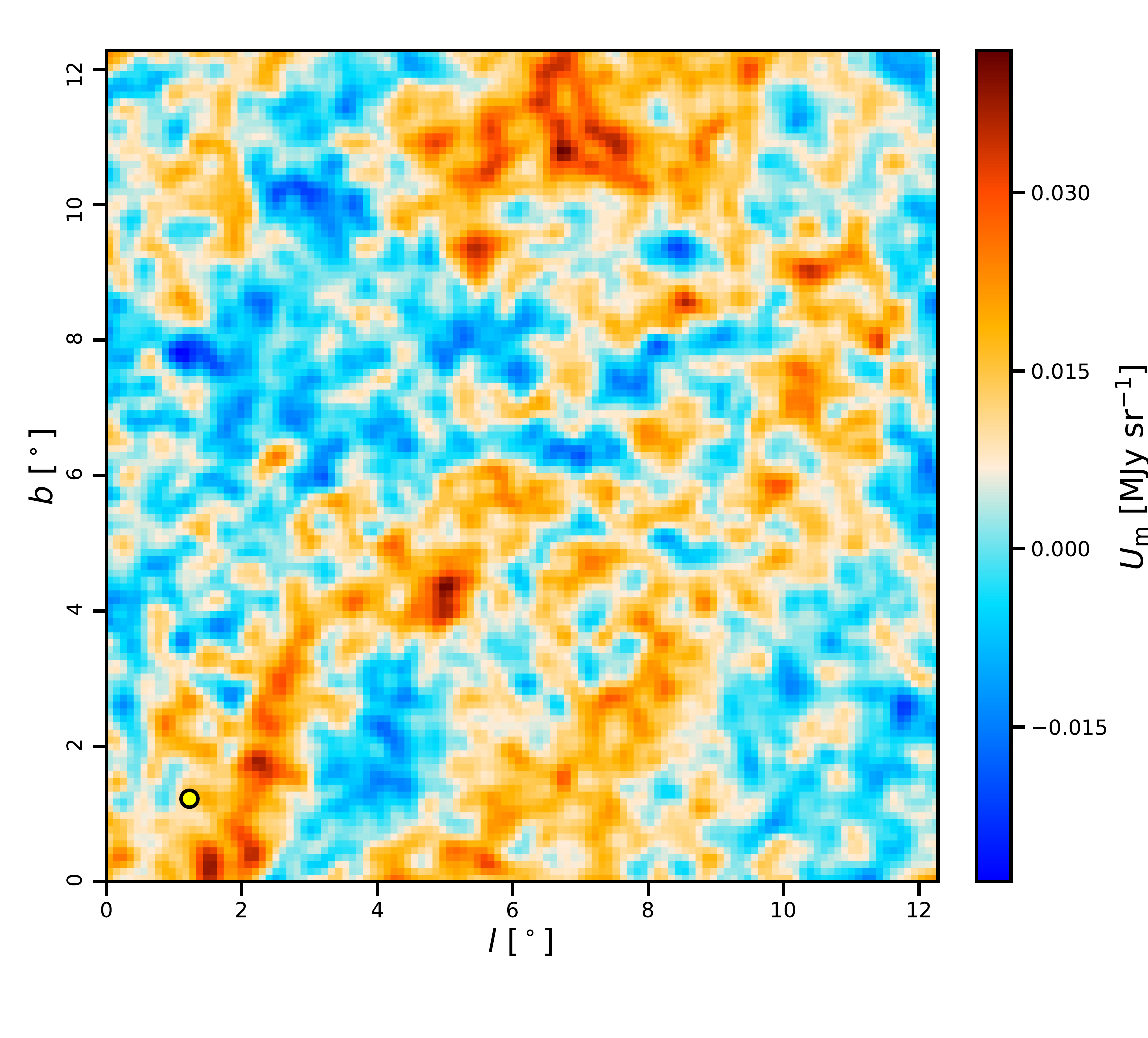}
\includegraphics[width=0.49\textwidth,trim=-20 -10 0 0,clip=true]{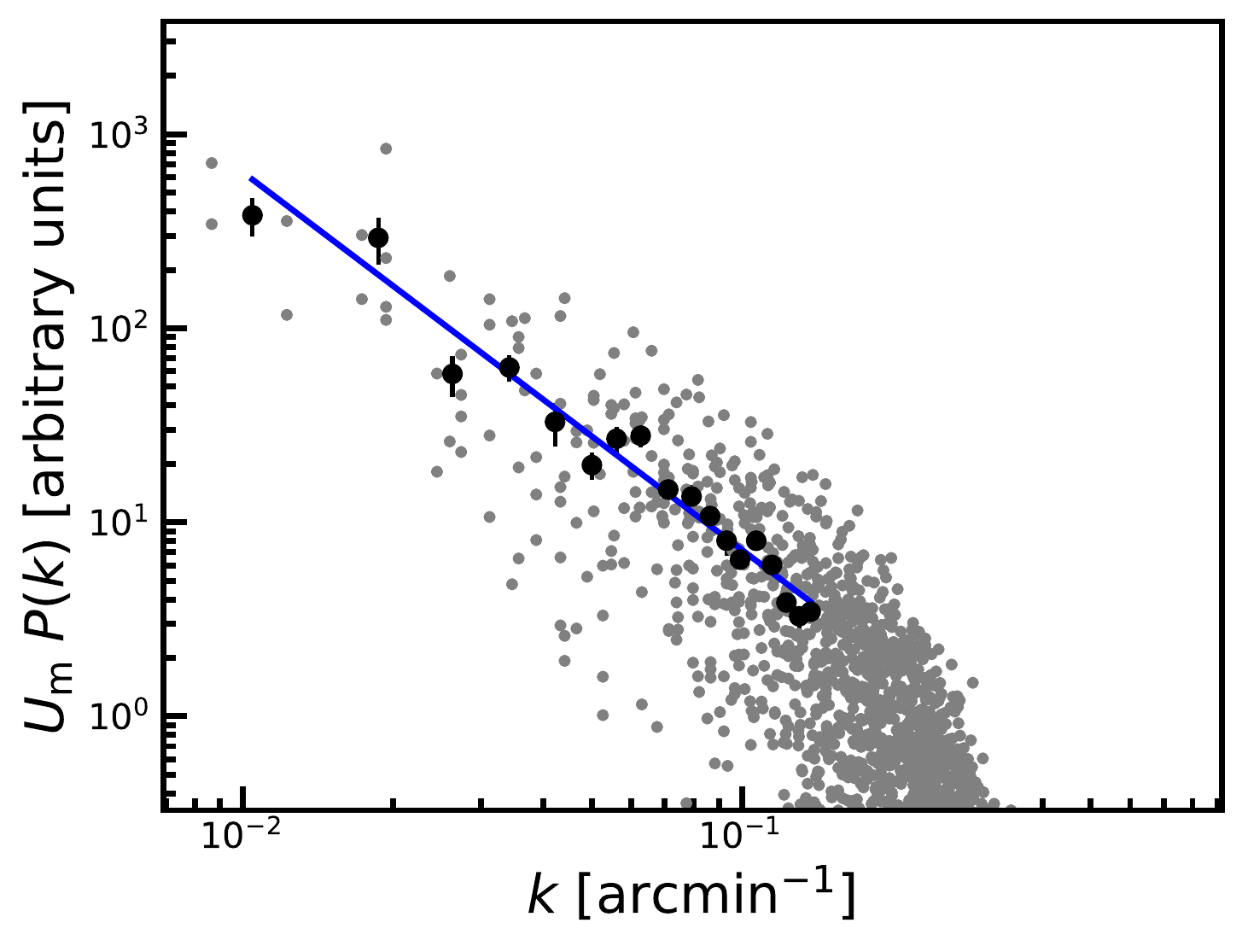}
}
\caption{{\it Left column row :} Example maps (from top to bottom : total intensity $I_\mathrm{m}$ on a logarithmic scale, Stokes $Q_\mathrm{m}$, and Stokes $U_\mathrm{m}$) from simulation A (see~\ref{sec:results} and Table~\ref{tab:results_sim}). {\it Right column : } Corresponding power spectra. The gray points represent the two-dimensional power spectra, while the black dots represent the azimuthal averages in Fourier space in a set of wavenumber bins, and the blue line is a power-law fit to the black points.}
\label{fig_maps_and_ps_simA}
\end{figure*}

\begin{figure*}[htbp]
\resizebox{\hsize}{!}{
\includegraphics{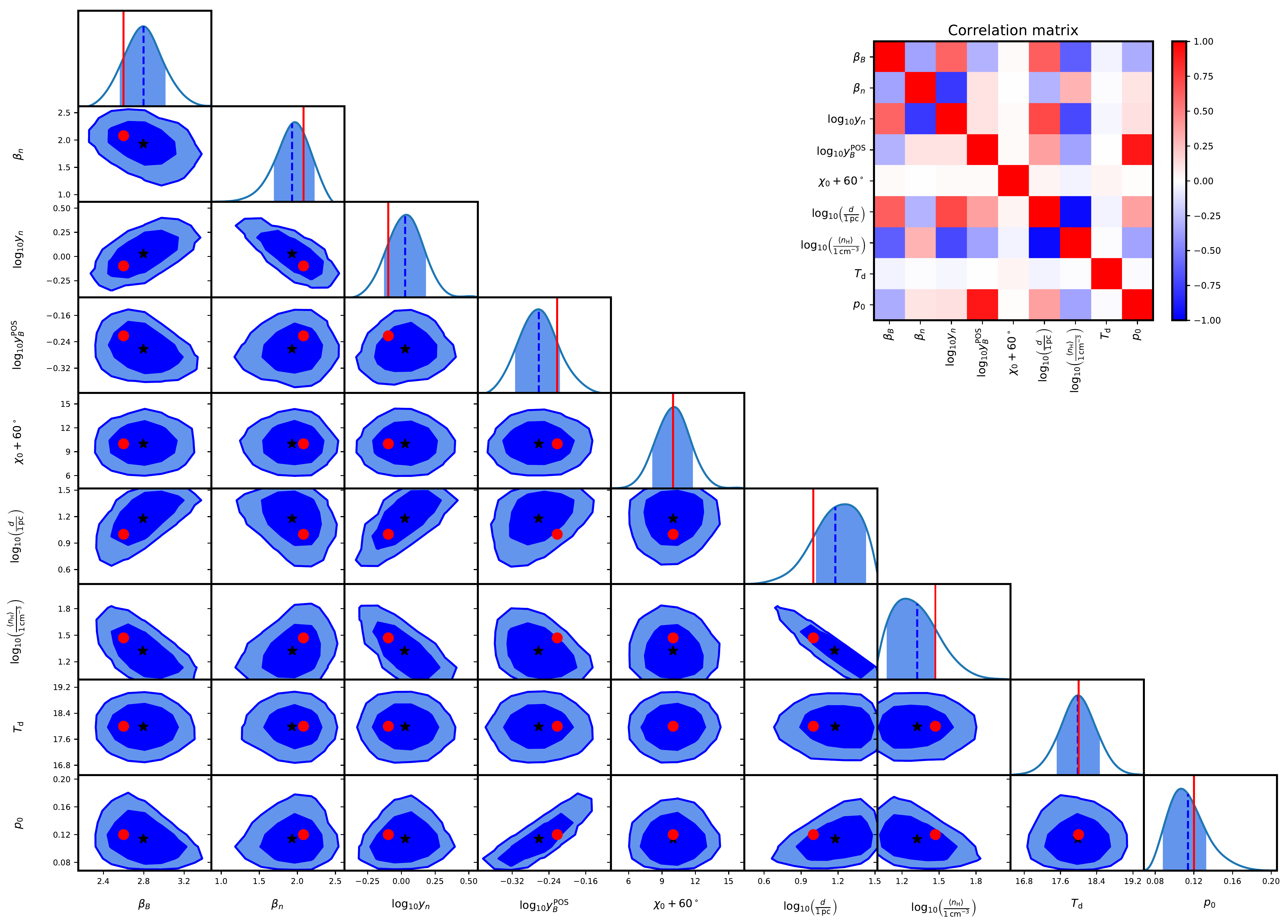}
}
\caption{Constraints (posterior probability contours and marginalized PDFs) on the statistical properties of the dust density and magnetic field for the simulation A maps. On the posterior probability contours, the filled dark and light blue regions respectively enclose 68.3\% and 95.4\% of the probability, the black stars indicate the averages over the two-dimensional posterior PDFs, and the red circles indicate the input values for the simulation. In the plots showing the marginalized posterior PDFs, the light blue regions enclose 68.3\% of the probability, the dashed blue lines indicate the averages over the posterior PDFs, and the solid red lines indicate the input values for the simulation. The upper right plot displays the correlation matrix between the fitted parameters.}
\label{fig_triangle_simA}
\end{figure*}

From the model maps above, we build an ensemble of derived maps, starting with the normalized Stokes maps
$$
i=\frac{I_\mathrm{m}}{\langle I_\mathrm{m}\rangle} \qquad q=\frac{Q_\mathrm{m}}{I_\mathrm{m}} \qquad u=\frac{U_\mathrm{m}}{I_\mathrm{m}}
$$
where $\langle I_\mathrm{m}\rangle$ is the spatial average of the model Stokes $I_\mathrm{m}$ map. Then we define the polarization fraction, which requires us to note that since our models include noise, we should not use the ``na\"ive'' estimator~\citep{montier-et-al-2015a,montier-et-al-2015b}
\begin{equation*}
p=\frac{\sqrt{Q_\mathrm{m}^2+U_\mathrm{m}^2}}{I_\mathrm{m}},
\end{equation*}
but rather the modified asymptotic (MAS) estimator proposed by~\cite{plaszczynski_et_al_14}
\begin{equation}
p_\mathrm{MAS}=p-b^2\frac{\displaystyle 1-e^{-p^2/b^2}}{2p}
\end{equation}
where the noise bias parameter $b^2$ derives from the elements of the noise covariance matrix $\boldsymbol{\Sigma}$~\citep[see][]{montier-et-al-2015b}. Next, we define the polarization angle 
$$
\psi=\frac{1}{2}\mathrm{atan}{\left(U_\mathrm{m},Q_\mathrm{m}\right)}
$$
where the two-argument $\mathrm{atan}$ function lifts the $\pi$-degeneracy of the usual $\mathrm{atan}$ function. Note that this expression means that the polarization angle is defined in the {\tt Healpix} convention.

We also build maps of the polarization angle dispersion function $\mathcal{S}$~\citep{planck2014-XIX,planck2014-XX,alina-et-al-2016}, which quantifies the local dispersion of polarization angles at a given lag $\delta$ and is defined by
$$
\mathcal{S}(\boldsymbol{r},\delta)=\sqrt{\frac{1}{\mathcal{N}}\sum_{i=1}^\mathcal{N}\left[\psi\left(\boldsymbol{r}+\boldsymbol{\delta}_i\right)-\psi\left(\boldsymbol{r}\right)\right]^2}
$$
where the sum is performed over the $\mathcal{N}$ pixels $\boldsymbol{r}+\boldsymbol{\delta}_i$ whose distance to the central pixel $\boldsymbol{r}$ lies between $\delta/2$ and $3\delta/2$. For the sake of consistency with the analysis performed on simulated polarization maps in~\cite{planck2014-XX}, we take $\delta=16\arcmin$.

Finally we build the column density and optical depth $\tau_{353}$ maps from the dust density cube using 
\begin{equation*}
\NH = \int\nH\ud z \quad\text{ and }\quad \tau_{353} = \sigma_{353}\left(\NH\right) \times\NH
\end{equation*}
with the $\sigma_{353}\left(\NH\right)$ conversion\footnote{The conversion factor is given for a map of $\NH$ at a resolution of $30\arcmin$. Thus, before applying it pixel by pixel, we smooth the simulated $\NH$ map to $30\arcmin$ resolution, apply the conversion, then resample the resulting $\tau_{353}$ map at the original resolution.} from \cite{planck2013-p06b}. A temperature map $T_{\rm obs}$ is also created using the anti-correlation with the column density $\NH$ observed in the data. This "dust temperature map" does not pretend to model reality but since $T_\mathrm{d}$ is one of the parameters of the model, the fitting algorithm requires a map whose mean value should yield $T_\mathrm{d}$.

\section{Exploring the parameter space}
\label{sec:method}

The goal of this paper is to constrain the physical parameters of molecular clouds, in particular the spectral indices of the dust density and of the turbulent magnetic field, using {\Planck} polarization maps and a grid of model maps built as explained in the previous section.

\subsection{Parameter space}

The nine physical parameters that are explored in this paper using fBm simulations are summarized in Table~\ref{tab:priors}. They are sufficient to describe the one-point and two-point statistical properties of the dust density and magnetic field models. Note that unlike what was done in~\cite{planck2016-XLIV}, the field of view of the maps analysed in the following (approximately $12^\circ$) is too small to contain remarkable features that could be used to constrain the angle $\gamma_0$ that the mean magnetic field makes with the POS. In such small fields of view, there is a degeneracy between $y_B$ and $\gamma_0$ which cannot be lifted. Consequently, we chose not to try to fit for $\gamma_0$ and $y_B$, but for the ratio of the turbulent magnetic field RMS to the mean magnetic field in the POS, i.e.,
$$
y_B^{\rm POS} = \frac{\sigma_B}{B_0^{\mathrm{POS}}}=\frac{\sqrt{\left<\B_t^2\right>}}{||\B_{0}||\cos\gamma_0} = \frac{y_B}{\cos\gamma_0}.
$$

This analysis was applied to the Polaris Flare (see section~\ref{sec:Polaris}), so the priors are chosen to be flat over a reasonably large range, to cover the expected physical values of the molecular cloud under consideration, but they are necessary for the analysis to converge. The cloud's average column density is of the order $\langle\NH\rangle\approx 10^{21}\,\mathrm{cm}^{-2}$~\citep{planck2014-XX}. This value was used to set the range for the prior on the depth $d$ of the cube, with the limits of this range chosen in such a way that the average gas density $\langle\nH\rangle$ lies between 10 and $500\,\mathrm{cm}^{-3}$, a reasonable assumption for the Polaris Flare molecular cloud. This translates to a total cube depth $d$ between 0.5 and $32.5\,\mathrm{pc}$. The range used for the prior on $\beta_n$ is justified by a number of observational studies~\citep[see, e.g., the review by][]{hennebelle-falgarone-2012}, and that on $\beta_B$ is chosen based on the results from~\cite{vansyngel-et-al-2017}, but also on numerical studies of MHD turbulence~\citep[see, e.g.,][]{perez-et-al-2012,beresnyak-2014}. The fluctuation ratios $y_n$ and $y_B^{\rm{POS}}$ are explored in a logarithmic scale, as we are mainly interested in order of magnitude estimates for these parameters. The polarization maps being statistically identical when the angle $\chi_0$ of the POS projection of the mean magnetic field is shifted by $180^\circ$, the prior on this parameter is such that this periodicity is applied when the Metropolis algorithm~(see~\ref{sec:MCMC}) draws values outside the given range. The priors chosen for $T_\mathrm{d}$ and $p_0$ are very large and do not play a role in the fitting procedure.

\subsection{Comparing models with data}\label{sec:chi2}

To set constraints on the parameters listed in Table~\ref{tab:priors}, {we build a likelihood function, which expresses the probability that a given set of synthetic polarization maps reproduce adequately actual observational data. From the model Stokes maps $I_\mathrm{m}$, $Q_\mathrm{m}$, and $U_\mathrm{m}$, we derive a set of observables that are used in the likelihood function. These observables are given in Table~\ref{tab:observables}. More precisely, we use i) the mean values for the optical depth $\tau_{353}$ and the dust temperature $T_\mathrm{obs}$, ii) the distribution functions (one-point statistics) of the $I_\mathrm{m}$, $Q_\mathrm{m}$, $U_\mathrm{m}$, $p_\mathrm{MAS}$, $\psi$, $\mathcal{S}$, and $\tau_{353}/\left\langle \tau_{353}\right\rangle$ maps, iii) the power spectra (two-point statistics) of the $I_\mathrm{m}$ ,$Q_\mathrm{m}$, and $U_\mathrm{m}$ maps, and iv) the pixel-by-pixel anti-correlation between $\mathcal{S}$  and $p_\mathrm{MAS}$ underlined by~\cite{planck2014-XIX}. Indeed, we have found that the shape of this two-dimensional distribution function also depends on the model parameters. Many other observables were tested but we have retained only those which bring constraints on the model parameters. 

\begin{table}[htb]
\caption[]{Observables from polarization maps used to fit data.}
\label{tab:observables}
\begin{center}
\begin{tabular}{cc} \hline \hline \\ [-1ex]
Type  & From \\  [1ex] \hline \\ [-1ex]
Mean values & $\tau_{353}$, $T_\mathrm{obs}$ \\
Distribution function & $I_\mathrm{m}$, $Q_\mathrm{m}$, $U_\mathrm{m}$, $p_\mathrm{MAS}$, $\psi$, $\mathcal{S}$, $\tau_{353}/\left\langle \tau_{353}\right\rangle$ \\
Power spectrum & $I_\mathrm{m}$, $Q_\mathrm{m}$, $U_\mathrm{m}$ \\
Correlation & $\left\lbrace\mathcal{S},p_\mathrm{MAS}\right\rbrace$\\   [1ex]  \hline \\ [-1ex]
\end{tabular}
\end{center}
\end{table}

On the simulation side, $N_r=60$ model realizations per set of parameter values are generated with their observables, to be compared with data. The $N_r$ models differ by the random phases $\phi_X$ and $\phi_{A_\lambda}$ used to build the dust density and magnetic field cubes (see Eqs.~\ref{eq:Xfbm} and~\ref{eq:Btlambda}), and by the random realization of the noise applied to the model (Eq.~\ref{eq:addnoise}). We checked that 60 simulations represent a statistically large enough sample to get robust averages and dispersions for the observables. The statistical properties of the observables derived from the observational polarization data are thus compared with the observables from those 60 models, through the evaluation of a parameter $D^2$ which quantifies the distance between data and one random realisation of the model, averaged over the $N_r$ random realisations, with contributions associated to the various observables listed in Table~\ref{tab:observables}, i.e.,
 \begin{equation}
 \label{eq:chi2tot}
D^2 = \frac{1}{N_r}\sum_{i=1}^{N_{\rm r}} \left[ D^2_{ \mu} + \sum_{o}D^2_{{\rm DF}(o)} + \sum_{o} D^2_{P(o)}  + D^2_{\mathcal{S}-p_\mathrm{MAS}}\right].
\end{equation}
This quantity is subtly different from the usual $\chi^2$ (see Appendices~\ref{sec:appendix:L2distance} and~\ref{sec:appendix:chi2}). The first term in Eq.~\ref{eq:chi2tot} covers the observables $\left\langle \tau_{353}\right\rangle$ and $\left\langle T_\mathrm{obs}\right\rangle$, and quantifies the difference between these values in the simulated maps and in the data. The second sum extends over the observable maps $o$ in the set $\left\{I_\mathrm{m},Q_\mathrm{m},U_\mathrm{m},p_\mathrm{MAS},\psi,\mathcal{S},\tau_{353}/\left\langle \tau_{353}\right\rangle\right\}$ and quantifies the difference between the distribution functions (DF) of these observables in synthetic maps and those of the same observables in the data. The third sum extends over the observable maps $o$ in the set $\left\{I_\mathrm{m},Q_\mathrm{m},U_\mathrm{m}\right\}$ and quantifies the difference between the power spectra of the simulated maps and those of the same maps in the data. Finally, the last term quantifies the discrepancy between the two-dimensional joint DFs of $\mathcal{S}$ and $p_\mathrm{MAS}$ in the data and in synthetic maps. We detail the computation of these various terms in Appendix~\ref{sec:appendix:L2distance}.

\subsection{MCMC chains}
\label{sec:MCMC}

Given the vast parameter space to explore, we built a Monte Carlo Markov Chain method (MCMC)~\citep[see, e.g.][]{MCMC} that has the advantage to sample specifically the regions of interest in this space. We used a simple Metropolis-Hastings algorithm to build five Markov chains which sample the posterior probability distribution of the parameters listed in Table~\ref{tab:priors}. The likelihood $\mathcal{L}$ of a set $s$ of parameters is evaluated thanks to the $D^2$ criterion described in section~\ref{sec:chi2} and Appendix~\ref{sec:appendix:L2distance} as 
$$
\mathcal{L}(s) \propto e^{-D^2(s)/2}\pi(s)
$$ 
with $\pi(s)$ the prior associated to the parameters. 

According to the Metropolis-Hastings algorithm, at each step $q$ of the chain, parameters are drawn according to a multivariate probability distribution function whose covariance is set to allow for an efficient exploration of the parameter space, with an average given by the parameter values $s_{q-1}$ at step $q-1$. If the likelihood for the new set of parameters, $s_q$, is larger than for the previous one, then the chain records the new set. Otherwise, the likelihood ratio $\mathcal{L}\left(s_q\right)/\mathcal{L}\left(s_{q-1}\right)< 1$ is compared to a number $\alpha$ drawn randomly from a uniform distribution over $[0,1]$. If the likelihood ratio is larger than $\alpha$, the $s_q$ set of parameters is kept, otherwise the chain duplicates the $s_{q-1}$ set, i.e., $s_q=s_{q-1}$. The posterior probability distribution function is then given by the occurence frequency of the parameters along the chains, after removal of the initial "burn-in" phase.

The priors used for each parameter are detailed in Table~\ref{tab:priors}. For all parameters, flat priors are set covering a reasonable range of physical interest. If the Metropolis algorithm draws values outside of these priors we set $\pi(s)=0$, except for the position angle of the mean magnetic field, $\chi_0$, for which the $180^\circ$ periodicity is used to bring back the angle inside its definition range when it is perchance drawn outside. 

The convergence of the Markov chains is tested using the Gelman-Rubin statistics $R$ \citep{gelman1992}, which is essentially the ratio of the variance of the chain means to the mean of the chain variances. We estimate that the chains converged when $R-1<0.03$ for the least-converged parameter. The convergence is also assessed by checking visually the $D^2$ and parameter evolutions along the chains.

\begin{figure*}[htbp]
\centering
\includegraphics[width=0.33\textwidth]{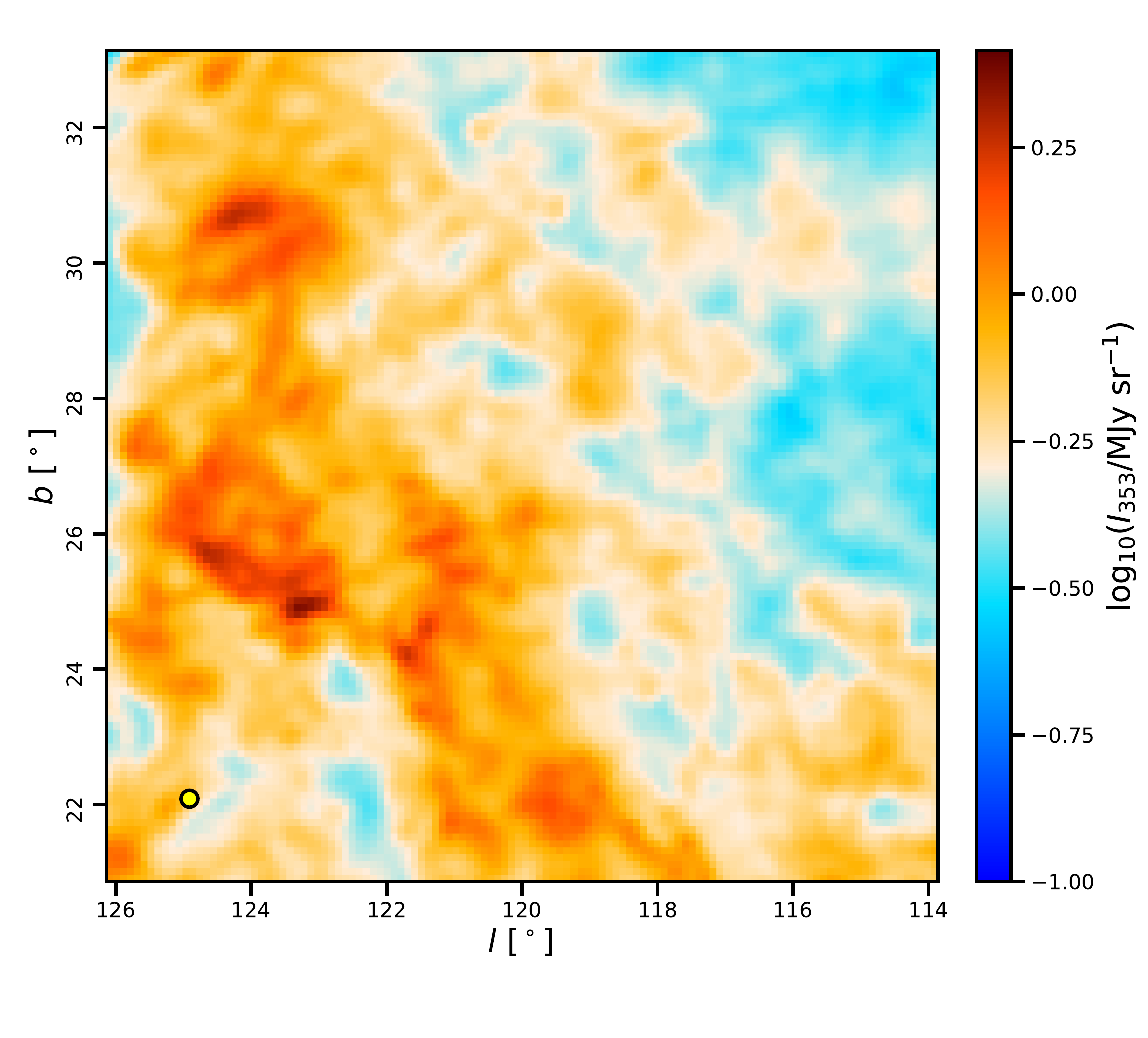}
\includegraphics[width=0.33\textwidth]{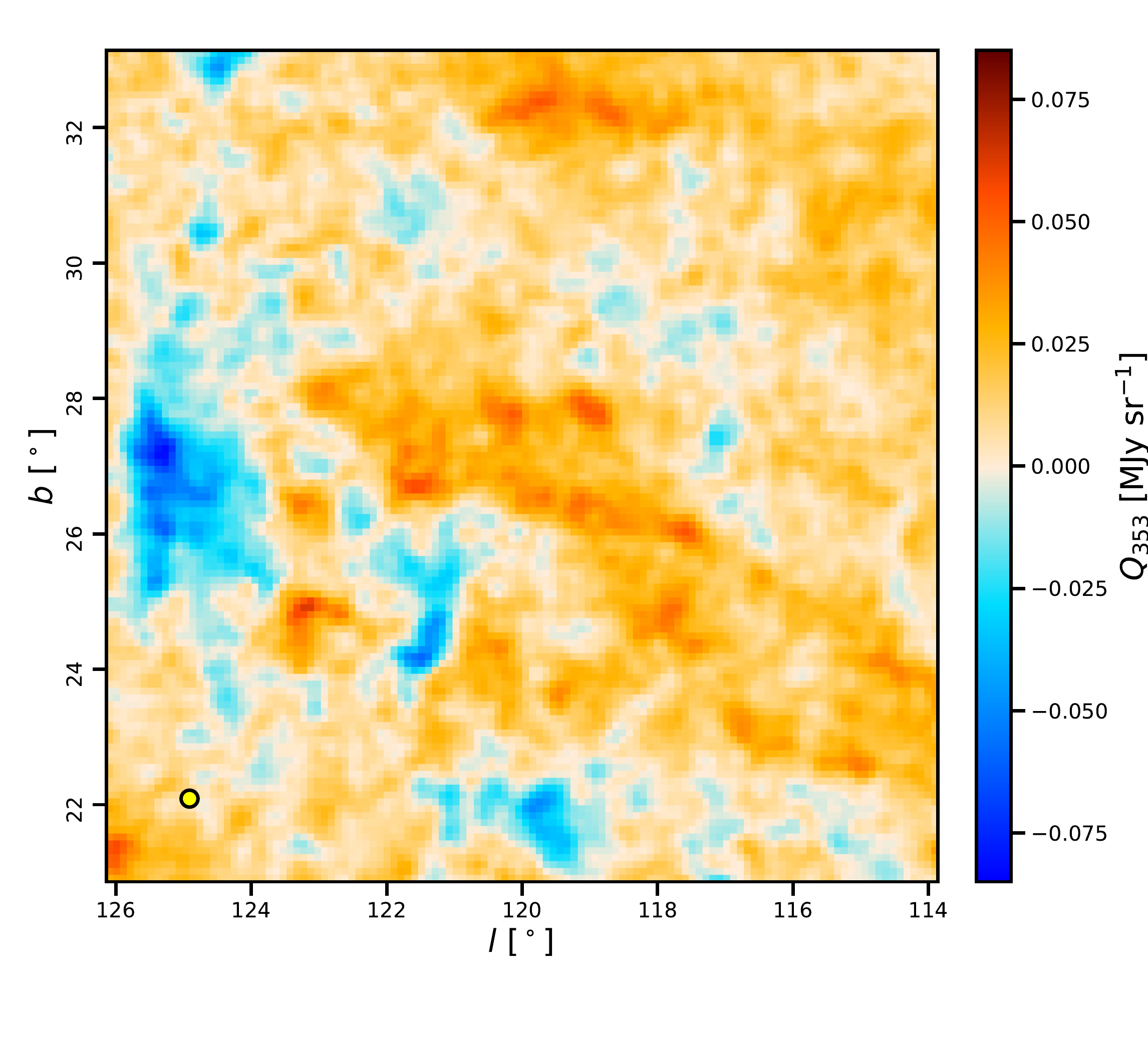}
\includegraphics[width=0.33\textwidth]{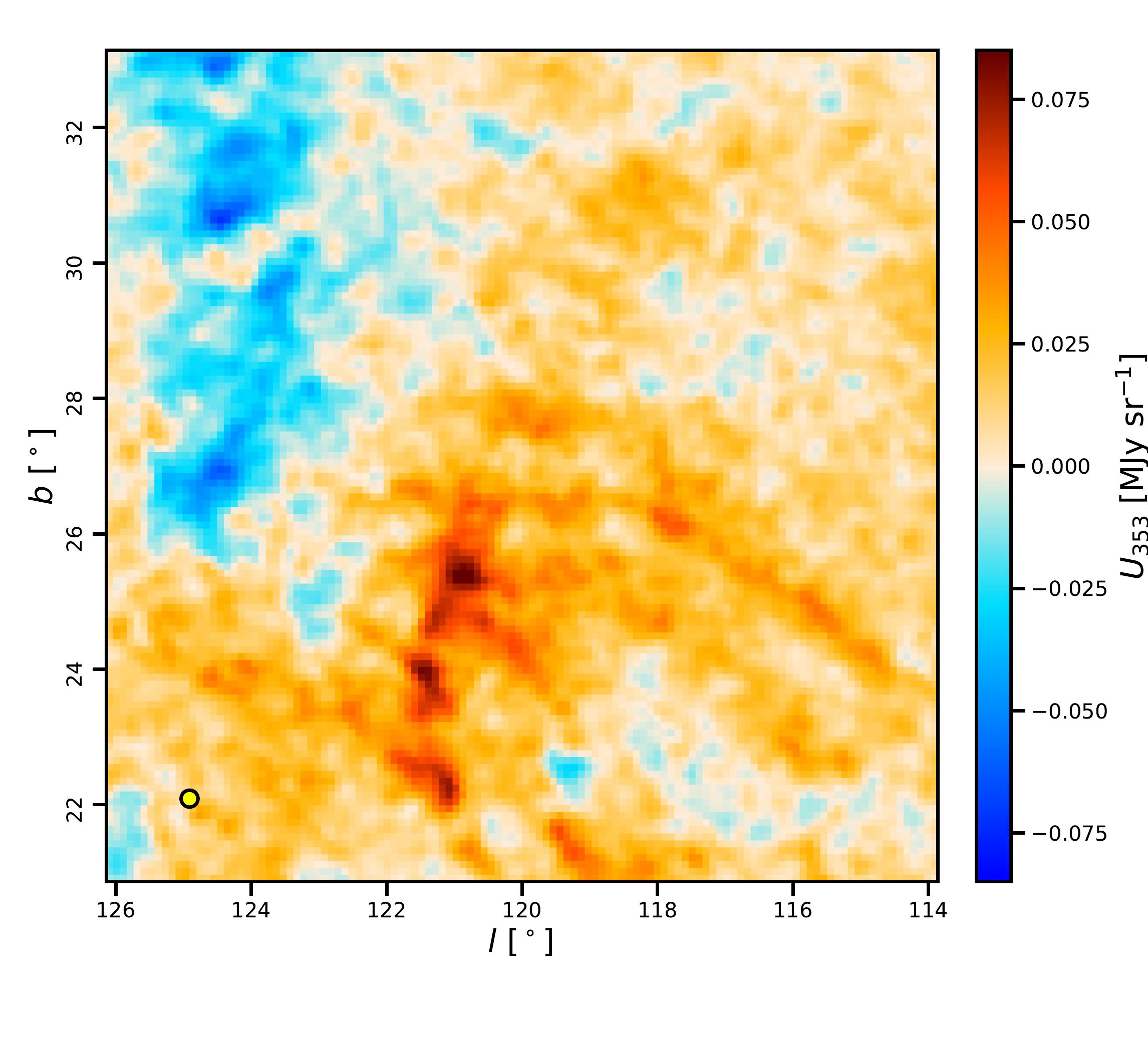}\\
\includegraphics[width=0.33\textwidth]{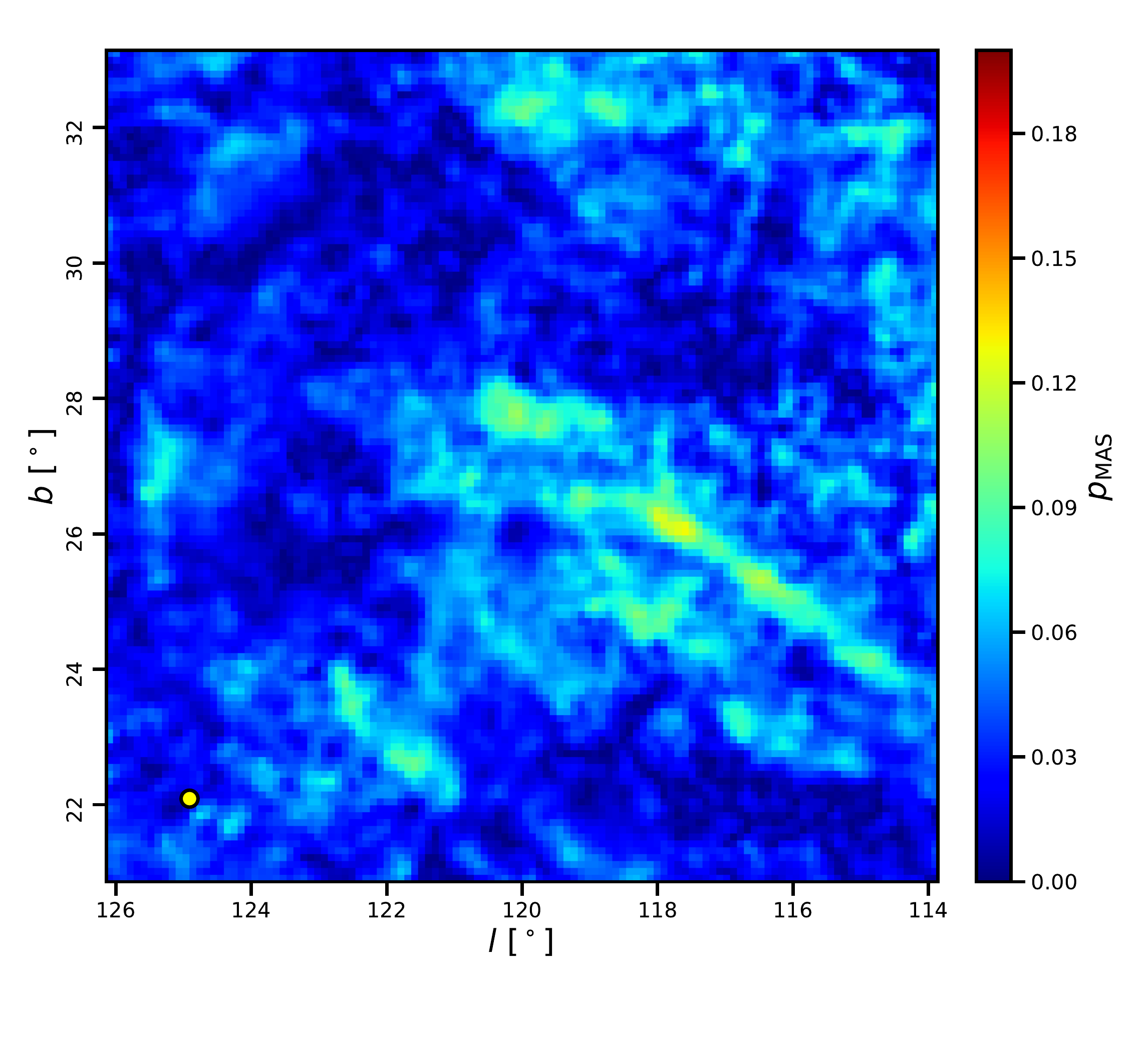}
\includegraphics[width=0.33\textwidth]{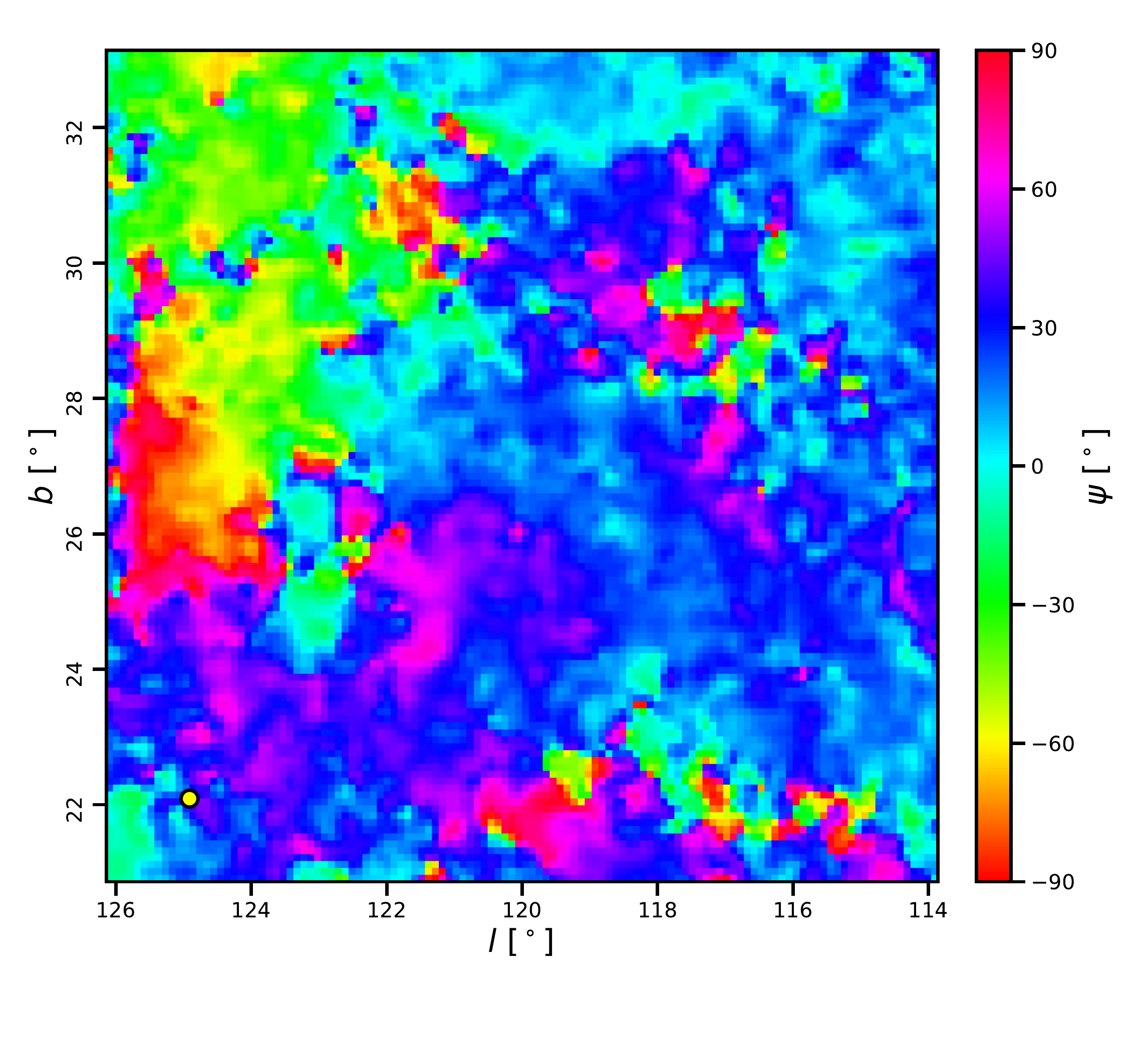}
\includegraphics[width=0.33\textwidth]{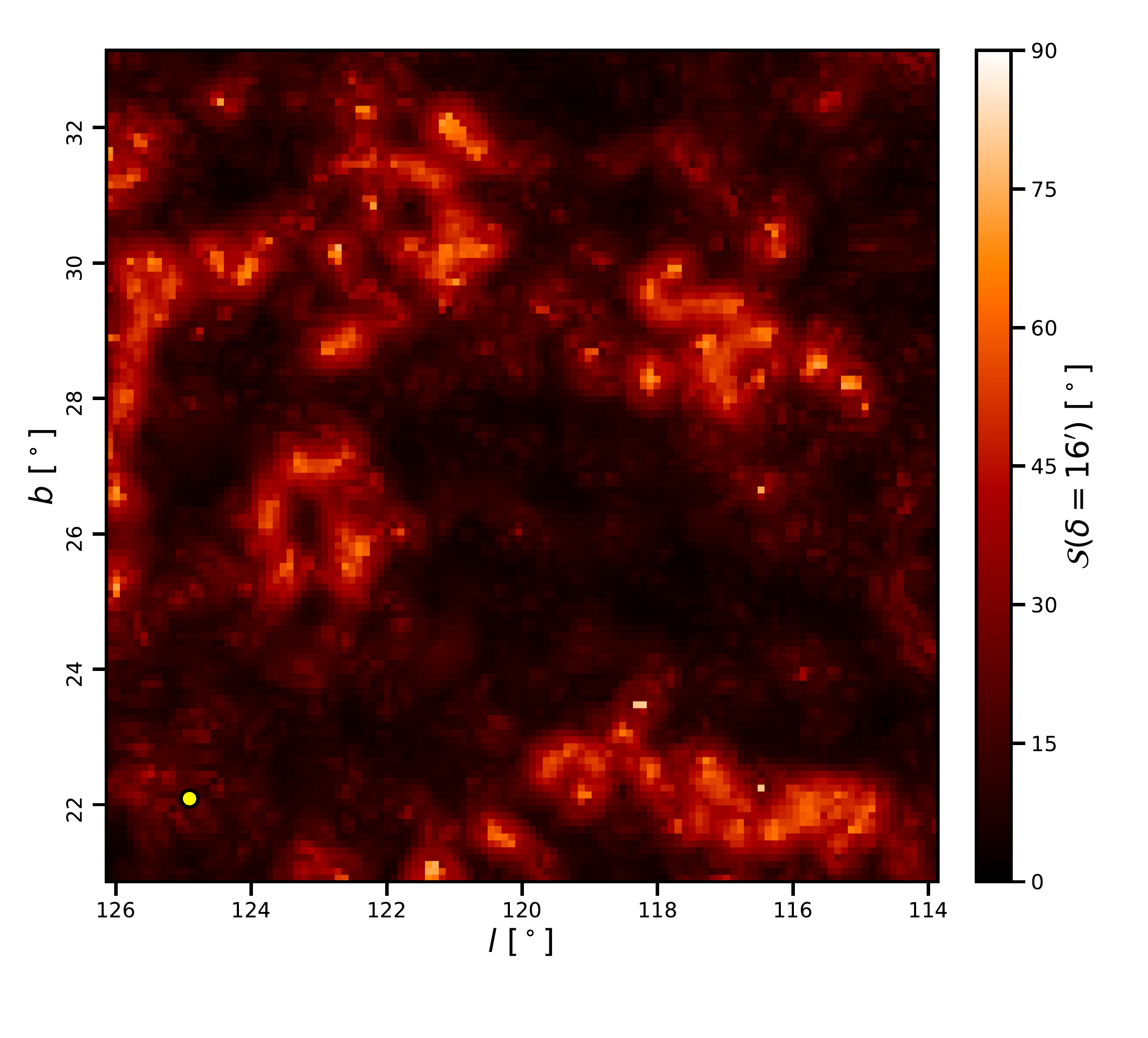}
\caption{{\Planck} 353\,GHz maps of the Polaris Flare molecular cloud. The top row shows, from left to right, the total intensity $I_{353}$ on a logarithmic scale, the Stokes $Q_{353}$ map and the Stokes $U_{353}$ map, while the bottom row shows the polarization fraction $p_{\rm MAS}$, the polarization angle $\psi$, and the polarization angle dispersion function $\mathcal{S}$. The $\tau_{353}$ and $T_{\rm obs}$ maps have the same aspects as the $I_{353}$ map but with their own scales.}
\label{fig_polaris_map}
\end{figure*}

\begin{figure*}[htbp]
\resizebox{\hsize}{!}{
\includegraphics{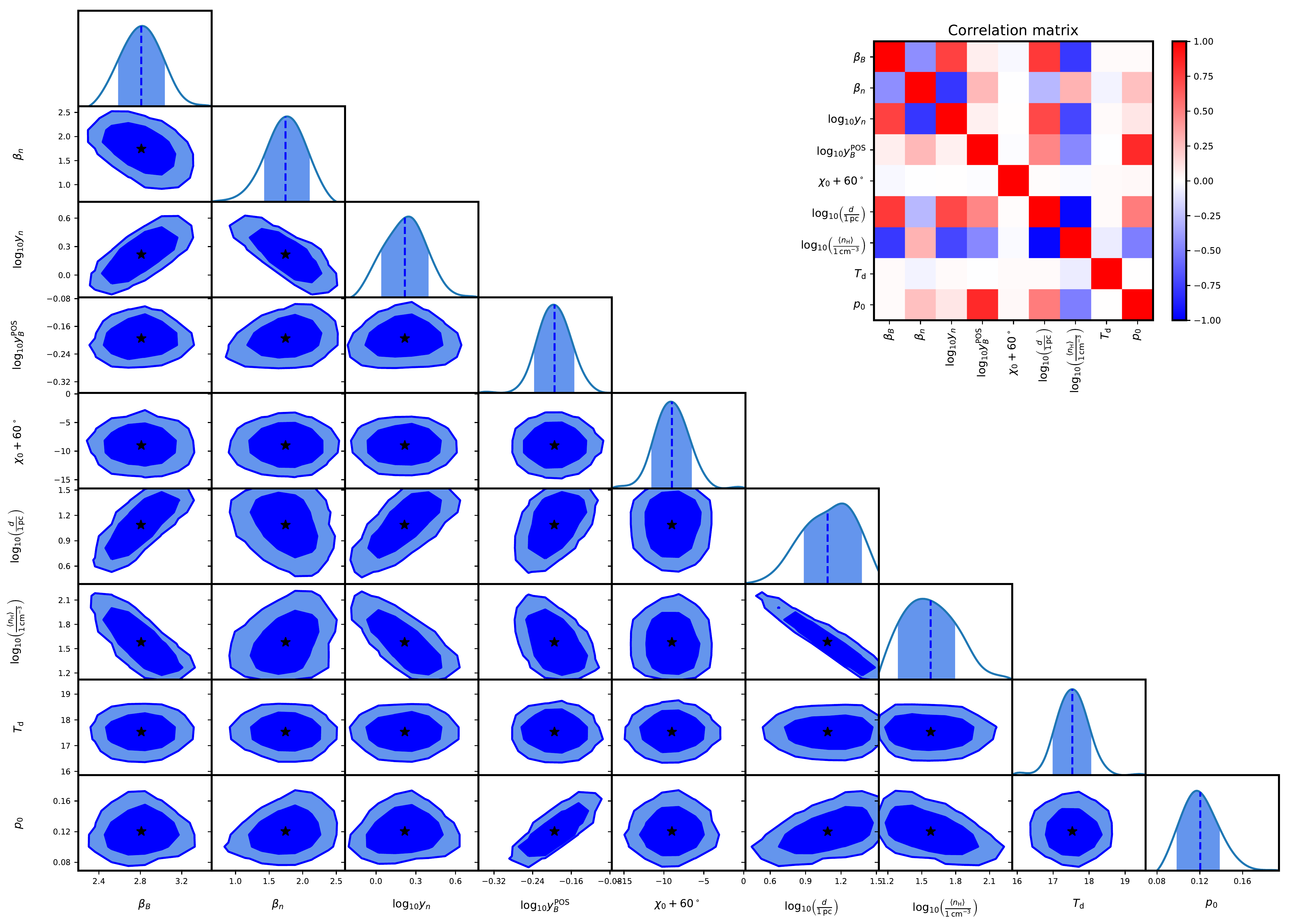}
}
\caption{Constraints (posterior probability contours and marginalized PDFs) on the statistical properties of the dust density and magnetic field for the {\Planck} maps of the Polaris Flare. On the posterior probability contours, the filled dark and light blue regions respectively enclose 68.3\% and 95.4\% of the probability, and the black stars indicate the averages over the two-dimensional posterior PDFs. In the plots showing the marginalized posterior PDFs, the light blue regions enclose 68.3\% of the probability, and the dashed blue lines indicate the averages over the posterior PDFs. The upper right plot displays the correlation matrix between the fitted parameters.}
\label{fig_triangle}
\end{figure*}

\begin{table*}
\caption{Best fit values from four fBm simulations using the observables from Table~\ref{tab:observables}. The column $\left<\chi^2_{\rm{best}}\right>$ shows the $\chi^2$ values for the best fit parameters averaged over 100 fits (see Appendix~\ref{sec:appendix:chi2}).}
\label{tab:results_sim}
\begin{center}
\resizebox{\textwidth}{!}{\begin{tabular}{ccccccccccc} \hline \hline \\ [-1ex]
Parameters  & $\beta_B$  & $\beta_n$ & $\log_{10}y_n$ & $\log_{10} y_B^{\rm POS}$ & $\chi_0$ [$^\circ$] & $\log_{10}\left(\frac{d}{1\,\mathrm{pc}}\right)$  & $\log_{10}\left(\frac{\langle\nH\rangle}{1\,\mathrm{cm}^{-3}}\right)$ & $T_\mathrm{d}$ [K] & $p_0$  & $\left<\chi^2_{\rm{best}}\right>$\\  [1ex] \hline \\ [-1ex]
\multicolumn{10}{c}{\bf Simulation A}  \\  [1ex] 
Input parameters & 2.6 & 2.08 & -0.10 & -0.22 & $-50$ & 1.00 & 1.48 & 18.0 & 0.12  \\  [1ex] 
Best fit values & $2.8^{+0.2}_{-0.2}$ & $1.9^{+0.3}_{-0.2}$ & $0.03^{+0.15}_{-0.15}$ & $-0.26^{+0.05}_{-0.05}$ & $-50^{+2}_{-2}$ & $1.2^{+0.2}_{-0.2}$ & $1.3^{+0.1}_{-0.2}$ & $18.0^{+0.5}_{-0.5}$ & $0.11^{+0.02}_{-0.03}$ & 1.3 \\  [1ex]  \hline \\ [-1ex]
\multicolumn{10}{c}{\bf Simulation B}  \\  [1ex] 
Input parameters & 2.6 & 2.09 & -0.10 & -0.22 & $-70$ & 0.70 & 2.00 & 20.0 & 0.15  \\  [1ex] 
Best fit values & $2.7^{+0.1}_{-0.2}$ & $1.9^{+0.3}_{-0.2}$ & $-0.01^{+0.13}_{-0.20}$ & $-0.24^{+0.03}_{-0.04}$ & $-70^{+2}_{-2}$ & $0.8^{+0.2}_{-0.3}$ & $2.0^{+0.3}_{-0.2}$ & $20.0^{+0.5}_{-0.5}$ & $0.13^{+0.02}_{-0.02}$ & 1.7\\  [1ex]  \hline \\ [-1ex]
\multicolumn{10}{c}{\bf Simulation C}  \\  [1ex] 
Input parameters & 3.0 & 2.8 & 0.0 & -0.10 & $-30$ & 1.18 & 1.30 & 22.0 & 0.2  \\  [1ex] 
Best fit values & 	$2.8^{+0.1}_{-0.2}$ & $2.6^{+0.3}_{-0.2}$ & $0.00^{+0.10}_{-0.11}$ & $-0.08^{+0.03}_{-0.04}$ & $-30^{+3}_{-4}$ & $1.1^{+0.2}_{-0.2}$ & $1.4^{+0.2}_{-0.2}$ & $22.1^{+0.5}_{-0.5}$ & $0.21^{+0.03}_{-0.03}$ & 1.8\\  [1ex]  \hline \\ [-1ex]
\multicolumn{10}{c}{\bf Simulation D}  \\  [1ex] 
Input parameters & 2.0 & 1.87 & -0.22 & -0.10 & $-10$ & 0.70 & 2.18 & 16.0 & 0.1  \\  [1ex] 
Best fit values & 	$2.2^{+0.2}_{-0.2}$ & $1.7^{+0.3}_{-0.3}$& $-0.1^{+0.2}_{-0.2}$ & $-0.08^{+0.06}_{-0.05}$ & $-8^{+2}_{-2}$ & $1.0^{+0.3}_{-0.3}$ & $1.9^{+0.3}_{-0.3}$ & $16.0^{+0.5}_{-0.5}$ & $0.11^{+0.03}_{-0.03}$ & 0.7 \\  [1ex]  \hline \\ [-1ex]
\end{tabular}}
\end{center}
\end{table*}

The obtained 9D posterior probability distribution is generally not a multivariate Gaussian distribution. To quote an estimate of the best fit value for any one of the nine parameters and the associated uncertainties, we first marginalize over the other eight parameters to obtain the one-dimensional posterior PDF for the remaining parameter. In the following, the quoted best fit value for a parameter is the mean over this posterior PDF (which is less sensitive to binning effect than the maximum likelihood). As the PDFs are usually not Gaussian, we quote asymmetric error bars following the minimum credible interval technique \citep[see, e.g.,][]{hamann2007}.

\section{Results}
\label{sec:results}

\subsection{Validation of the method}
\label{sec:validation}

To validate the fitting method, we simulated four sets of model cubes and computed the corresponding $I_\mathrm{m}$, $Q_\mathrm{m}$, and $U_\mathrm{m}$ maps, including noise, with different values of the input parameters (these are labelled simulations A, B, C and D hereafter). The MCMC fitting procedure was run on these mock polarization data sets to check if it was able to recover the statistical properties of the input dust density and magnetic field cubes through the selected observables. The results are presented in Table~\ref{tab:results_sim}. For the four sets of maps, the fitting method recovered the input values within the quoted uncertainties, after the convergence criteria for all the chains were reached\footnote{And after removing the burn-in phase, which is quite short in our case ($\lesssim 30\%$ of the chain lengths in general).
}. This shows that this choice of observables is relevant to extract the input values from polarized thermal dust emission data within our model. To assess the goodness of fit of the model to the data, we use an {\it a posteriori} $\chi^2$ test, as explained in Appendix~\ref{sec:appendix:chi2}.} In all four cases, we find that the match is very good, since $\left<\chi^2_{\rm{best}}\right>\approx 1$. For illustration, the posterior probability contours for simulation A are presented in Figure~\ref{fig_triangle_simA}. We note that the MCMC procedure reveals correlations between the model parameters, which is not unexpected, e.g., between $y_B^\mathrm{POS}$ and $p_0$, or between $y_n$ and $\langle\nH\rangle$. These trends are best visualized with the correlation matrix, shown in the upper right corner of Fig.~\ref{fig_triangle_simA}.

\begin{figure*}[htbp]
\centering
\includegraphics[width=0.33\textwidth]{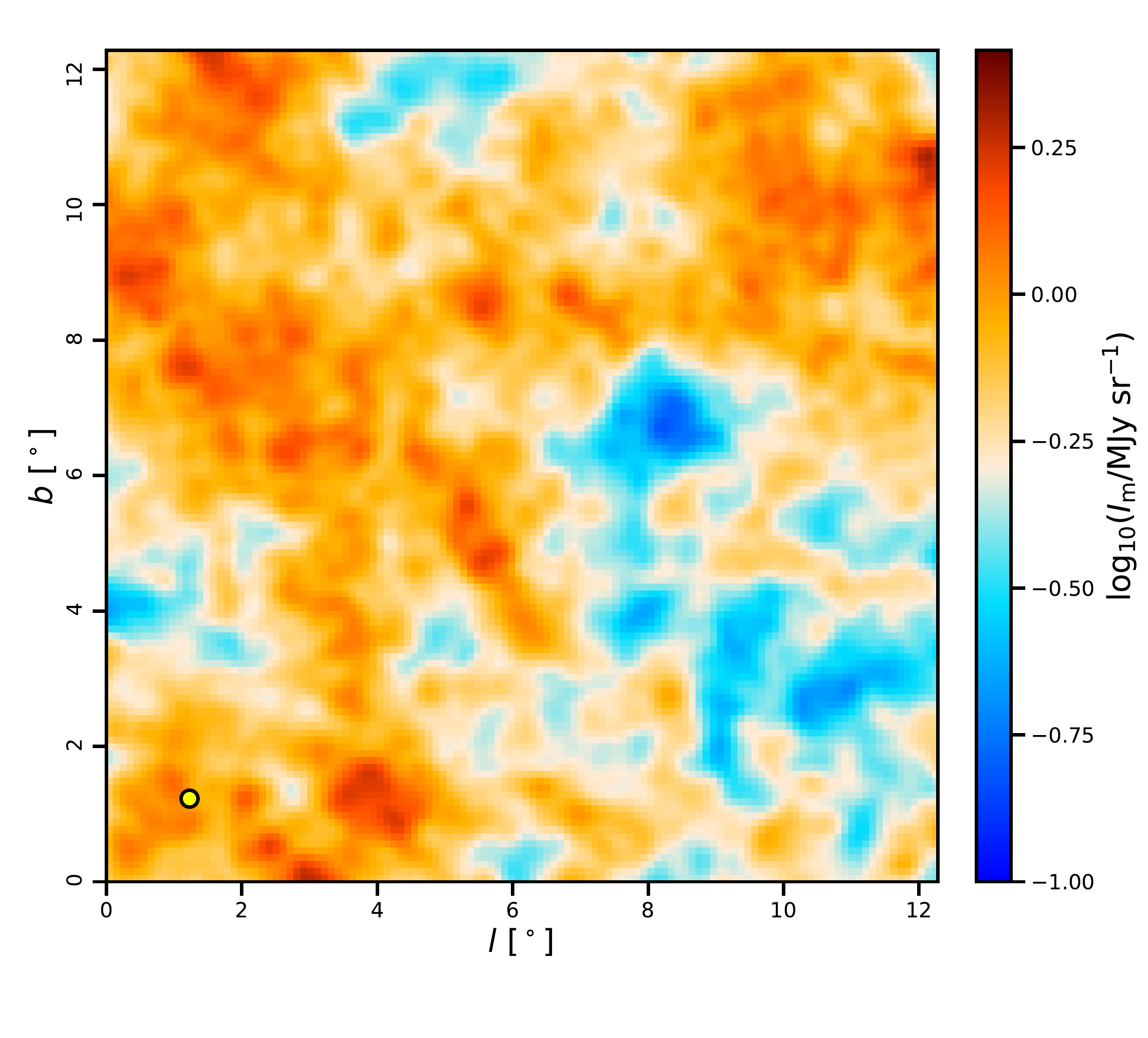}
\includegraphics[width=0.33\textwidth]{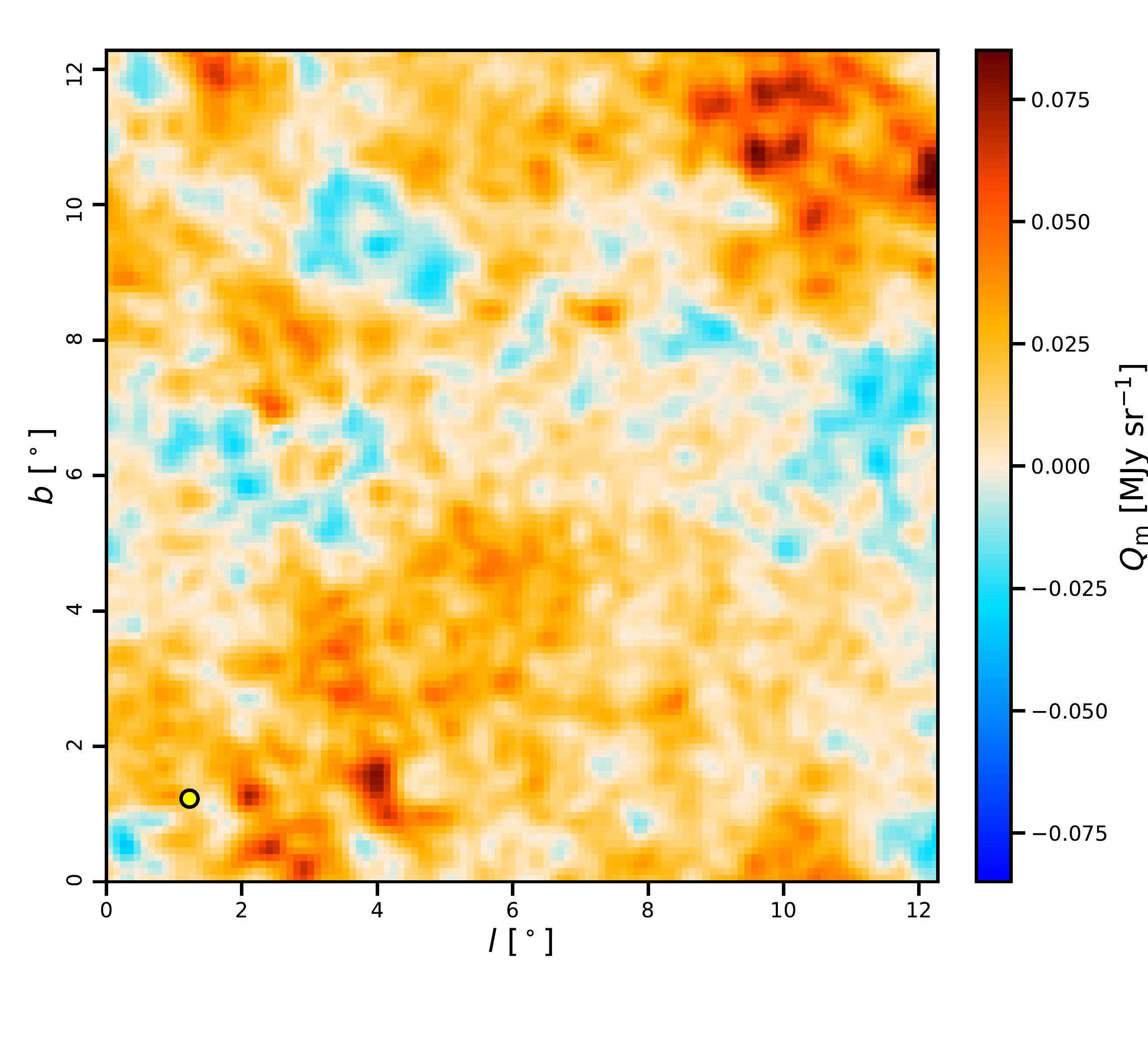}
\includegraphics[width=0.33\textwidth]{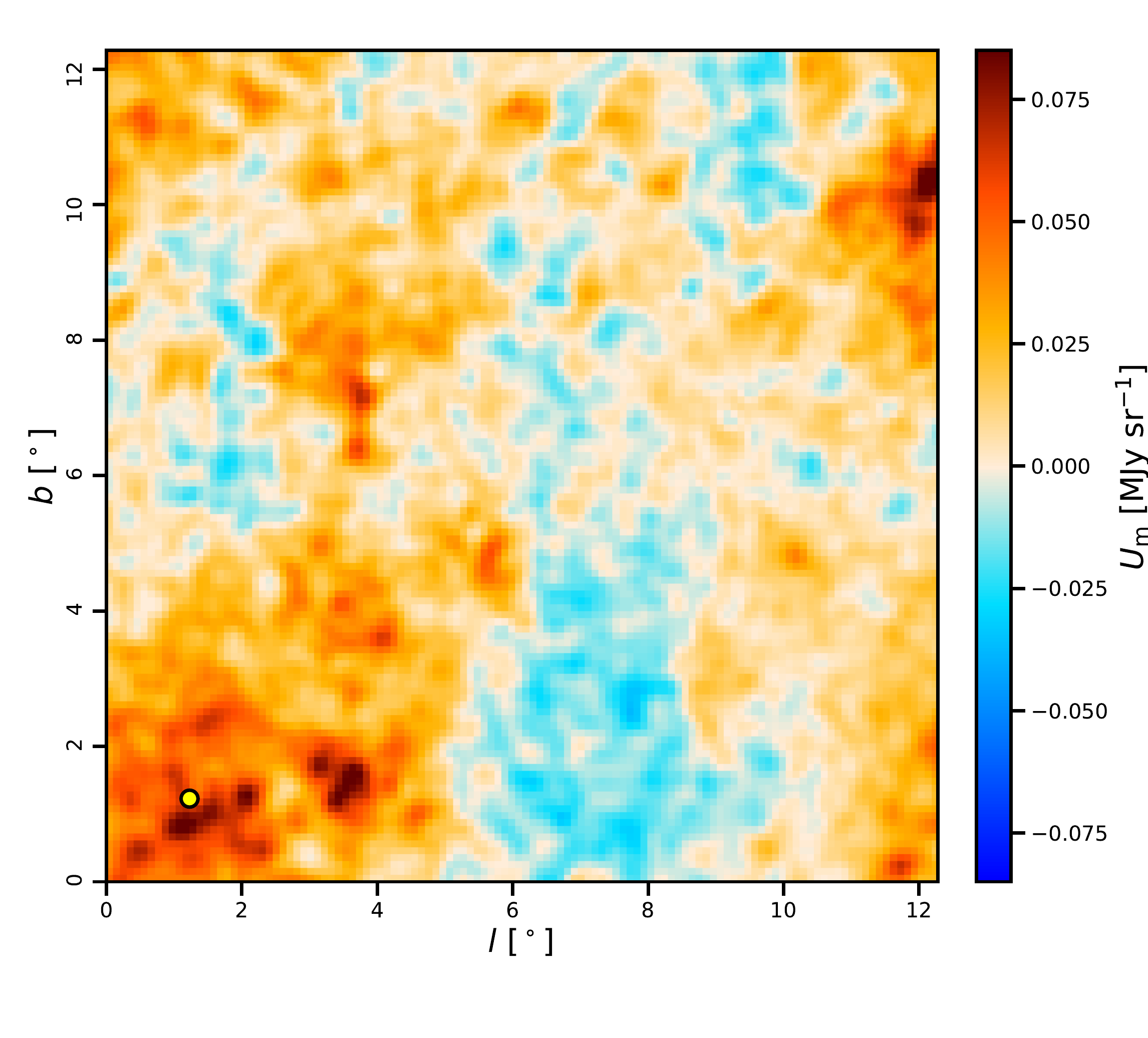}\\
\includegraphics[width=0.33\textwidth]{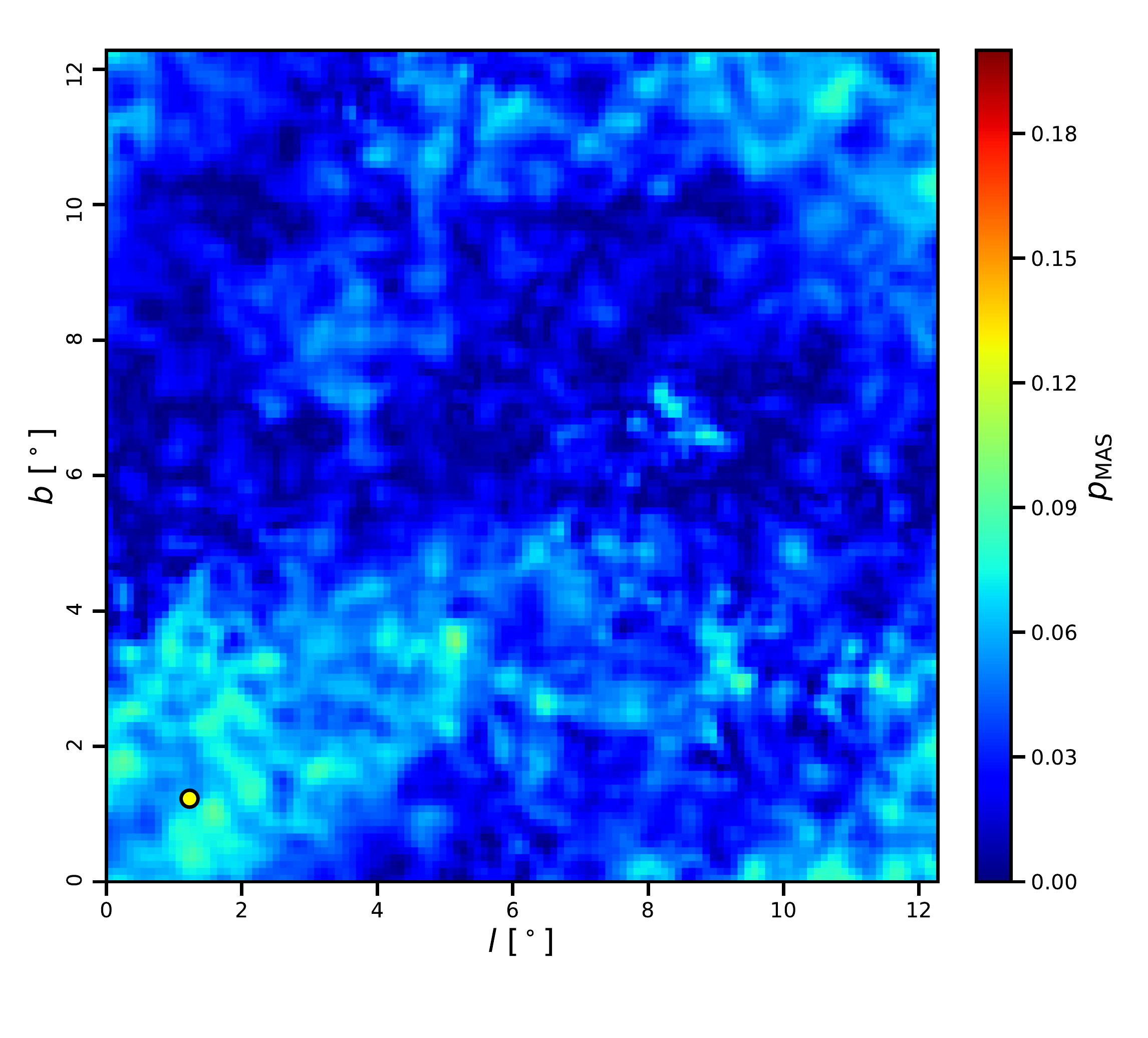}
\includegraphics[width=0.33\textwidth]{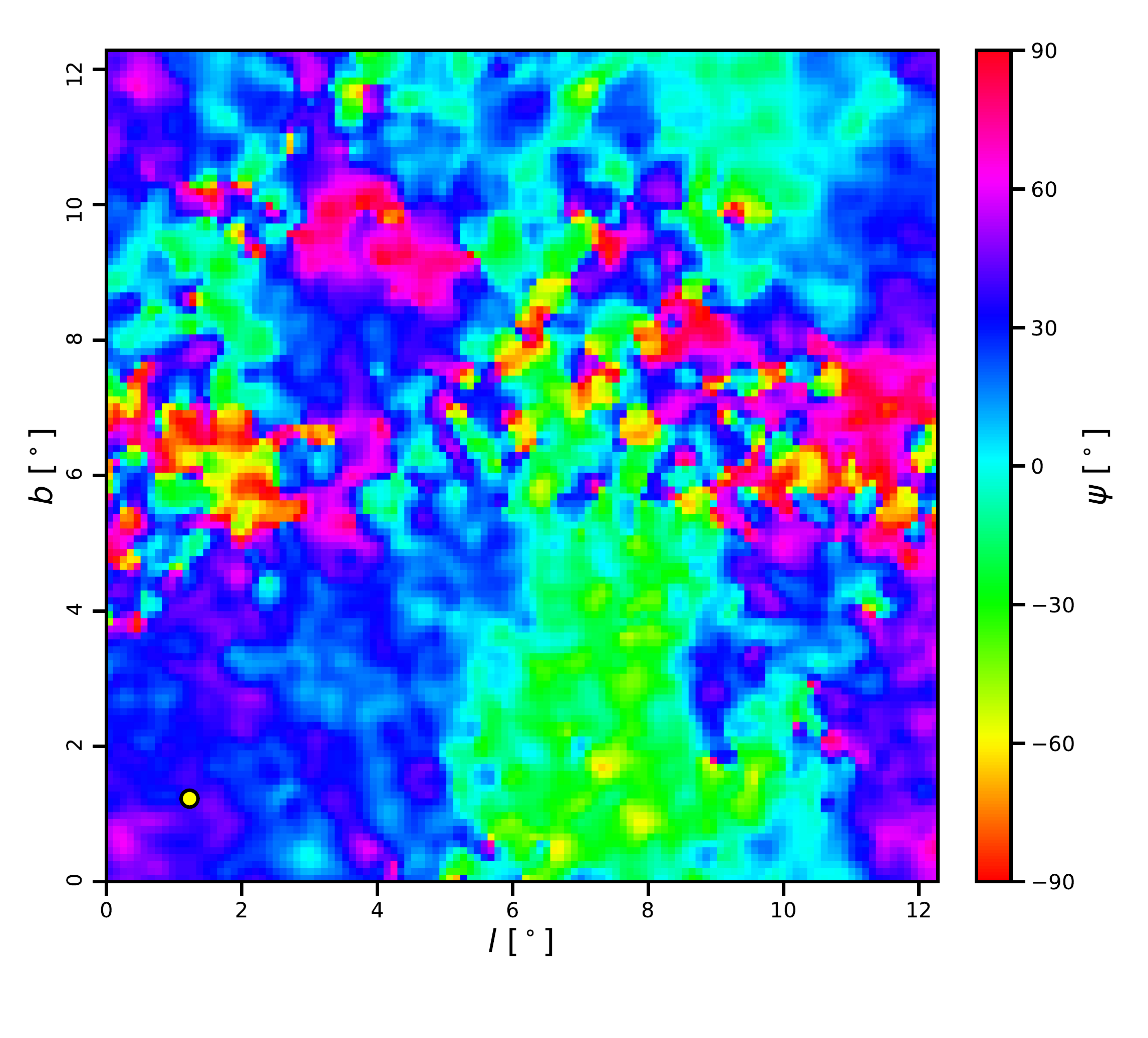}
\includegraphics[width=0.33\textwidth]{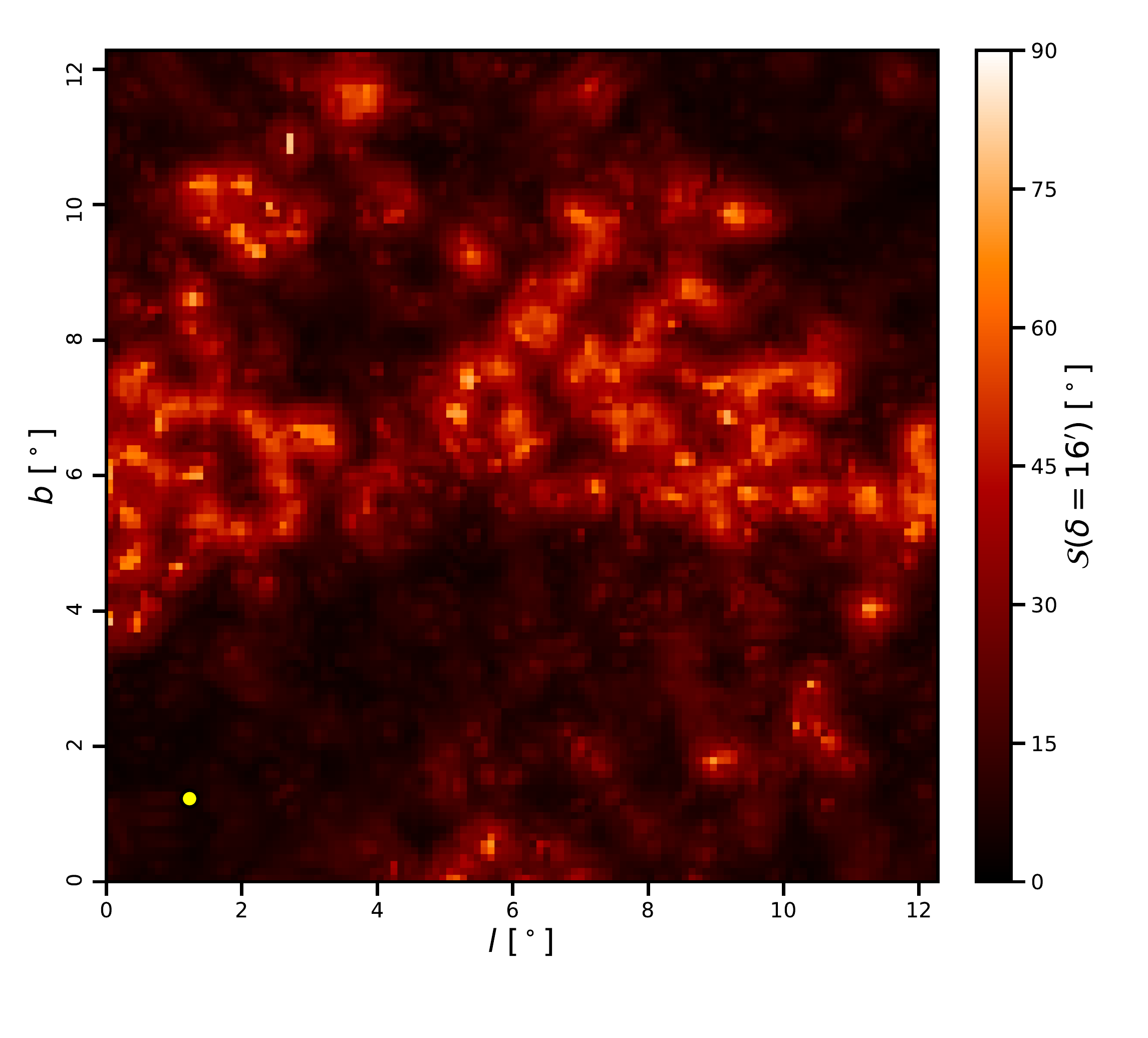}
\caption{Same as Fig.~\ref{fig_polaris_map} with the same color scales, but for model maps using the best fitting parameters to the Polaris Flare data.}
\label{fig_best_fit_map}
\end{figure*}

\begin{figure*}[htbp]
\centerline{
\includegraphics[width=0.4\textwidth]{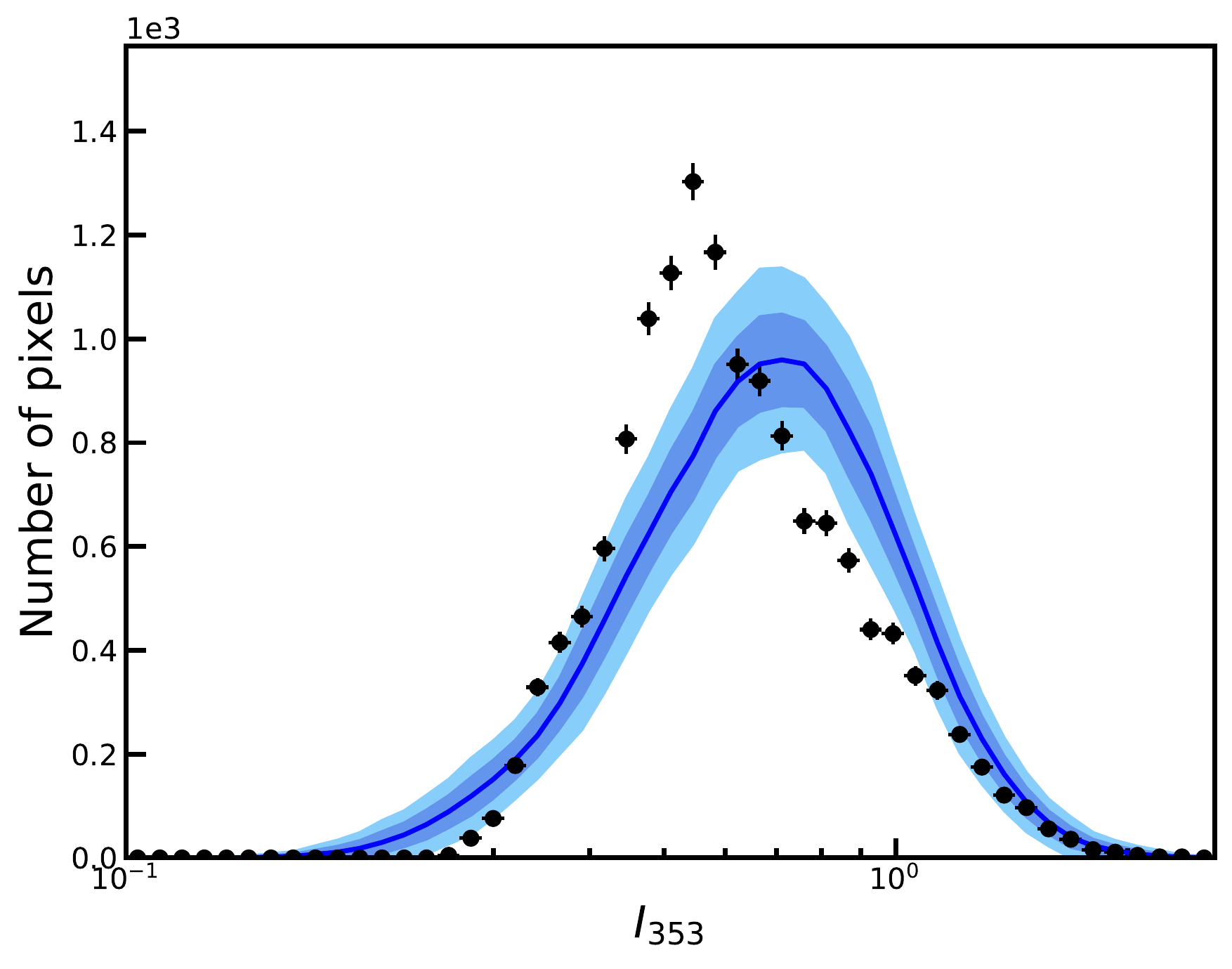}
\includegraphics[width=0.4\textwidth]{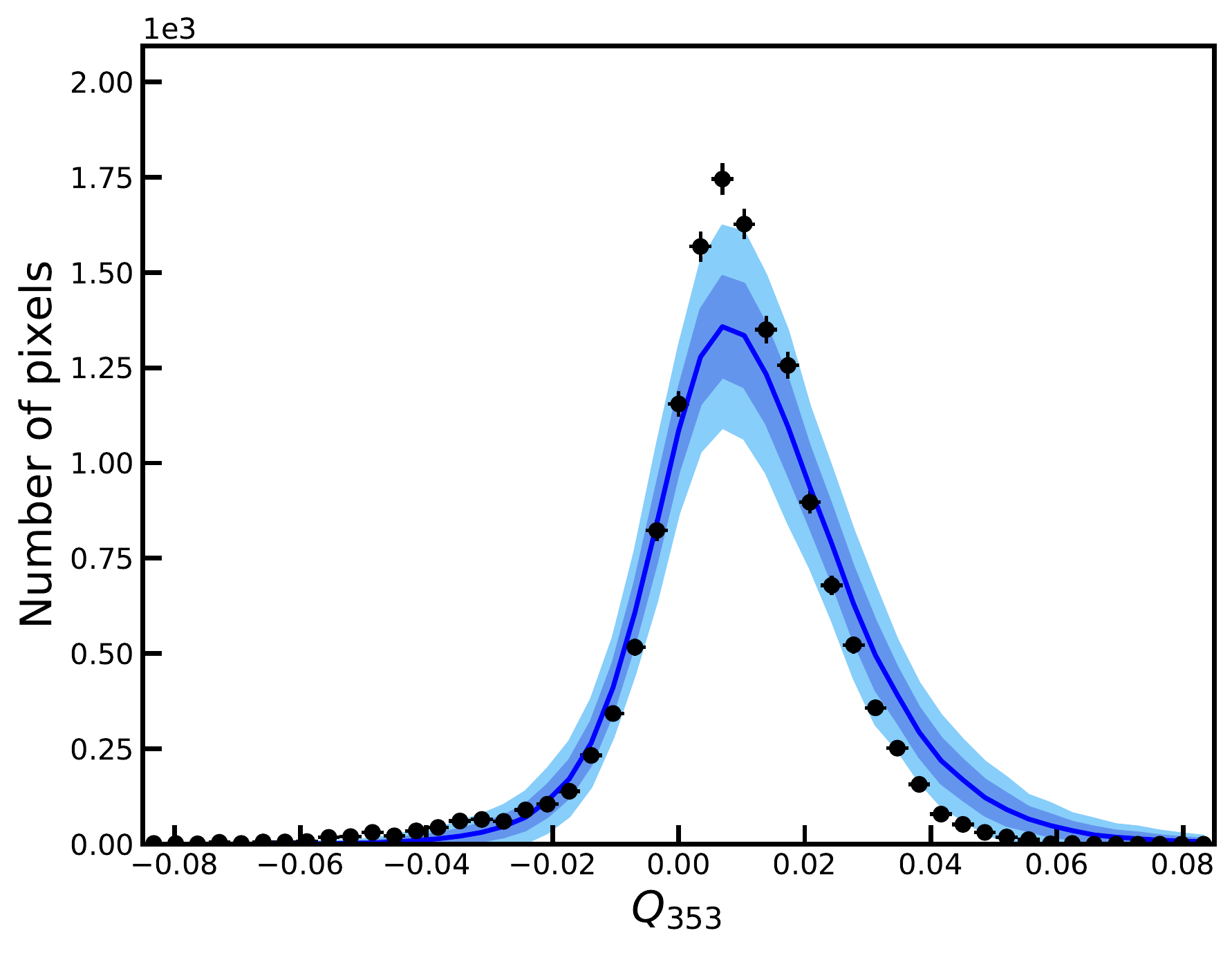}
}
\centerline{
\includegraphics[width=0.4\textwidth]{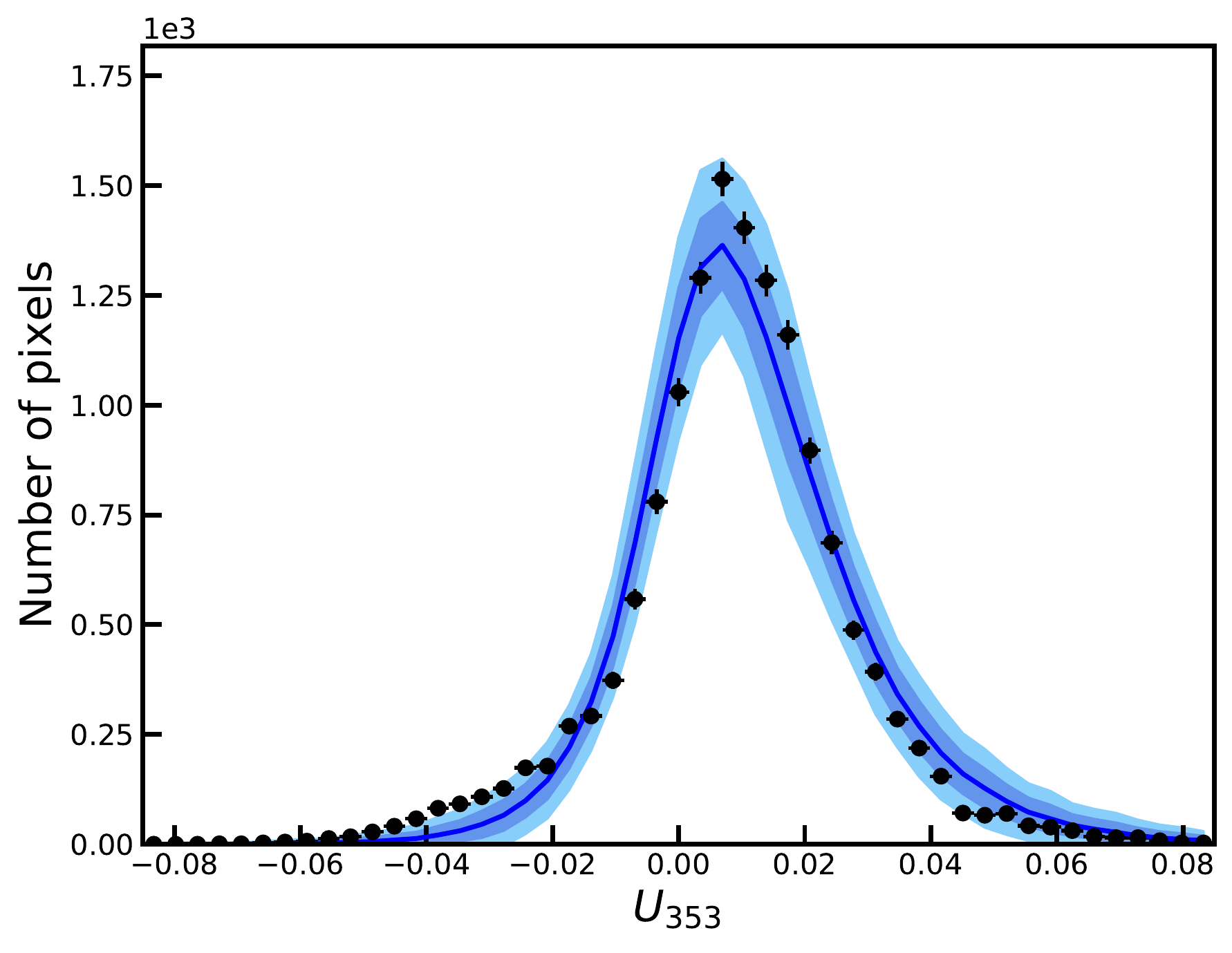} 
\includegraphics[width=0.4\textwidth]{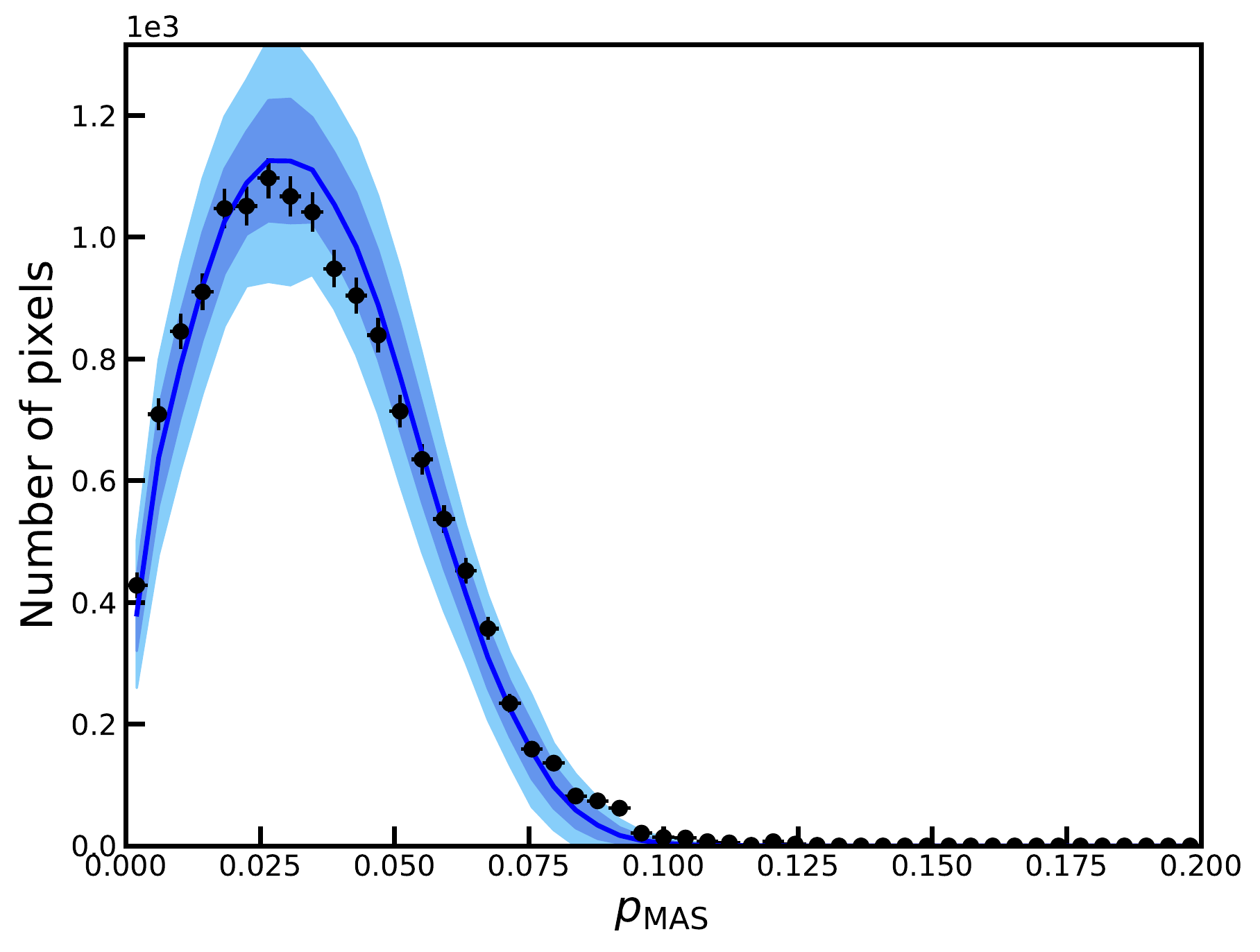}
}
\centerline{
\includegraphics[width=0.4\textwidth]{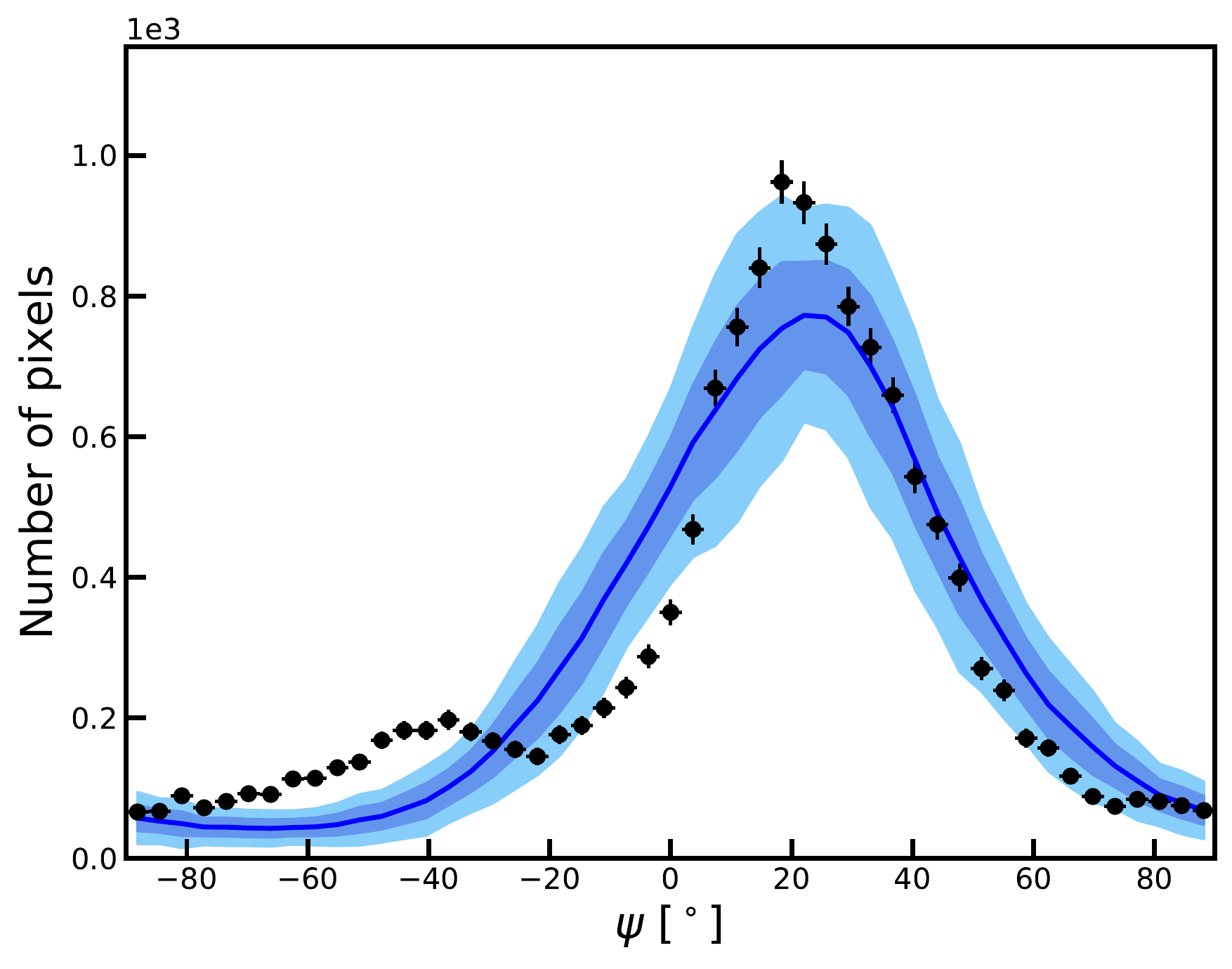}
\includegraphics[width=0.4\textwidth]{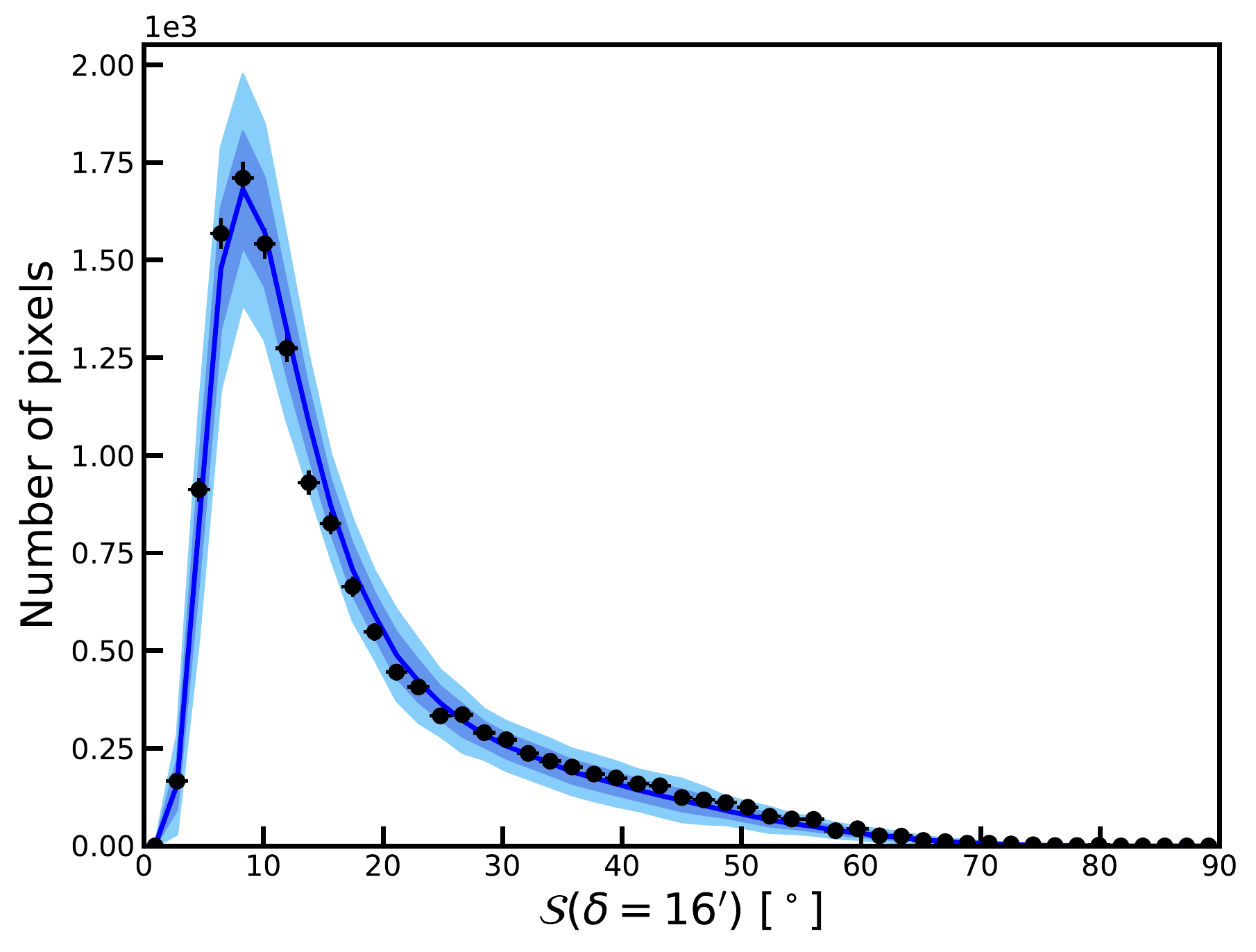}
}
\centerline{
\includegraphics[width=0.4\textwidth]{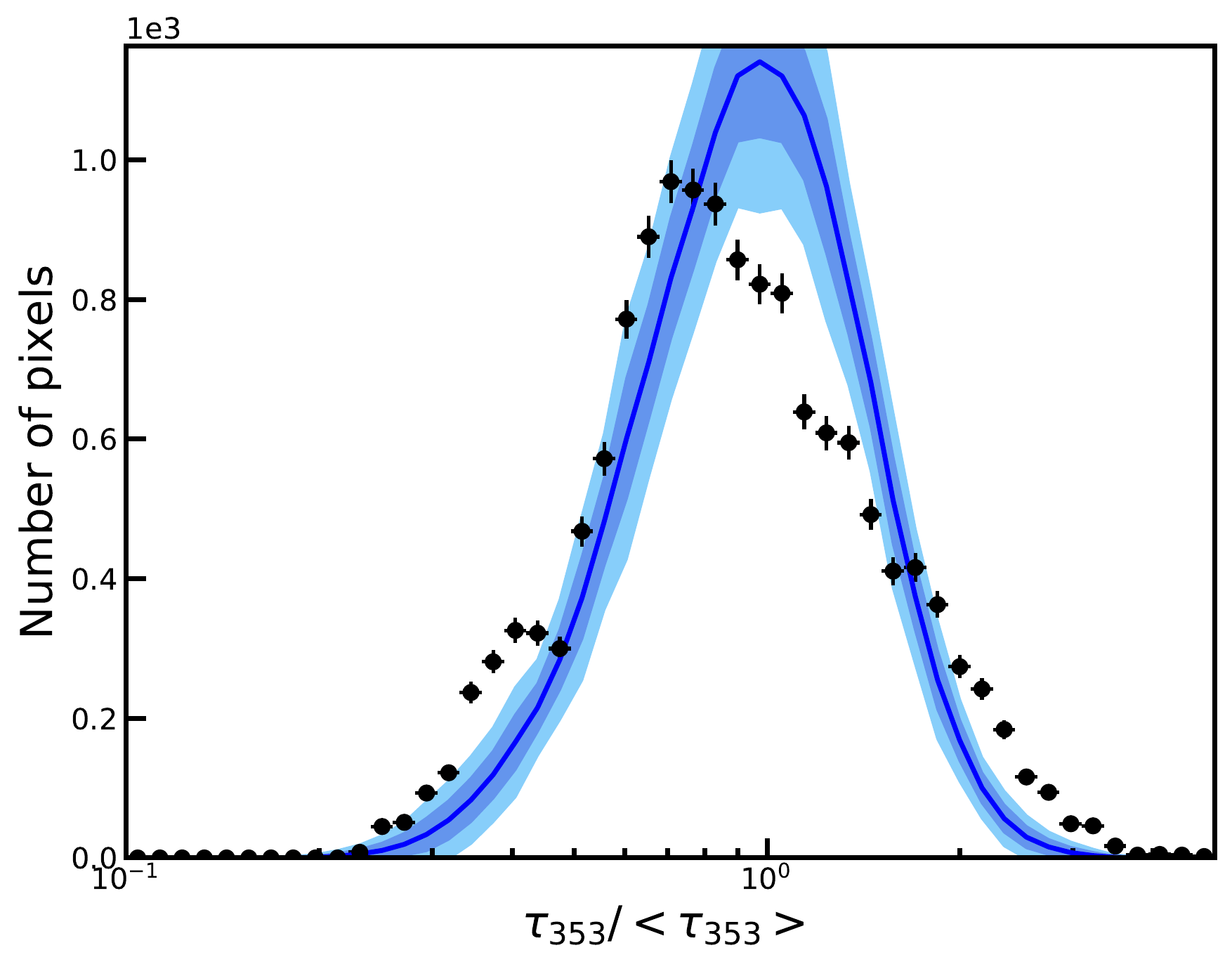}
}
\caption{Comparison of the DFs extracted from the {\Planck} Polaris Flare maps (black points) with the observables computed from simulations using the best fitting parameters (blue curves). The latter curves are averaged over 60 realizations, as described in section~\ref{sec:chi2} : the average is given by the central blue curve and the shaded bands give the $1\sigma$ and $2\sigma$ standard deviation in each bin.}
\label{fig_best_fit_pdf}
\end{figure*}

\begin{figure*}[htbp]
\centering
\includegraphics[width=0.33\textwidth]{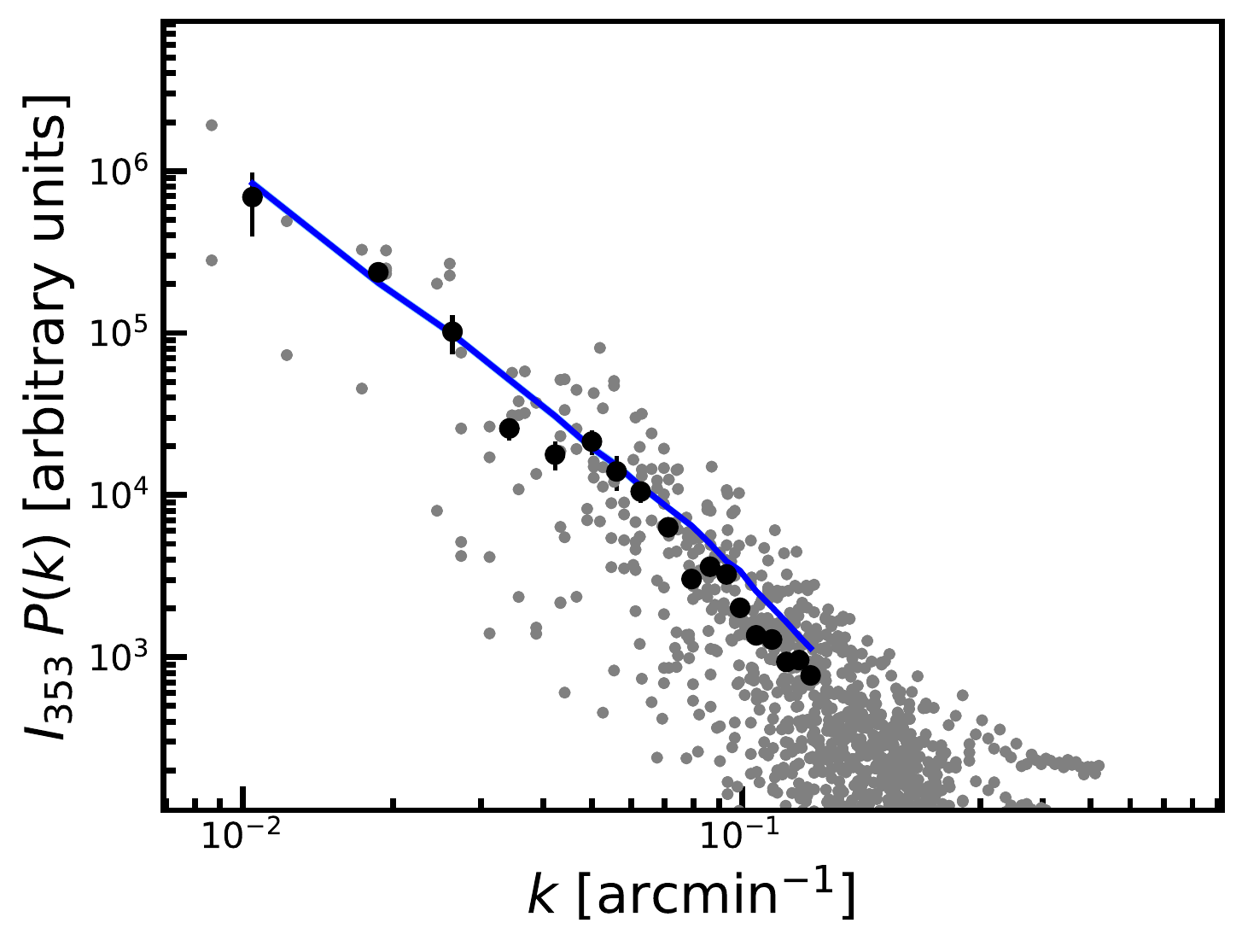}
\includegraphics[width=0.33\textwidth]{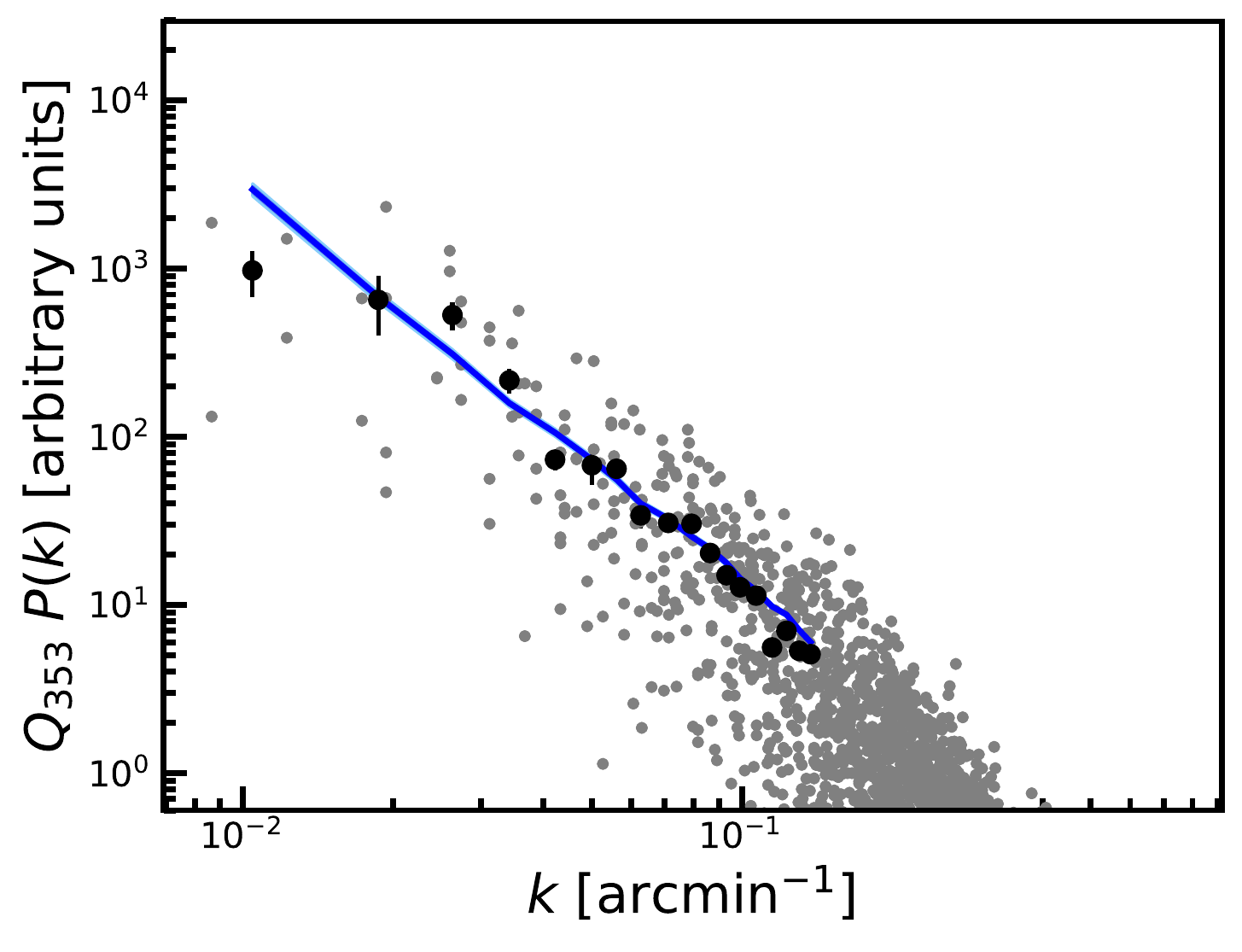}
\includegraphics[width=0.33\textwidth]{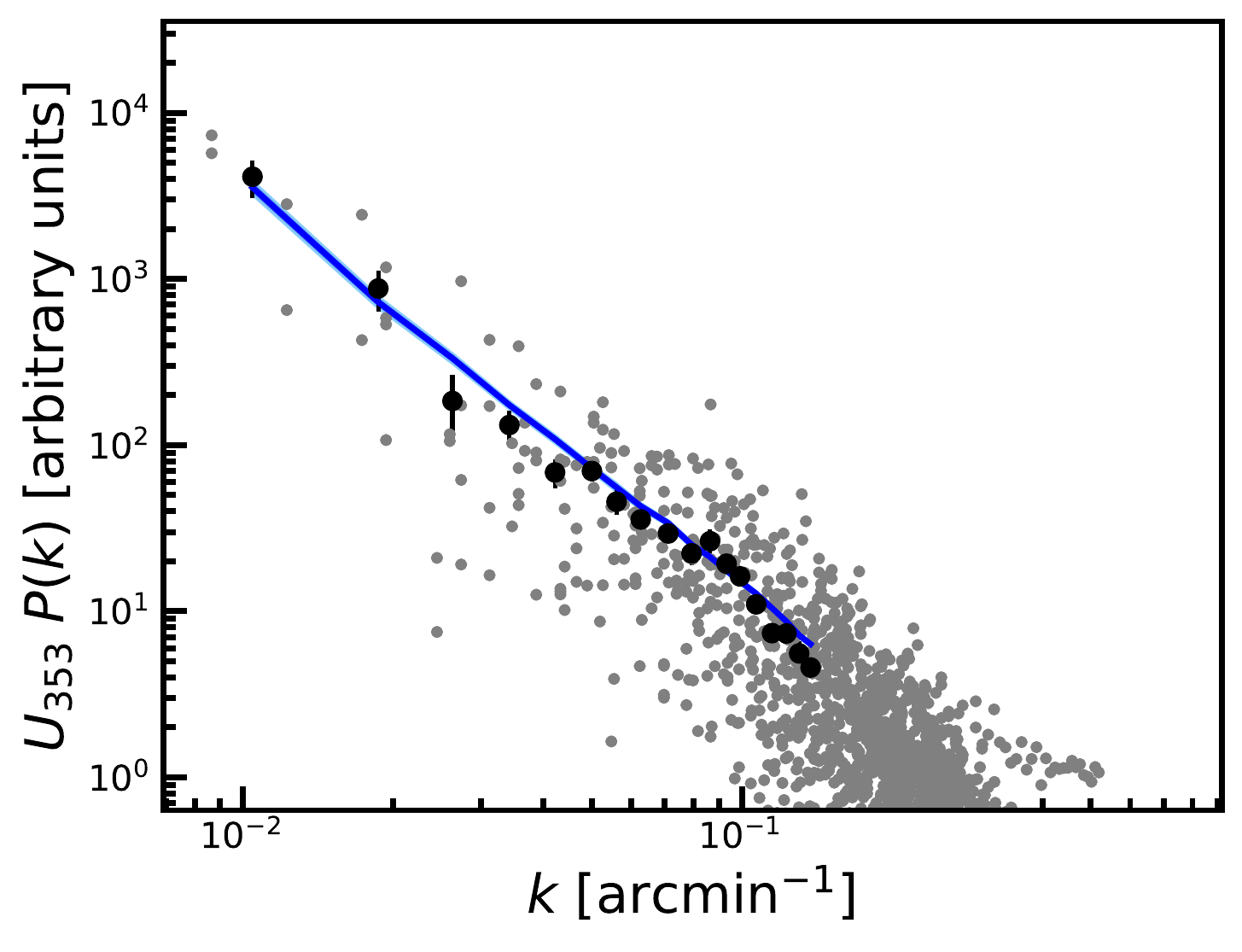} \\
\caption{Comparison of the $I_\mathrm{m}$, $Q_\mathrm{m}$ and $U_\mathrm{m}$ power spectra extracted from the {\Planck} Polaris Flare maps (gray points representing the two-dimensional power spectra, and black dots representing the azimuthal averages in Fourier space) with the observables computed from simulations using the best fitting parameters (blue curves). The latter curves are averaged over 60 realizations as described in section~\ref{sec:chi2} : the average is given by the central blue curve and the narrow shaded bands give the $1\sigma$ and $2\sigma$ standard deviation in each bin.}
\label{fig_best_fit_ps}
\end{figure*}

\begin{figure*}[htbp]
\centering
\includegraphics[width=0.33\textwidth]{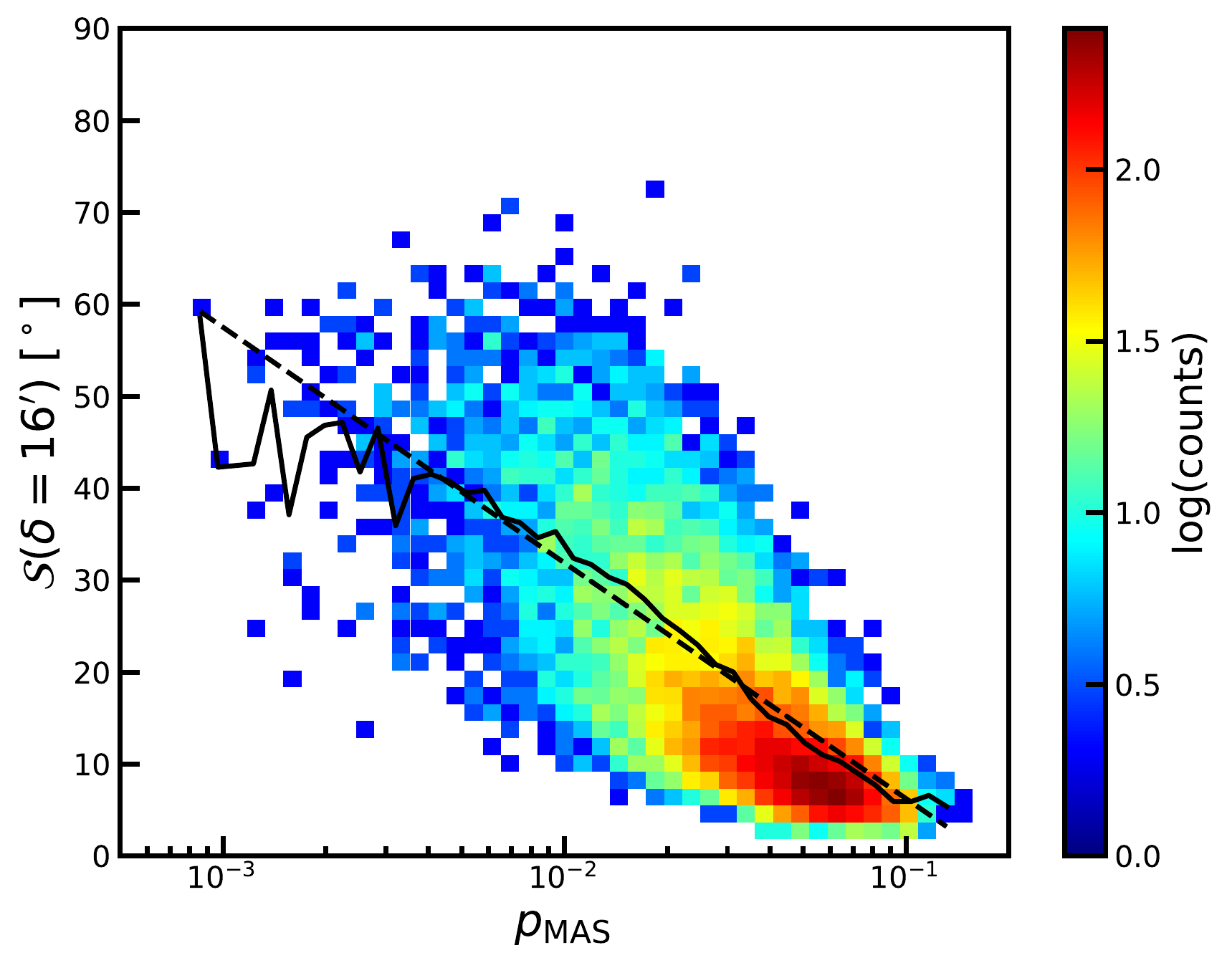}
\includegraphics[width=0.33\textwidth]{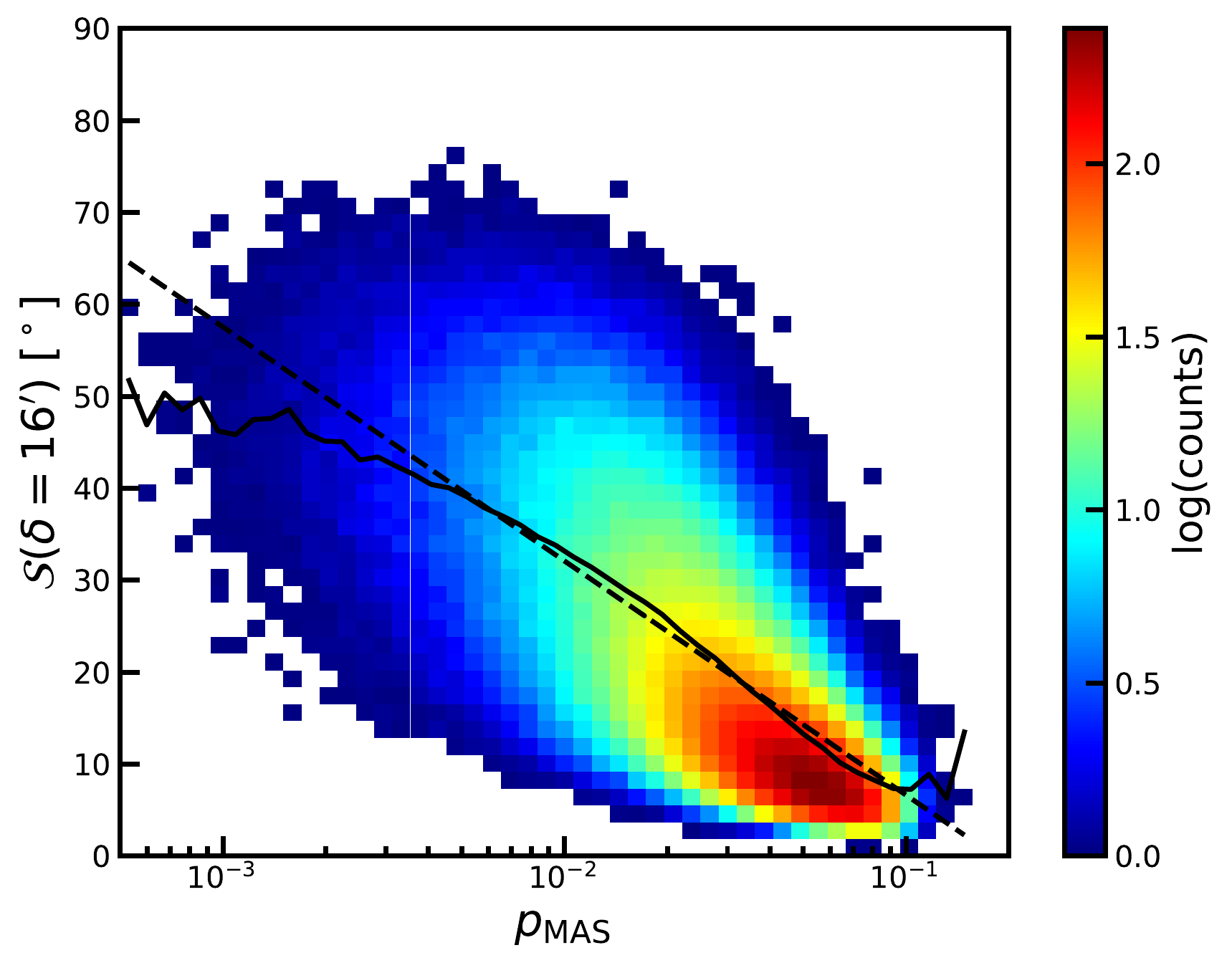}
\includegraphics[width=0.33\textwidth]{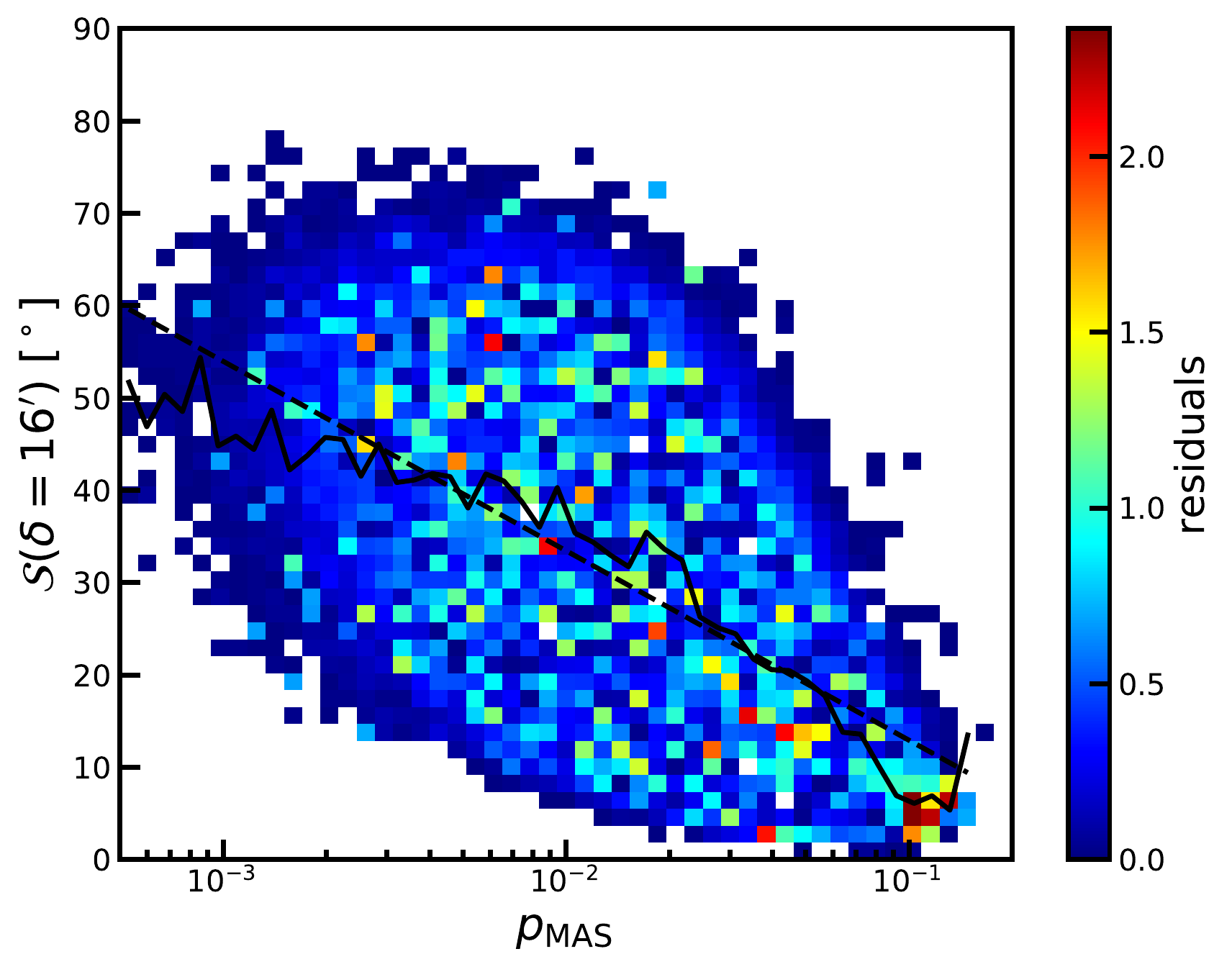}
\caption{Two-dimensional distribution function of $\mathcal{S}$ and polarization fraction $p_\mathrm{MAS}$ for the Polaris Flare maps (left), for the model maps using the best fitting parameters averaged over 60 realizations (middle), and residuals (right). The polarization angle dispersion function $\mathcal{S}$ is computed at a lag $\delta=16\arcmin$. The solid black line shows the mean $\mathcal{S}$ for each bin in $p_\mathrm{MAS}$ and the dashed black line is a linear fit of that curve, restricted to bins in $p_\mathrm{MAS}$ which contain at least 1\% of the total number of points (120$\times$120).}
\label{fig_best_fit_corr}
\end{figure*}

\subsection{Application to the Polaris Flare}
\label{sec:Polaris}

As an application of our method, we wish to constrain statistical properties of the turbulent magnetic field in the Polaris Flare, a diffuse, highly dynamical, non-starforming molecular cloud. There are several reasons for choosing this particular field. First, it has been widely observed : the structures of matter were studied in dust thermal continuum emission by, e.g.,~\cite{miville-deschenes-et-al-2010}; the velocity field of the molecular gas was studied down to very small scales through CO rotational lines~\citep{falgarone-et-al-1998,hily-blant-et-al-2009}; and the magnetic field was probed by optical stellar polarization data in~\cite{panopoulou-et-al-2016}. Second, as this field does not show signs of star formation, the dynamics of the gas and dust are presumably dominated by magnetized turbulence processes, without contamination by feedback from young stellar objects (YSOs). It therefore seems like an ideal test case for our method.

To this aim, we use the full-mission \Planck\ maps of Stokes parameters $(I_{353},Q_{353},U_{353})$ at 353\,GHz and, as already mentioned, the associated covariance matrices from the {\Planck} Legacy Archive. We also use the thermal dust model maps $\tau_{353}$ and $T_{\rm obs}$ from the 2013 public release~\citep{planck2013-p06b}. All maps are at a native $4\arcmin.8$ resolution in the {\tt Healpix} format with $N_\mathrm{side}=2048$, and the Polaris Flare maps are obtained by projecting these onto a Cartesian grid with 6$\arcmin$ pixels, centered on Galactic coordinates $(l,b)=(120^\circ,27^\circ)$, with a field of view $\Delta l=\Delta b=12^\circ$. The maps of $I_{353}$, $Q_{353}$, $U_{353}$, and $\tau_{353}$ are then smoothed using a circular Gaussian beam, to obtain maps at a 15$\arcmin$ FWHM resolution. The covariance matrix maps are computed at the same resolution, using a set of Monte-Carlo simulations of pure noise maps, drawn from the original full-resolution covariance maps and smoothed at 15\arcmin. The maps of $I_{353}$, $Q_{353}$, $U_{353}$, $p_\mathrm{MAS}$, $\psi$, and $\mathcal{S}$ obtained in this way are shown in Fig.~\ref{fig_polaris_map}. 

Note that the features of simulated Stokes maps are not located in the same regions as in the Polaris Flare maps. As the noise covariance matrices are the same for all the simulated maps, this means that the signal-to-noise ratio per pixel in the model $I_{\rm m}$, $Q_{\rm m}$ and $U_{\rm m}$ maps is different for each set of parameters. However, the MCMC procedure is able to choose the parameter sets that give signal-to-noise ratios similar to those in the {\Planck} maps.

\begin{table*}
\caption{Best fit values for the {\Planck} Polaris Flare maps, using the observables from Table~\ref{tab:observables}. The column $\left<\chi^2_{\rm{best}}\right>$ shows the $\chi^2$ values for the best fit parameters averaged over 100 fits (see Appendix~\ref{sec:appendix:chi2}).}
\label{tab:results}
\begin{center}
\resizebox{\textwidth}{!}{\begin{tabular}{ccccccccccc} \hline \hline \\ [-1ex]
Parameters  & $\beta_B$  & $\beta_n$ & $\log_{10}y_n$ & $\log_{10} y_B^{\rm POS}$ & $\chi_0$ [$^\circ$] & $\log_{10}\left(\frac{d}{1\,\mathrm{pc}}\right)$  & $\log_{10}\left(\frac{\langle\nH\rangle}{1\,\mathrm{cm}^{-3}}\right)$ & $T_\mathrm{d}$ [K] & $p_0$ & $\left<\chi^2_{\rm{best}}\right>$\ \\  [1ex] \hline \\ [-1ex]
Best fit values & 		$2.8^{+0.2}_{-0.2}$ & $1.7^{+0.4}_{-0.3}$ & $0.2^{+0.2}_{-0.2}$ & $-0.19^{+0.04}_{-0.04}$ & $-69^{+2}_{-3}$ & $1.1^{+0.3}_{-0.2}$ & $1.6^{+0.2}_{-0.3}$ & $17.5^{+0.5}_{-0.5}$ & $0.12^{+0.02}_{-0.02}$  & 2.9\\  [1ex]  \hline \\ [-1ex]
\end{tabular}}
\end{center}
\end{table*}

The results of the analysis of the {\Planck} polarized thermal dust emission data towards the Polaris Flare are presented in Table~\ref{tab:results}, and the posterior probability distribution contours are shown in Figure~\ref{fig_triangle}. We find in particular that the spectral index of the turbulent component of the magnetic field is $\beta_B=2.8\pm 0.2$, and that the spectral index of the dust density field is around $\beta_n=1.7$ with a rather large uncertainty. The fluctuation ratio of the density field is about unity, $y_n\approx 1.6$, and the magnitude of the large scale magnetic field in the POS dominates slightly the RMS of the turbulent component, $y_B^{\rm POS}\approx 0.65$, with a position angle $\chi_0 \approx -69^\circ$. The constraint on the depth of the cloud seems to indicate that $d\approx 13\,\mathrm{pc}$, with $\langle\nH\rangle\approx 40\,\mathrm{cm}^{-3}$. The temperature $T_\mathrm{d}$ is 17.5\,K equal to the average of the $T_{\rm obs}$ {\Planck} map, and the polarization fraction is $p_0\approx 0.12$. The parameter set for simulation A was chosen a posteriori to give similar best fit parameters and to test our likelihood method in the conditions driven by the Polaris Flare data.
 
Using the best fitting parameters from Table~\ref{tab:results} we performed simulations to visually check the agreement between the model and {\Planck} data. Figure~\ref{fig_best_fit_map} shows the polarization maps from a simulation using these best fitting parameters. The overall similarity with the data maps from Figure~\ref{fig_polaris_map} is reasonably good, although spatially coherent structures appear in the data maps which cannot be reproduced by the model maps. The agreement between the best fitting simulation and the data is quantified through plots of the different observables that were used in the fitting procedure (Figs.~\ref{fig_best_fit_pdf}, \ref{fig_best_fit_ps} and \ref{fig_best_fit_corr}). The agreement  is excellent for most observables, although substantial deviations are visible in the DFs of the intensity $I_{\rm m}$, of the normalized optical depth $\tau_{353}/\left\langle\tau_{353}\right\rangle$ and of the polarization angle $\psi$. These deviations are due to the simplifying assumptions of our fBm model. It may be that in the Polaris Flare the large scale magnetic field has two major components that superimposed, with global orientations $\chi_0 \approx -70^\circ$ and $\chi_0 \approx 50^\circ$. Note that the DF in Fig.~\ref{fig_best_fit_pdf} is that of the $\psi$ angle, which differs from $\chi_0$ by 90$^\circ$. Also the exponentiation procedure to model the dust field is a good but incomplete approximation of the reality and it is not able to totally reproduce the shapes of the $I_{\rm m}$ and $\tau_{353}/\left\langle\tau_{353}\right\rangle$ DFs together.  These deviations impact the reduced best fit $\left<\chi^2_{\rm{best}}\right> \approx 2.9$, which is somewhat larger than for mock data ($\approx 1$), but still reasonably good.

Concerning the mean values used as observables, the Polaris Flare has a mean optical depth of $\left\langle\tau_{353}\right\rangle = \left(1.25 \pm 0.05 \right)\times 10^{-5}$ and a mean temperature of $\left\langle T_{\rm obs} \right\rangle = 17.5 \pm 0.4$\,K. Using the best fitting parameters from Table~\ref{tab:results} we get optical depth maps with an average of $\left\langle\tau_{353}\right\rangle = \left(1.82 \pm 0.05 \right)\times 10^{-5}$ over 60 realizations which is in tension with the data value, as mentioned above for the $\tau_{353}/\left\langle\tau_{353}\right\rangle$ discrepancy. However the best fitting parameter for temperature is $T_\mathrm{d} = 17.5\pm 0.5$\,K which is exactly the same as in data with a width reflecting the data uncertainties. 

The observables we used to extract the statistical properties of the Polaris Flare field are by themselves unable to constrain the $\gamma_0$ angle of the large scale magnetic field on the LOS. However, the \cite{planck2016-XLIV} analysis was able to fit the $\chi_0$ and $\gamma_0$ angle is the southern Galactic cap and found an intrinsic polarization fraction of the gas of $p_\mathrm{int}\approx 0.26$. If we believe this latter value is true also in the Polaris Flare, then it is related to our fit as $p_0 \approx p_\mathrm{int} \left\langle\cos^2 \gamma \right\rangle \approx p_\mathrm{int} \cos^2 \gamma_0$. We can thus constrain the $\gamma_0$ angle to be around $45^\circ$.

\section{Discussion and summary}
\label{sec:discussion}

We have presented an analysis framework for maps of polarized thermal dust emission in the diffuse ISM aimed at constraining the statistical properties of the dust density and magnetic field responsible for this emission. Our framework rests on a set of synthetic models for the dust density and magnetic field, for which we precisely control the one- and two-point statistics, and on a least-squares analysis in which the space of parameters is explored via a MCMC method. The application of the method to {\Planck} maps of the Polaris Flare molecular cloud leads to a spectral index of the turbulent component of the magnetic field $\beta_B=2.8\pm 0.2$, which is in very good agreement with the findings of~\cite{planck2016-XLIV} and~\cite{vansyngel-et-al-2017}, who used a very different approach over a much larger fraction of the sky. The dust density field exhibits a much flatter spectrum, $\beta_n=1.7$. This latter exponent is remarkably close to the Kolmogorov index for the velocity field in incompressible hydrodynamical turbulence, but this comparison should be taken  with caution, as closer examination of the power spectrum of
the model density field shows a spectral break with an exponent closer to 2.2 at the largest scales ($k\lesssim 1\,\mathrm{pc}^{-1}$) while the smaller scales ($k\gtrsim 1\,\mathrm{pc}^{-1}$) have a 1.7 exponent\footnote{Incidentally, from the \Planck\ maps, we can measure the spectral index of the total intensity for the Polaris Flare to be $\beta_I=2.84\pm0.10$, in excellent agreement with the measurement by~\cite{stutzki-et-al-1998} on CO integrated emission at a similar angular resolution.}. What is clear is that the magnetic field power spectrum is much steeper, which underlines the role that the large scale magnetic field plays in the structure of polarized emission maps. We find that the fluctuation ratio of the dust density field and the ratio of turbulent-to-uniform magnetic field are both around unity. Finally, our analysis is able to give a constraint on the polarization fraction, $p_0 \approx 0.12$, and on the depth of the Polaris Flare molecular cloud, $d\approx 13\,\mathrm{pc}$, which is about half the transverse extent of the field-of-view, with $\langle\nH\rangle \approx 40 \,\mathrm{cm}^{-3}$. The good visual agreement between the Polaris Flare maps and model maps for the best-fitting parameters (Figs.~\ref{fig_polaris_map} and~\ref{fig_best_fit_map}), and the excellent agreement between the two sets of maps for most of the observables used in the analysis (Figs.~\ref{fig_best_fit_pdf}, \ref{fig_best_fit_ps} and \ref{fig_best_fit_corr}), all lead us to conclude that our fBm-based model, although limited, provides a reasonable description of the magnetized, turbulent, diffuse ISM.

In fact, it is quite remarkable to find such a good agreement with the data, considering the limitations of the model. First, it is statistically isotropic, and therefore cannot reproduce the interstellar filamentary structures observed at many scales and over a large range in column densities~\citep[see, e.g.][]{miville-deschenes-et-al-2010,arzoumanian-et-al-2011}. Second, our model dust density and magnetic fields are completely uncorrelated, which is clearly not realistic, as it was found that there is a preferential relative orientation between structures of matter and magnetic field, both in molecular clouds~\citep{planck2015-XXXV} and in the diffuse, high-latitude sky~\citep{planck2014-XXXII,planck2015-XXXVIII}. The change in relative orientation, from mostly parallel to mostly perpendicular, as the total gas column density $\NH$ increases, is also not reproducible with our fully-synthetic models. Third, it is now commonly acknowledged that two-point statistics
such as power spectra are not sufficient to properly describe the structure of interstellar matter. Improving our synthetic models along these three directions will be the subject of future work.

For completeness, we have also looked into applying our MCMC approach based on fBm models to synthetic polarization maps built from a numerical simulation of MHD turbulence. We used simulation cubes from {\tt http://www.mhdturbulence.com}~\citep{cho-lazarian-2003,burkhart-et-al-2009,burkhart-et-al-2014}, basing our choice on the simulation parameters, which seemed more or less consistent with the parameters found for the Polaris Flare data. We built simulated Stokes $I$, $Q$, and $U$ maps using the same resolution and noise parameters, and launched the MCMC analysis on these simulated Stokes maps. It turns out that the Markov chains have a much harder time converging than when applying the method to the {\Planck} data. It is not yet completely clear why that is so, but we suspect that part of the reason may lie with the limited range of spatial scales over which the fields in the MHD simulation can be accurately described by scale-invariant processes. Indeed, while the fBm models exhibit power-law power spectra over the full range of accessible scales (basically one decade in our case), the MHD simulations are hampered by effects of numerical dissipation at small scales (possibly over nearly 10 pixels), and the properties at large scales are dependent on the forcing, which is user-defined. The data, on the other hand, exhibit a much larger « inertial range ». In that respect, our fBm models, despite all their drawbacks, and despite the fact that they lack the physically realistic content of MHD simulations, provide a better framework for assessing the statistical properties of the {\Planck} data than current MHD simulations can. Of course, this conclusion is based on just one simulation, and there would definitely be a point in applying the MCMC approach to assess various instances of MHD simulations with respect to the observational data, based on the same observables, but independently of the grid of fBm models. This project, however, is clearly beyond the scope of this paper.

\begin{acknowledgements}
We gratefully acknowledge fruitful discussions with S. Plaszczynski and O. Perdereau.
\end{acknowledgements}
\bibliographystyle{aa}
\bibliography{biblio,Planck_bib}

\appendix

\section{Statistical properties of $\nH$ models}
\label{sec:appendix:nHproperties}

\subsection{Probability distribution function}
The PDF $f(\nH)$ of the density field $\nH$ built using Eq.~\ref{eq:expfbm} derives from the Gaussian PDF of $X$, which in all generality has a mean $\langle X\rangle$ and variance $\sigma_X^2$. We thus have a log-normal PDF
\begin{equation}
\label{eq:PDFn}
f(\nH)=\frac{1}{\sqrt{2\pi}\sigma_X}\frac{X_r}{\nH}\exp{\left\{-\frac{1}{2\sigma_X^2}\left[X_r\ln{\left(\frac{\nH}{n_0}\right)}-\langle X\rangle\right]^2\right\}},
\end{equation}
which is defined for $\nH>0$. Figure~\ref{fig:densitypdfs} presents the distribution function of the $\nH$ field\footnote{In that case, $\langle X\rangle=0$, $\sigma_X^2=1$, $X_r=1.2$, and $n_0=20\,\mathrm{cm}^{-3}$. } used to build Fig.~\ref{fig:densitymaps}, with the theoretical PDF expected from Eq.~\ref{eq:PDFn}. It should be noted that the distribution function for a single realisation over a finite grid such as the ones used here may deviate from the theoretical PDF, especially for large values of $\beta_X$, but the mean distribution function over a sufficiently large sample converges to the lognormal form (Eq.~\ref{eq:PDFn}).

\subsection{Moments and fluctuation level}
\begin{figure}[htbp]
\includegraphics[width=9.5cm]{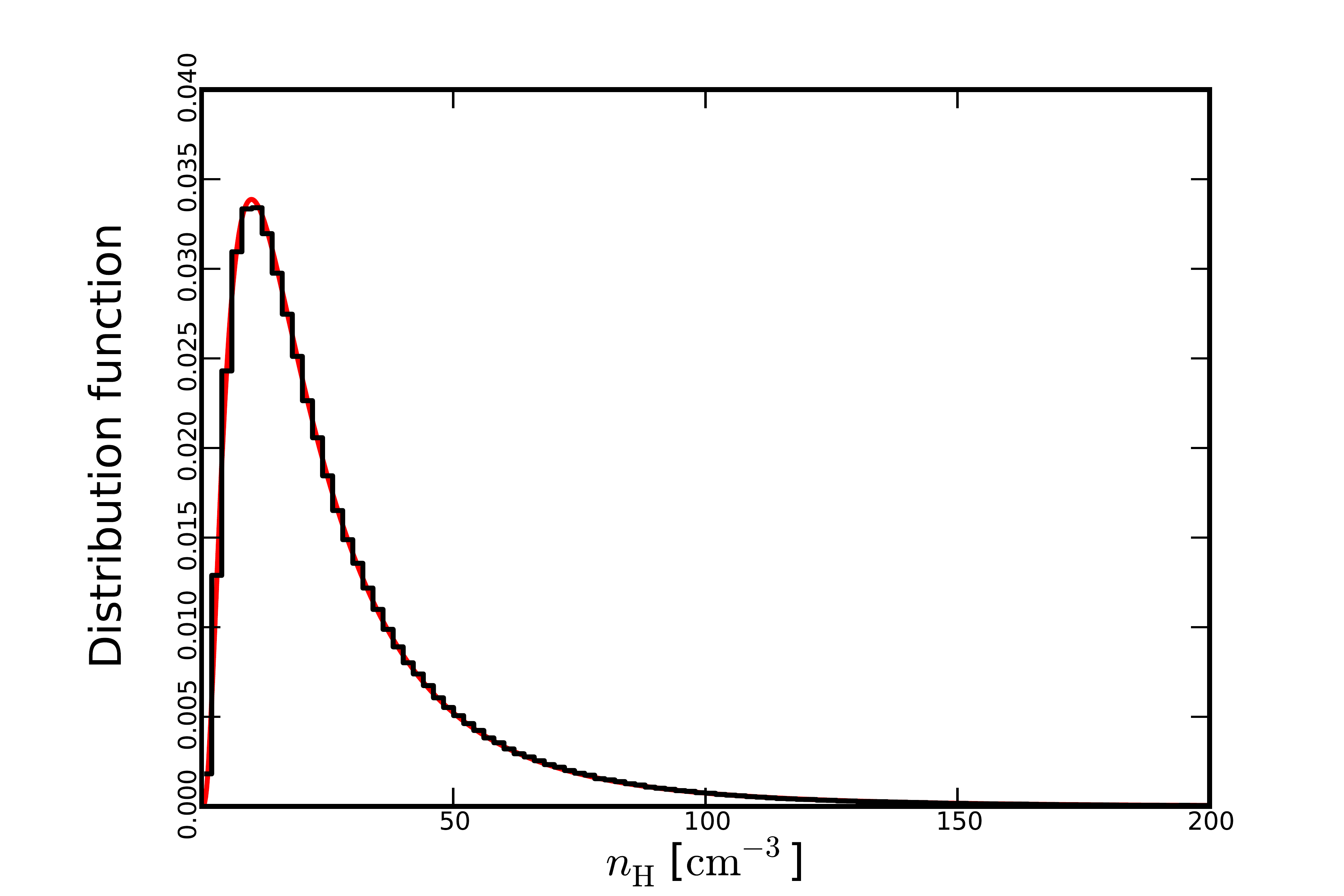}
\caption{Distribution function of the synthetic field $\nH$ used for Fig.~\ref{fig:densitymaps} (black histogram). The red curve shows the theoretical probability distribution function $f(\nH)$ for the chosen set of parameters.}   
\label{fig:densitypdfs}
\end{figure}

From Eq.~\ref{eq:PDFn}, we may compute moments of any order $p$ of the PDF of the total gas density $\nH$
\begin{equation*}
\left<\nH^p\right>=\int\nolimits_{0}^{\infty} \nH^p f(\nH)\ud \nH=n_0^p\exp{\left(p\frac{\left<X\right>}{X_r}+p^2\frac{\sigma_X^2}{2X_r^2}\right)},
\end{equation*}
which allows, in particular, to compute its mean value
$$
\left<\nH\right>=\int\nolimits_{0}^{\infty} \nH f(\nH)\ud \nH=n_0\exp{\left(\frac{\left<X\right>}{X_r}+\frac{\sigma_X^2}{2X_r^2}\right)},
$$
as well as its variance
\begin{equation*}
\sigma_{\nH}^2=n_0^2\exp{\left(2\frac{\left<X\right>}{X_r}\right)}\left[\exp{\left(\frac{2\sigma_X^2}{X_r^2}\right)}-\exp{\left(\frac{\sigma_X^2}{X_r^2}\right)}\right].
\end{equation*}
The density fluctuation level, which is one of the parameters of our model, is therefore
\begin{equation*}
y_n=\frac{\sigma_{\nH}}{\langle\nH\rangle}=\sqrt{2}\exp{\left(\frac{\sigma_X^2}{4X_r^2}\right)}\left[\sinh{\left(\frac{\sigma_X^2}{2X_r^2}\right)}\right]^{1/2}.
\end{equation*}
For instance, the $\nH$ field whose distribution function is shown in Fig.~\ref{fig:densitypdfs} has values ranging from 0.3\,$\mathrm{cm}^{-3}$ to $880\,\mathrm{cm}^{-3}$, with a mean and standard deviation of $\langle\nH\rangle=\sigma_{\nH}=28.3\,\mathrm{cm}^{-3}$, resulting in the desired fluctuation level $y_n=1$. 

\subsection{Power spectra}
\begin{figure}[htbp]
\resizebox{\hsize}{!}{
\includegraphics{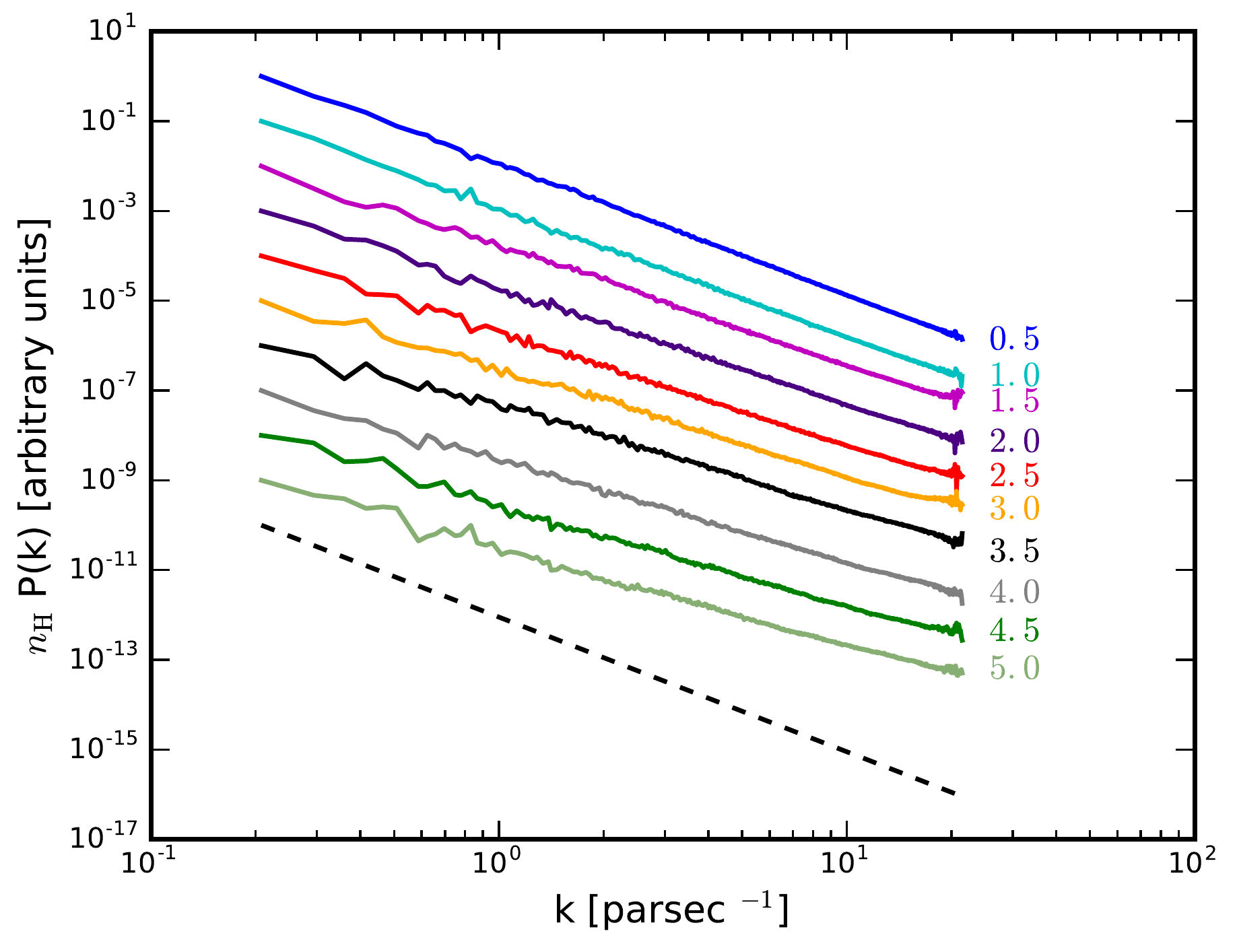}
}
\caption{Power spectra $P_{\nH}(k)$ of synthetic $\nH$ fields obtained by exponentiation of a $120 \times 120 \times 120$ pixels fractional Brownian motion with spectral index $\beta_X=3$. The fluctuation levels $y_n$ are specified next to each curve, and the original field's power spectrum $P_{\nH}(k)$ is represented as a dashed line.}
\label{fig_11}
\end{figure}

Figure~\ref{fig_11} shows the azimuthally-averaged power spectra of $\nH$ fields obtained through Eq.~\ref{eq:expfbm}, from a $120 \times 120 \times 120$ pixels fractional Brownian motion $X$ with spectral index $\beta_X=3$, for various fluctuation levels $y_n$. The spectra were normalized differently so as to allow comparison between them. 

\begin{figure}[htbp]
\includegraphics[width=9.5cm]{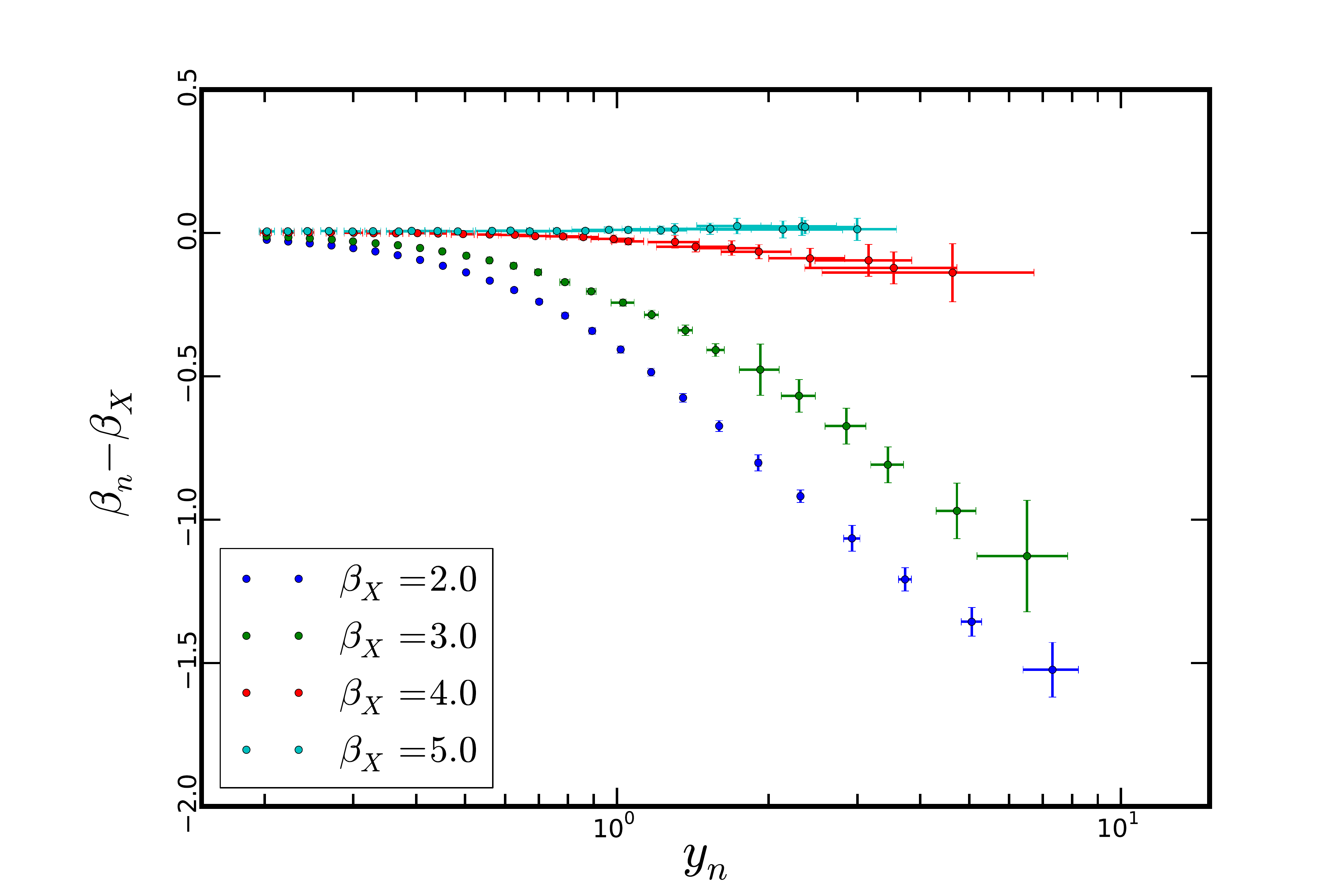}
\caption{Evolution of the differences $\beta_n-\beta_X$ between the spectral indices of the original fBm field $X$ and that of the model density field $\nH$ with fluctuation level $y_n$. Each point corresponds to the mean of 20 realisations of the model density field $\nH$, and the error bars represent the standard deviation of the fitted spectral indices $\beta_n$ and fluctuation levels $y_n$. }
\label{fig:beta_nH_with_yn}
\end{figure}

The power-law behaviour is apparent, even at large fluctuation levels, but the spectral index decreases (i.e., the spectra flatten) as the fluctuation level increases. This is quantified in Fig.~\ref{fig:beta_nH_with_yn}, which shows the differences $\beta_n-\beta_X$ between the spectral indices of the original fBm field $X$ and that of the model density field $\nH$. The power spectra of the latter are indeed flatter than the original ones ($\beta_n<\beta_X$), with differences that may become large when $\beta_X$ is low, but remain negligible for higher $\beta_X$. We interpret this trend as the exponentiation process amplifying the two-point differences $X(\boldsymbol{r}+\boldsymbol{\delta})-X(\boldsymbol{r})$, for small separations $||\boldsymbol{\delta}||$, that exist in the original field when $\beta_X$ is low, leading to an increase of the small-scale power in the model $\nH$ field, i.e. $\beta_n<\beta_X$. This effect is all the more important than exponentiation stretches these field differences more strongly, i.e. when $y_n$ increases. On the other hand, for high $\beta_X$, the fBm fields are much smoother, so the exponentiation process has little impact on these two-point statistics at small scales, leading to $\beta_n\simeq \beta_X$. Note that at low fluctuation levels ($y_n\leqslant 0.3$), the differences are smaller than 0.1 for all values of $\beta_X$. 

\section{Likelihood terms}
\label{sec:appendix:L2distance}

\subsection{$D^2$ terms for mean values}
\label{sec:chi2mean}

The first term in Eq.~\ref{eq:chi2tot}  is given by
\begin{equation*}
D^2_\mu =\frac{\left(\left\langle \tau_{353}^{\rm m} \right\rangle - \left\langle \tau_{353}^{\rm d} \right\rangle \right)^2}{\sigma_{\left\langle \tau_{353}^{\rm m} \right\rangle}^2 + {\sigma_{\left\langle \tau_{353}^{\rm d} \right\rangle}^2}}+\frac{\left( \left\langle T_{\rm obs}\right\rangle - T_\mathrm{d} \right)^2}{\sigma_{\left\langle T_{\rm obs}\right\rangle}^2}.
\end{equation*}

The mean value of the optical depth $\tau_{353}$ is evaluated on the simulated $ \tau_{353}^{\rm m}$  and {\Planck} data $ \tau_{353}^{\rm d} $  maps. These are compared through a standard $\chi^2$ test. The denominator includes the uncertainty $\sigma_{\left\langle \tau_{353}^{\rm d} \right\rangle}$ on the mean $\tau^{\rm d}_{353}$ value propagated from the uncertainty map provided by the {\Planck} collaboration~\citep{planck2013-p06b}, and the uncertainty $\sigma_{\left\langle \tau_{353}^{\rm m} \right\rangle}$ coming from the conversion factor $\sigma_{353}\left(\NH\right)$ used to build the simulated map~\citep{planck2013-p06b}. 

In the second term, $\left\langle T_{\rm obs}\right\rangle$ is the mean value of the temperature map $T_{\rm obs}$ from {\Planck} data, and  $\sigma_{\left\langle T_{\rm obs}\right\rangle}$ represents the uncertainty on the averaged value propagated from the uncertainty map provided by the {\Planck} collaboration. $T_\mathrm{d}$ is directly the model parameter for the dust temperature.

\subsection{$D^2$ terms for distribution functions}
\label{sec:chi2DF}

For a given observable map $o$, we compute its DF over an ensemble of $N_{\rm b}$ bins\footnote{We set $N_{\rm b}=50$ for the $120 \times 120$ pixel maps.}. When considering the {\Planck} data, we write this DF as $h_{o,i}^\mathrm{d}$, where $i$ is the bin number, and we estimate the uncertainty on the value of the DF in bin $i$ through the associated Poisson noise $\sigma_{h_{o,i}^\mathrm{d}}$. When considering the model, we write $h_{o,i}^\mathrm{m}$ and $\sigma_{h_{o,i}^\mathrm{m}}$ to be respectively the bin value and the Poisson noise of the DF in bin $i$, independently for each of the $N_r=60$ model realizations.

The contribution $D^2_{{\rm DF}(o)}$ of the observable's DF to the total $D^2$ in Eq.~\ref{eq:chi2tot} is then built as an average over the $N_{\rm b}$ bins
\begin{equation}
\label{eq:D2DF}
D^2_{{\rm DF}(o)} = \frac{1}{N_{\rm b}}  \sum_{i=1}^{N_{\rm b}}  \frac{\left(h_{o,i}^\mathrm{d} - h_{o,i}^\mathrm{m} \right)^2}{(\sigma_{h_{o,i}^\mathrm{d}})^2 + (\sigma_{h_{o,i}^\mathrm{m}})^2}
\end{equation}
and averaged over the $N_{\rm r}=60$ realizations. The inner sum is normalized to the number of bins so that $D^2_{{\rm DF}(o)}$ is less sensitive to the binning choice and the map noise. If the model fits the data correctly as far as the DF of observable $o$ is concerned, then $D^2_{{\rm DF}(o)}$ is minimum. 

This quantity is different from a standard $\chi^2$ test as it compares data with one random\footnote{We recall that the randomness comes from the fBm itself and the noise addition to the Stokes maps.} realisation for a given set of model parameters, and this comparison is repeated and averaged $N_r$ times. A standard $\chi^2$ test would compare data with a model prediction that would be the average of the random realizations (see Eq~\ref{eq:chi2DF} in Appendix~\ref{sec:appendix:chi2}). The latter could not be used in our MCMC analysis due to the mathematical relation between the power spectrum of the map and the variance in each bin of the DF : with steep power spectra (high values of the spectral index $\beta$), only the few large-scale modes effectively contribute to the power, leading to a large variance in each DF bin\footnote{This is akin to the cosmic variance problem in cosmology.}. Thus the $\chi^2$ test tends to favor these large values of $\beta$, as they yield large denominators and thus allow for a "good" fit. This drives the fit towards a region of parameter space yielding mean DFs that fit the data well but with a huge dispersion: one realization of such a model gives a DF with bin values highly scattered even though the data DF is quite smooth. This means that the data cannot reasonably be interpreted as a random realization using these parameter values, and explains why we had to switch to the $D^2$ function, which directly compares data with one model random realization. In this fashion, we are able to reach the region of parameter space correctly describing the data.

\subsection{$D^2$ terms for power spectra}

For a given observable map $o$, we first compute its two-dimensional power spectrum 
$$
P_o(\boldsymbol{k})=\left|\,\widetilde{o}(\boldsymbol{k})\right|^2
$$
then average these within $N'_{\rm b}$ annuli in Fourier space, centred on a set of wavenumbers $\left\{k_i\right\}$. We write $P_{o,i}^\mathrm{d}$ for this azimuthal average in bin $i$ when considering the {\Planck} data, and $P_{o,i}^\mathrm{m}$ when considering model fields. 
The uncertainties affecting these quantities strongly depend on the number $N_i$ of wavevectors $\boldsymbol{k}_n$ in each bin. The best estimate for the standard deviation $\sigma_{P_{o,i}^\mathrm{d}}$ of the power spectrum of the data is
\begin{equation*}
\sigma_{P_{o,i}^\mathrm{d}} = t_{N_i}\sqrt{\frac{1}{N_i\left(N_i-1\right)}\sum_{n=1}^{N_i} \left[ P_{o}^{d} (\boldsymbol{k}_n) - P_{o,i}^d \right]^2}
\end{equation*}
where the factor $t_N$ is the Student coefficient. We thus obtain the best estimate of the true standard deviation in bins with only a few modes (i.e., at large scale). The standard deviation $\sigma_{P_{o,i}^\mathrm{m}}$ for the 60 model realizations is computed in the same way as for data.

The contribution $D^2_{P(o)}$ of the observable's power spectrum to the total $D^2$ in Eq.~\ref{eq:chi2tot} is then computed as a sum over the $N'_{\rm b}$ bins\footnote{To have reliable and smooth power spectra with $120\times 120$ pixel maps, we set initially $N'_{\rm b}=100$ but later cut off wavenumbers larger than $k_{\rm{max}}=2\pi/(3\times 15\arcmin)$, which corresponds to scales smaller than 3 beam sizes. Indeed, for bins $k_i > k_{\rm{max}}$ the power spectrum is completely washed out by the beam convolution (see Fig.~\ref{fig_maps_and_ps_simA}) and contains no information about the underlying interesting parameters. This uninformative part is thus removed from $D^2_{P(o)}$,  and thus $N'_{\rm b}<100$.} in wavenumber space, 
\begin{equation*}
\label{eq:chi2PS}
D^2_{P(o)} =\frac{1}{N'_{\rm b}} \sum_{i=1}^{N'_{\rm b}} \frac{\left(P_{o,i}^\mathrm{d} - P_{o,i}^\mathrm{m} \right)^2}{(\sigma_{P_{o,i}^\mathrm{d}})^2 + (\sigma_{P_{o,i}^\mathrm{m}})^2}.
\end{equation*}

\subsection{$D^2$ term for the $\left\lbrace\mathcal{S},p_\mathrm{MAS}\right\rbrace$ anti-correlation}

To use the $\left\lbrace\mathcal{S},p_\mathrm{MAS}\right\rbrace$ anti-correlation~\citep{planck2014-XIX,planck2014-XX}, we compute the joint distribution function of the $\mathcal{S}$ and $p_\mathrm{MAS}$ maps, which we write $h_{ij}^\mathrm{d}$ and $h_{ij}^\mathrm{m}$ for the {\Planck} data and model maps respectively, with $1\leqslant i\leqslant N_{{\rm b},1}$ and $1\leqslant j\leqslant N_{{\rm b},2}$ the binning scheme used for the two maps. The standard deviations $\sigma_{ij}^\mathrm{d}$ and $\sigma_{ij}^\mathrm{m}$ are defined in the same way as for the one-dimensional DFs in \ref{sec:chi2DF}, and the contribution $D^2_{\mathcal{S}-p_\mathrm{MAS}}$ to the total $D^2$ is then
\begin{equation*}
D^2_{\mathcal{S}-p_\mathrm{MAS}} = \frac{1}{N_{\rm b,tot}} \sum_{i=1}^{N_{\rm b,1}} \sum_{j=1}^{N_{\rm b,2}} \frac{\left(h_{ij}^\mathrm{d} - h_{ij}^\mathrm{m} \right)^2}{(\sigma_{ij}^\mathrm{d})^2 + (\sigma_{ij}^\mathrm{m})^2}.
\end{equation*}
In this expression, it should be noted that the total number $N_{\rm b,tot}$ of two-dimensional bins considered is less than the product $N_{\rm b,1}N_{\rm b,2}$ of the number of bins in each dimension, which we set to $N_{\rm b,1}=N_{\rm b,2}=50$. The reason for this is that we discard the empty bins and those with a signal-to-noise ratio below three\footnote{i.e., bins where $h_{ij}^\mathrm{d}=0$, $h_{ij}^\mathrm{m}=0$,  $h_{ij}^\mathrm{d}/\sigma_{ij}^\mathrm{d}<3$, or $h_{ij}^\mathrm{m}/\sigma_{ij}^\mathrm{m}<3$.}. We thus keep only the significantly populated bins that can drive the fit and contribute to the total $D^2$.

\section{Goodness-of-fit}
\label{sec:appendix:chi2}
To assess the goodness of the fit, we use an {\it a posteriori} $\chi^2$ test, which we define as 
 \begin{equation}
 \label{eq:chi2tot_gof}
\chi^2 = \frac{1}{N_o}\left[ \chi^2_{ \mu} + \sum_{o}\chi^2_{{\rm DF}(o)} + \sum_{o} \chi^2_{P(o)}  + \chi^2_{\mathcal{S}-p_\mathrm{MAS}}\right]
\end{equation}
with $N_o=13$ the total number of observables. Each term is a $\chi^2$ test comparing data with the mean of the $N_r=60$ realisations. The first term from Eq~\ref{eq:chi2tot_gof} is
\begin{equation*}
\chi^2_\mu = \frac{\left(\overline{\left\langle \tau_{353}^{\rm m} \right\rangle} - \left\langle \tau_{353}^{\rm d} \right\rangle \right)^2}{\Sigma_{\left\langle \tau_{353}^{\rm m} \right\rangle}^2 + {\sigma_{\left\langle \tau_{353}^{\rm d} \right\rangle}^2}} +\frac{\left( \left\langle T_{\rm obs}\right\rangle - T_\mathrm{d} \right)^2}{\sigma_{\left\langle T_{\rm obs}\right\rangle}^2}
\end{equation*}
where $\overline{\left\langle \tau_{353}^{\rm m} \right\rangle} $ is the ensemble average over the 60 model realisations of the (spatial) mean of the optical depth. In the following, the brackets stand for an average on the pixels while the upper bar represents the average over the $N_r$ realisations. The $\Sigma_{\left\langle \tau_{353}^{\rm m} \right\rangle}^2$ quantity is the variance of $\left\langle \tau_{353}^{\rm m} \right\rangle$ over the $N_r$ realisations (capital $\Sigma$ denotes the variance over the random realisations). 

The second term is 
\begin{equation}\label{eq:chi2DF}
\chi^2_{\rm{DF}(o)} = \frac{1}{N_{\rm b}} \sum_{i=1}^{N_{\rm b}} \frac{\left(h_{o,i}^\mathrm{d} - \overline{ h_{o,i}^\mathrm{m}} \right)^2}{(\sigma_{h_{o,i}}^\mathrm{d})^2 + (\Sigma_{h_{o,i}}^\mathrm{m})^2}
\end{equation}
where 
\begin{equation*}
\overline{ h_{o,i}^\mathrm{m}}=\frac{1}{N_r}\sum_{k=1}^{N_r}h_{o,i}^\mathrm{m}
\end{equation*}
is the ensemble average of the $i^\mathrm{th}$ bin of the DF for the observable $o$, over the $N_r=60$ model realizations, and $(\Sigma_{ h_{o,i}}^\mathrm{m})^2$ is the associated variance. Note that while $D^2_{\rm DF(o)}$ in Eq.~\ref{eq:D2DF} is the average of the observable $\chi^2$, $\chi^2_{\rm{DF}(o)}$ is the $\chi^2$ of the averaged observable.

The third term is
\begin{equation*}
\chi^2_{P(o)} = \frac{1}{N'_{\rm b}} \sum_{i=1}^{N'_{\rm b}} \frac{\left(P_{o,i}^\mathrm{d} - \overline{P_{o,i}^\mathrm{m}} \right)^2}{(\sigma_{P_{o,i}^\mathrm{d}})^2 + (\Sigma_{{P_{o,i}^\mathrm{m}}})^2}
\end{equation*}
where $\overline{P_{o,i}^\mathrm{m}}$ is the averaged power spectrum in bin $i$ and $(\Sigma_{{P_{o,i}^\mathrm{m}}})^2$ its variance.

Finally, the fourth term is
\begin{equation*}
\chi^2_{\mathcal{S}-p_\mathrm{MAS}} = \frac{1}{N_{\rm b,tot}} \sum_{i=1}^{N_{\rm b,1}} \sum_{j=1}^{N_{\rm b,2}} \frac{\left(h_{ij}^\mathrm{d} - \overline{h_{ij}^\mathrm{m}} \right)^2}{(\Sigma_{ij}^\mathrm{d})^2 + (\sigma_{ij}^\mathrm{m})^2}
\end{equation*}
with the same notation conventions as above.

To quantify the goodness of fit, once the MCMC procedure has converged, we perform 100 fits for the set of best fitting parameters, each of these fits comprising 60 model realizations and providing a value of the $\chi^2$ quantity defined in equation~\ref{eq:chi2tot_gof}. The average of these 100 $\chi^2$ values\footnote{Each simulation is a random realization of the fBm and the noise, then the $\chi^2$ values have an intrinsic dispersion even when compiling 60 simulations. To check the agreement with data, we thus average 100 computations of the best $\chi^2$ value.}  is listed as $\left<\chi^2_{\rm{best}}\right>$ in Tables~\ref{tab:results_sim} and~\ref{tab:results}.

\end{document}